\documentclass[iicol,sn-mathphys-num, pdflatex]{sn-jnl}

\usepackage{graphicx}%
\usepackage{multirow}%
\usepackage{amsmath,amssymb,amsfonts}%
\usepackage{amsthm}%
\usepackage{mathrsfs}%
\usepackage[title]{appendix}%
\usepackage{xcolor}%
\usepackage{textcomp}%
\usepackage{manyfoot}%
\usepackage{booktabs}%
\usepackage{algorithm}%
\usepackage{algorithmicx}%
\usepackage{algpseudocode}%
\usepackage{listings}%
\usepackage{lmodern}
\usepackage{siunitx}
\usepackage{etoolbox}
\usepackage[T1]{fontenc}
\usepackage{tabularx}
\usepackage{float}
\usepackage[all]{hypcap}
\usepackage{hyperref}
\usepackage[maxfloats=256]{morefloats}
\usepackage{tabularx} 
\begin{document}

\title[Article Title]{Towards A Generalizable Pathology Foundation Model via Unified Knowledge Distillation
}
\author[1]{\fnm{Jiabo} \sur{Ma}}\email{jmabq@connect.ust.hk}
\equalcont{These authors contributed equally to this work.}
\author[1]{\fnm{Zhengrui} \sur{Guo}}\email{zguobc@connect.ust.hk}
\equalcont{These authors contributed equally to this work.}
\author[1]{\fnm{Fengtao} \sur{Zhou}}\email{fzhouaf@connect.ust.hk}
\author[1]{\fnm{Yihui} \sur{Wang}}\email{ywangrm@connect.ust.hk}
\author[1]{\fnm{Yingxue} \sur{Xu}}\email{yxueb@connect.ust.hk}
\author[2,3]{\fnm{Jinbang} \sur{Li}} \email{lzcy2008@126.com}
\author[4]{\fnm{Fang} \sur{Yan}} \email{yanfang@pjlab.org.cn}
\author[5]{\fnm{Yu} \sur{Cai}}\email{yu.cai@connect.ust.hk}
\author[6]{\fnm{Zhengjie}\sur{Zhu}}\email{zzhuar@connect.ust.hk}
\author[1]{\fnm{Cheng} \sur{Jin}}\email{cheng.jin@connect.ust.hk}
\author[1]{\fnm{Yi} \sur{Lin}}\email{yi.lin@connect.ust.hk}
\author[1]{\fnm{Xinrui} \sur{Jiang}}\email{csexrjiang@ust.hk}
\author[2,3,7]{\fnm{Chenglong} \sur{Zhao}} \email{zcl.125@163.com}
\author[2,3]{\fnm{Danyi} \sur{Li}} \email{lidanyi26@163.com}
\author[8]{\fnm{Anjia}\sur{Han}} \email{hananjia@mail.sysu.edu.cn}
\author[9]{\fnm{Zhenhui} \sur{Li}}\email{lizhenhui@kmmu.edu.cn}
\author[10]{\fnm{Ronald Cheong Kin}\sur{Chan}} \email{ronaldckchan@cuhk.edu.hk}
\author[11,12]{\fnm{Jiguang} \sur{Wang}} \email{jgwang@ust.hk}
\author[13]{\fnm{Peng} \sur{Fei}} \email{feipeng@hust.edu.cn }
\author[1,5]{\fnm{Kwang-Ting} \sur{Cheng}} \email{timcheng@ust.hk}
\author*[4,14]{\fnm{Shaoting} \sur{Zhang}} \email{zhangshaoting@pjlab.org.cn}
\author*[2,3,15]{\fnm{Li} \sur{Liang}}\email{lli@smu.edu.cn}
\author*[1,11,12,16,17]{\fnm{Hao} \sur{Chen}}\email{jhc@cse.ust.hk}

\affil[1]{\orgdiv{Department of Computer Science and Engineering}, \orgname{The Hong Kong University of Science and Technology}, \orgaddress{\state{Hong Kong SAR}, \country{China}}}
\affil[2]{\orgdiv{Department of Pathology, Nanfang Hospital and School of Basic Medical Sciences}, \orgname{Southern Medical University},
\orgaddress{\state{Guangzhou}, \country{China}}}
\affil[3]{\orgname{Guangdong Provincial Key Laboratory of Molecular Tumor Pathology},
\orgaddress{\state{Guangzhou}, \country{China}}}
\affil[4]{\orgname{Shanghai Artificial Intelligence Laboratory}, \orgaddress{\state{Shanghai}, \country{China}}}
\affil[5]{\orgdiv{Department of Electronic and Computer Engineering}, 
\orgname{The Hong Kong University of Science and Technology},
\orgaddress{\state{Hong Kong SAR}, \country{China}}}
\affil[6]{\orgdiv{Information Hub}, 
\orgname{The Hong Kong University of Science and Technology (Guangzhou)},
\orgaddress{\state{Guangzhou}, \country{China}}}
\affil[7]{\orgdiv{Department of Pathology}, 
\orgname{The First Affiliated Hospital of Shandong First Medical University and Shandong Provincial Qianfoshan Hospital},
\orgaddress{\state{Jinan}, \country{China}}}
\affil[8]{\orgdiv{Department of Pathology, The First Affiliated Hospital}, 
\orgname{Sun Yat-sen University},
\orgaddress{\state{Guangzhou}, \country{China}}}
\affil[9]{\orgdiv{Department of Radiology}, 
\orgname{The Third Affiliated Hospital of Kunming Medical University},
\orgname{Yunnan Cancer Hospital},
\orgaddress{\state{Kunming}, \country{China}}}
\affil[10]{\orgdiv{Department of Anatomical and Cellular Pathology}, 
\orgname{The Chinese University of Hong Kong},
\orgaddress{\state{Hong Kong SAR}, \country{China}}}
\affil[11]{\orgdiv{Department of Chemical and Biological Engineering}, \orgname{The Hong Kong University of Science and Technology}, \orgaddress{\state{Hong Kong SAR}, \country{China}}}
\affil[12]{\orgdiv{Division of Life Science}, \orgname{The Hong Kong University of Science and Technology}, \orgaddress{\state{Hong Kong SAR}, \country{China}}}
\affil[13]{\orgdiv{School of Optical Electronic Information}, \orgname{Huazhong University of Science and Technology}, \orgaddress{\state{Wuhan}, \country{China}}}
\affil[14]{\orgdiv{Qing Yuan Research Institute}, \orgname{Shanghai Jiao Tong University}, \orgaddress{\state{Shanghai}, \country{China}}}
\affil[15]{\orgname{Jinfeng Laboratory},
\orgaddress{\state{Chongqing}, \country{China}}}
\affil[16]{\orgdiv{State Key Laboratory of Molecular Neuroscience}, \orgname{The Hong Kong University of Science and Technology}, \orgaddress{\state{Hong Kong SAR}, \country{China}}}
\affil[17]{\orgdiv{Shenzhen-Hong Kong Collaborative Innovation Research Institute}, \orgname{The Hong Kong University of Science and Technology}, \orgaddress{\state{Shenzhen}, \country{China}}}


\abstract{
Foundation models pretrained on large-scale datasets are revolutionizing the field of computational pathology (CPath).
The generalization ability of foundation models is crucial for the success in various downstream clinical tasks. 
However, current foundation models have only been evaluated on a limited type and number of tasks, leaving their generalization ability and overall performance unclear. 
To address this gap, we established a most comprehensive benchmark to evaluate the performance of off-the-shelf foundation models across six distinct clinical task types, encompassing a total of 72 specific tasks, including slide-level classification, survival prediction, ROI-tissue classification, ROI retrieval, visual question answering, and report generation. 
Our findings reveal that existing foundation models excel at certain task types but struggle to effectively handle the full breadth of clinical tasks.
To improve the generalization of pathology foundation models, 
we propose a unified knowledge distillation framework consisting of both expert and self-knowledge distillation, where the former allows the model to learn from the knowledge of multiple expert models, while the latter leverages self-distillation to enable image representation learning via local-global alignment.
Based on this framework, we curated a dataset of 96,000 whole slide images (WSIs) and developed a Generalizable Pathology Foundation Model (GPFM). 
This advanced model was trained on a substantial dataset comprising 190 million images extracted from approximately 72,000 publicly available slides, encompassing 34 major tissue types. 
Evaluated on the established benchmark, GPFM achieves an impressive average rank of 1.6, with 42 tasks ranked 1st, while the second-best model, UNI, attains an average rank of 3.7, with only 6 tasks ranked 1st.
The superior generalization of GPFM demonstrates its exceptional modeling capabilities across a wide range of clinical tasks, positioning it as a new cornerstone for feature representation in CPath. 
}




\maketitle
\section{Introduction}\label{sec1}
Pathology plays a crucial and evolving role in modern medicine, providing essential insights for the diagnosis, treatment, and prognosis of diseases \cite{bejnordi2017diagnostic,bandi2018detection,tolkach2020high,bulten2020automated,bera2019artificial,echle2021deep,hahn2021expanded}.
In recent decades, the shift to digital pathology, particularly through whole slide imaging, has modernized the workflow of clinicians and improved access to slide data \cite{niazi2019digital}. 
This has paved the way for CPath, an emerging field that leverages digital whole slide images (WSIs) and computational methods for clinical decision-making \cite{deng2020deep,srinidhi2021deep,song2023artificial}. 
Specifically, CPath introduces advanced capabilities such as gene mutation prediction \cite{coudray2018classification,bilal2021development,zamanitajeddin2024social}, direct prognosis \cite{wulczyn2021interpretable,mobadersany2018predicting,courtiol2019deep}, and treatment response assessment \cite{vanguri2022multimodal,zhang2024histopathology,hu2021using} directly from WSIs, demonstrating profound clinical significance. 
However, the diversity of clinical pathology tasks, combined with the limited data and annotations, poses significant challenges when training robust models for each individual task from scratch. 
This process is not only time-consuming but also impractical in real-world scenarios \cite{song2023artificial}. 
Consequently, the CPath community is actively seeking solutions that can effectively address this diverse range of tasks simultaneously \cite{kang2023benchmarking,li2021dual,lazard2023giga,schirris2022deepsmile,vu2023handcrafted,claudio2021adversarial,jiang2023masked}.
\begin{figure*}[]
\centering
\includegraphics[width=1.0\textwidth]{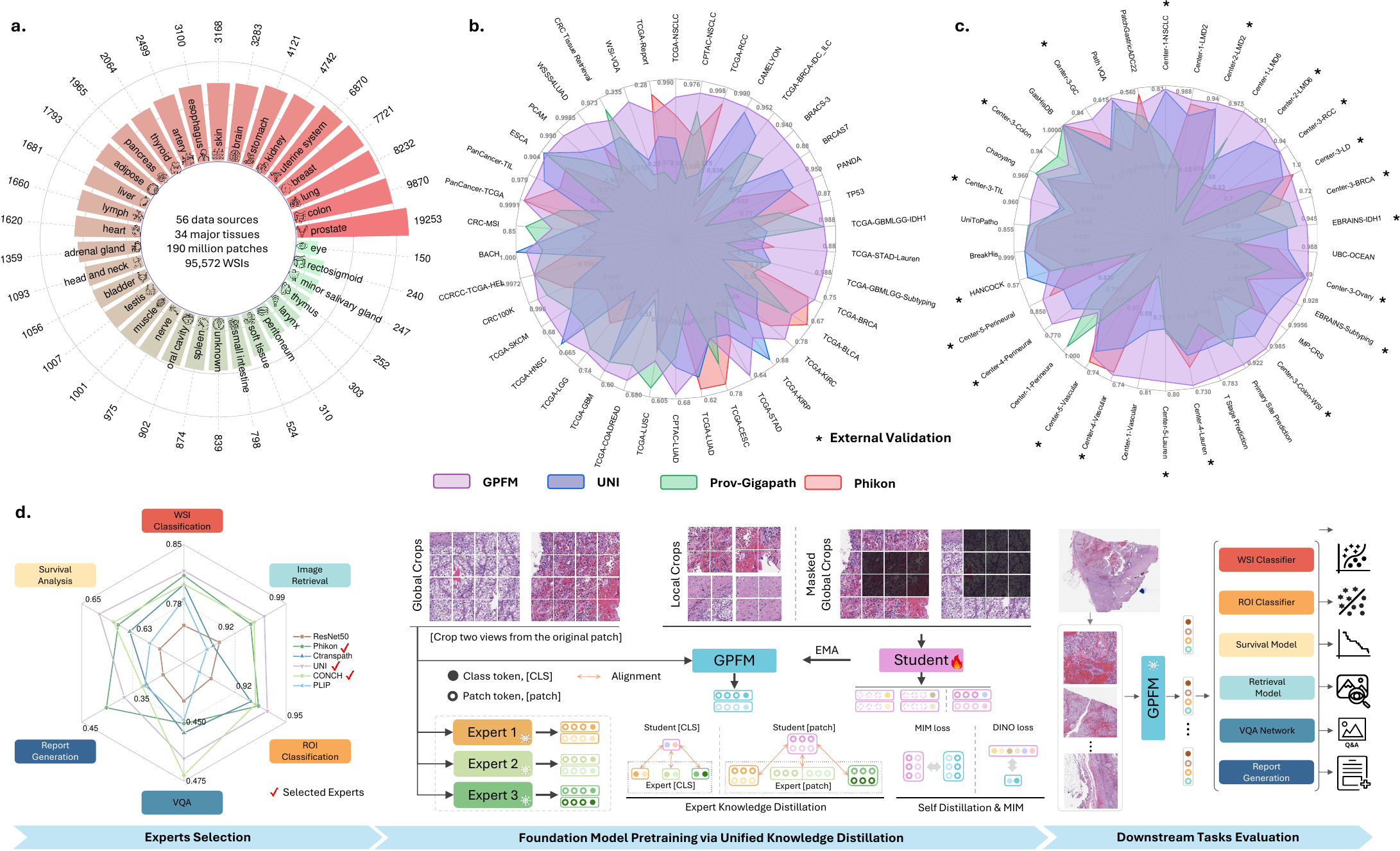}
\caption{\textbf{Overview of the GPFM.} GPFM is a state-of-the-art pretrained FM that demonstrates exceptional performance across 72 diverse tasks.
\textbf{a.} The GPFM dataset comprises a large-scale collection of 95,572 slides spanning 34 major tissue types, enabling comprehensive model training and evaluation.
\textbf{b-c.} Performance evaluation of foundation models (FMs) across a diverse set of tasks: 52 internal tasks and 20 external tasks. Only the top 4 models are presented here. For a more comprehensive analysis, including additional FMs, please refer to \textbf{Fig.} \ref{fig:overall_fig}.
\textbf{d.} The overview of unified knowledge distillation for GPFM.
The experts used for Expert Knowledge Distillation will be selected based on their average performance on six different clinical tasks.
The pretraining algorithm includes three key components: 1) Mask Image Modeling (MIM), 2) Self-Distillation, and 3) Expert Knowledge Distillation.
The parameters of GPFM are updated through Exponential Moving Average (EMA).
}
\label{fig:main}
\end{figure*}

\begin{figure*}
    \centering
    \includegraphics[width=1\linewidth]{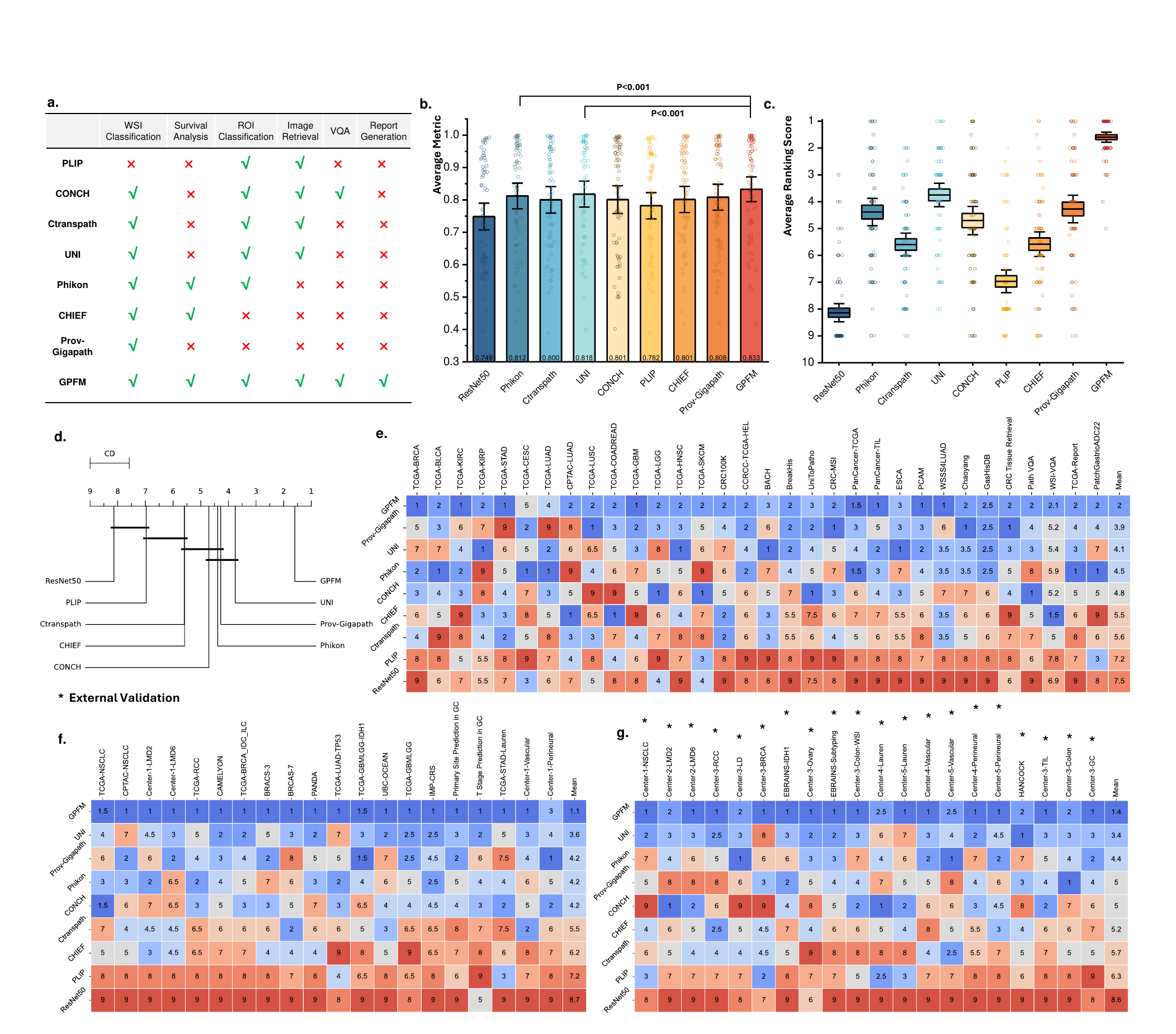}
    \caption{\textbf{Comprehensive Comparison of FMs across 72 Tasks.}
\textbf{a.} Task types evaluated by different FMs.
\textbf{b.} Average performance of FMs across 72 tasks:
WSI classification and tissue classification tasks are measured by AUC; survival analysis tasks are measured by C-index; the VQA task is measured by overall accuracy; the report generation task is measured by the average metric of BLEU, METEOR, and ROUGE-L; the image retrieval task is measured by average accuracy. The Wilcoxon signed-rank two-side test is employed to detect significant differences between off-the-shelf FMs and the proposed GPFM. The error bars in \textbf{b} and \textbf{c} indicate the 95\% CI.
The figure demonstrates that GPFM achieved the highest average performance.
\textbf{c.} Average rank of FMs across 72 downstream tasks.
The box limits represent the standard error.
\textbf{d.}Critical differences (CD) diagram of average ranking score with the Nemenyi test.
In the CD figure, there are no significant differences between the models covered by the black line. 
\textbf{e-f.} Ranking order of FMs across 32 and 20 internal tasks, respectively. 
\textbf{g.} Ranking order of FMs on 20 external validation datasets. 
If a model achieves the best performance, its rank value is set to 1. If two models have the same metric value, indicating a tie, the average rank value is assigned to all the tied models. 
For WSI-VQA, the rank is determined by the average of linguistic evaluation metrics and closed accuracy.
The evaluation metrics utilized to derive the ranking scores for the remaining tasks are consistent with those applied in subfigure \textbf{b}.
}
    \label{fig:overall_fig}
\end{figure*}

In recent years, there has been a notable progress in the fields of computer vision and natural language processing driven by self-supervised learning on large-scale datasets. These pretrained models, commonly referred to as foundation models (FM), have garnered significant attention and have exhibited remarkable success across various tasks \cite{bommasani2021opportunities,zhou2023comprehensive,moor2023foundation}. 
In the field of CPath, some efforts \cite{wang2022transformer,filiot2023scaling,uni_paper,vorontsov2023virchow,lu2024visual,huang2023visual, ctranspath_paper} have been dedicated to pretraining FMs that can learn inherent representations of histopathology images, catering to the diverse array of tasks encountered in clinical pathology practice. 
However, the current FMs have only been evaluated on a limited type of tasks (Fig. \ref{fig:overall_fig}a), leaving their overall performance unclear.
To comprehensively evaluate these models, we built a most comprehensive benchmark spanning six major clinical task categories (Fig. \ref{fig:main}d), comprising 72 specific tasks.
Our findings revealed that the generalization ability of these models is still limited and no single model can effectively address all the tasks (Fig. \ref{fig:main}d).
It can be seen that UNI \cite{uni_paper} achieves the best performance in WSI classification, image retrieval, survival analysis, and patch-level (ROI) tissue classification tasks, Phikon \cite{filiot2023scaling} performs best in report generation tasks, and CONCH \cite{lu2024visual} obtains highest performance in visual question answering (VQA) tasks.
This can be attributed to the fact that each FM is trained using distinct datasets and pretraining strategies, leading to specific advantages for each model within particular datasets.
These findings highlight the need for further research to develop more generalizable FMs that can consistently perform well across the diverse types of clinical tasks encountered in CPath. 
By addressing this challenge, we can unlock the full potential of the FM in CPath.

To improve the generalization of pathology FM and enhance the overall performance, an intuitive idea is to leverage the specific strengths of existing models by employing knowledge distillation techniques \cite{hinton2015distilling,gou2021knowledge}.
Accordingly, we proposed a novel self-supervised learning framework with expert and self knowledge distillation to develop a Generalizable Pathology Foundation Model (GPFM).
Based on the aforementioned pretraining method, we collected a dataset comprising 95,572 slides, encompassing 34 major tissue types, for the purpose of training and evaluating the GPFM. 
From this collection, we extracted 190 million patches derived from 72,280 slides to facilitate the pre-training (Fig. \ref{fig:main}a).
With the collected diverse tissues and the indirectly using of the images that used to pretrain expert models (e.g., UNI and CONCH), GPFM exhibits outstanding performance across the established benchmarks (Fig. \ref{fig:main}b-c), achieving an average rank of 1.6, while the second-best performing model, UNI, achieves an average rank of 3.7
(Fig. \ref{fig:overall_fig}c). 
These results demonstrate the efficacy of GPFM as a generalizable FM in CPath, showcasing its potential to significantly advance the field. 
The consistent performance of GPFM across a diverse range of clinical tasks underscores the advantages of employing knowledge distillation to integrate the strengths of specialized expert models. This approach facilitates the development of more robust and versatile foundation models (FMs), thereby enhancing their utility in supporting clinical decision-making and advancing patient care outcomes. 
\section{Results}\label{sec2}
We evaluated various FMs across 72 tasks, encompassing 36 WSI classification tasks, 15 survival analysis tasks, 16 patch-level (ROI) tissue classification tasks, 2 pathological visual question answering task, 2 report generation tasks, and 1 pathological image retrieval task (Fig. \ref{fig:overall_fig}e-g).
Since the tasks involved different types of evaluation metrics, we assessed the overall performance of the FMs using an average ranking approach and reported the critical difference (CD) diagram \cite{wilcoxn, maier2018rankings, antonelli2022medical}.
The model with the best performance was ranked 1st, while the model with the lowest performance was ranked 9th.
Across all tasks, the GPFM model achieved the top average rank score of 1.6 (ranked first in 42 tasks), outperforming the second-best model, UNI, which had a ranking score of 3.7 (ranked first in 6 tasks).
To evaluate the significance of GPFM's ranking score relative to other FMs, we performed the Nemenyi statistical test \cite{wilcoxn} (Fig. \ref{fig:overall_fig}d). 
The results demonstrate that GPFM exhibited a statistically significant critical difference compared to the other eight models.

We calculated the average evaluation metrics across all 72 tasks (Fig. \ref{fig:overall_fig}b), revealing that GPFM achieved the highest average score of 0.833, surpassing the second-best model, UNI, which scored 0.818. To assess statistical significance, we conducted a Wilcoxon signed-rank two-sided test \cite{wilcoxn} comparing GPFM with the second- and third-best models. The results showed that all \textit{p}-values were below 0.001, confirming that GPFM consistently and significantly outperformed the existing FMs.
Considering both the ranking perspective and the average metric aspect, the results clearly indicate that GPFM achieves state-of-the-art performance and is much more generalizable compared to the other FMs. 
\subsection{WSI Classification}
\begin{figure*}[]
    \centering
    \includegraphics[width=1\linewidth]{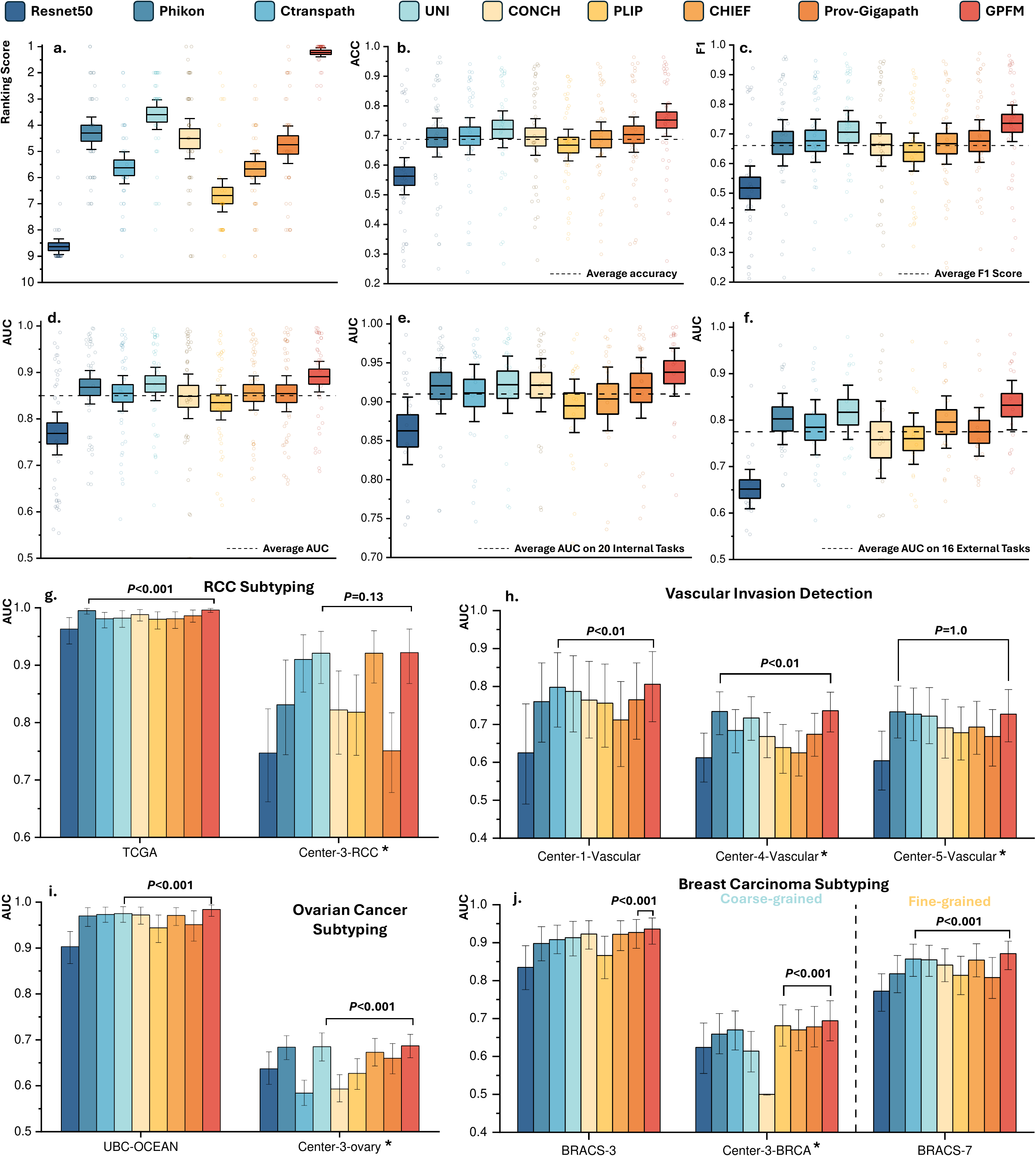}
\caption{\textbf{Performance of FMs on WSI Classification Tasks.}
\textbf{a.} Average ranking of FMs based on AUC across 36 WSI classification tasks.
\textbf{b-d.} Average balanced accuracy (ACC), and weighted F1 score (F1), and AUC of FMs across 36 WSI classification tasks.
\textbf{e.} Average AUC of FMs on 20 internal WSI classification tasks.
\textbf{f.} Average AUC of FMs on 16 external validation cohorts.
\textbf{g-h.} Model performance on specific tasks: RCC subtyping, vascular invasion detection, ovarian cancer subtyping, and breast carcinoma subtyping.
* represents external validation cohorts.
Error bars represent 95\% CI.
The box limits represent the standard error.
Additional results are shown in Extended Data  \textbf{Fig.} \ref{fig:WSI_ext1} and \textbf{Fig.} \ref{fig:WSI_ext2}.
}
    \label{fig:wsi_cls}
\end{figure*}
WSI classification is pivotal in accurate cancer diagnosis.
It aids in categorizing the specific subtype of cancer, which can be significantly improved by utilizing FMs. 
Therefore, it is important to evaluate the representation learning capabilities of different FMs. 
We conducted experiments on a total of 36 tasks, including 20 internal validation datasets and 16 external validation datasets. The detailed experimental results are presented in Extended Data Table \ref{lab:wsi:avg_cls}-\ref{lab:WSI:perineural}.

Across 36 WSI classification tasks, ranked according to the Area Under the Curve (AUC) metric, GPFM achieved an outstanding average ranking score of 1.22, significantly surpassing the second-best model, UNI, which attained an average ranking score of 3.60 (Fig. \ref{fig:wsi_cls}a). 
We assessed overall performance using average metrics: AUC, balanced accuracy, and weighted F1 score. 
Specifically, GPFM achieved the highest average AUC of 0.891, a 1.6\% improvement over UNI (P < 0.001; Fig. \ref{fig:wsi_cls}d). 
Similarly, GPFM outperformed UNI in balanced accuracy (0.752, +3.1\%, P < 0.001; Fig. \ref{fig:wsi_cls}b) and weighted F1 score (0.736, +3.0\%, P < 0.001; Fig. \ref{fig:wsi_cls}c). 
Additionally, GPFM achieved the best performance in both internal and external tasks, with AUCs of 0.938 (+1.6\% over UNI, Fig. \ref{fig:wsi_cls}e) and 0.832 (+1.5\% over UNI, Fig. \ref{fig:wsi_cls}f). These results across multiple metrics highlight GPFM's strong generalization capability and potential for WSI classification tasks.
\\\\\textbf{GPFM Enhances Diagnostic Accuracy Across Multiple Cancer Types.} 
GPFM demonstrates superior diagnostic accuracy across a range of cancer types and tasks. In breast cancer, GPFM outperforms other models in all six evaluated tasks, including five subtyping tasks (Fig. \ref{fig:wsi_cls}j, Extended Data Fig. \ref{fig:WSI_ext1}a) and one metastasis detection task (Extended Data Fig. \ref{fig:WSI_ext1}d).
For lung cancer, GPFM excels in three subtyping tasks, two metastasis detection tasks, and two primary site prediction tasks (Extended Data Fig. \ref{fig:WSI_ext1}b, f, and h), with the exception of one external validation for lung cancer metastasis detection, where UNI performs slightly better.
In gastric cancer, GPFM achieves the best performance in six out of nine tasks, including vascular invasion detection (Fig. \ref{fig:wsi_cls}h), perineural invasion detection, and Lauren subtyping (Extended Data Fig. \ref{fig:WSI_ext1}g and i).
Furthermore, GPFM consistently delivers top performance in tasks involving other organs, such as brain tumor subtyping, head and neck cancer primary site and T stage prediction, colon lesion grading, prostate cancer grade assessment, ovarian cancer subtyping (Extended Data Fig. \ref{fig:wsi_cls}c, d, b, c, i, and g), and renal cell carcinoma classification (Fig. \ref{fig:wsi_cls}g).
Overall, GPFM establishes itself as a leading model in cancer diagnosis across diverse tasks and cancer types.
\\\\\textbf{GPFM advances gene mutation prediction.} 
We conducted experiments on lung cancer and brain cancer slides. 
GPFM achieved the best results in both TP53 mutation prediction for lung cancer, with an AUC of 0.855 (+1.3\% over Phikon; Extended Data Fig. \ref{fig:WSI_ext1}e), and IDH1 mutation prediction for glioma, with an internal AUC of 0.986 and an external AUC of 0.943 (Extended Data Fig. \ref{fig:WSI_ext2}a).

These results, along with the cancer diagnosis findings, highlight GPFM's superior generalizability compared to existing FMs. A key factor in this success is GPFM's ability to integrate knowledge from expert models through a unified knowledge distillation mechanism. Unlike previous FMs that did not employ knowledge distillation, GPFM leverages this approach to learn from a broader range of data and perspectives, significantly enhancing its performance. This capability underscores GPFM's advanced adaptability and effectiveness across diverse tasks.
\subsection{Survival Analysis}
\begin{figure*}[]
    \centering
    \includegraphics[width=1\linewidth]{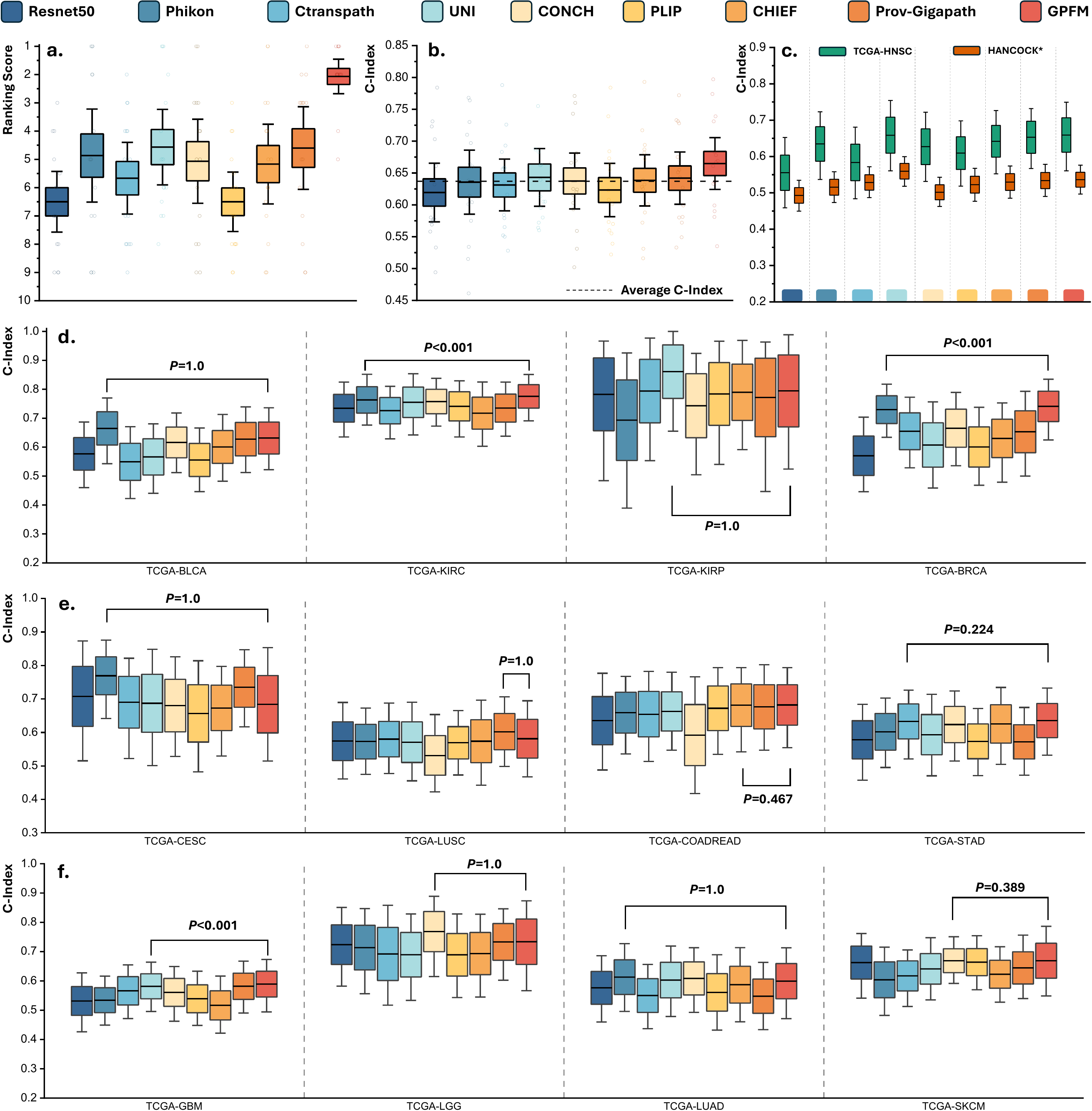}
\caption{\textbf{Performance of FMs across 15 Survival Analysis Tasks.}  
    \textbf{a.} Average ranking of FMs in 15 survival analysis tasks.  
    \textbf{b.} Average C-Index of various FMs across 15 tasks.  
    \textbf{c.} Results on TCGA-HNSC data and the HANCOCK cohort. The survival prediction model was trained on the TCGA-HNSC cohort and subsequently tested on the HANCOCK cohort.  
    \textbf{d-f.} C-Index of FMs across 12 survival analysis tasks.  
    In all subfigures, error bars indicate 95\% CI. For box plots, the center line represents the mean, and the box limits represent the standard error.  
}
    \label{fig:survival}
\end{figure*}
Accurate prediction of a patient's survival risk can enable more targeted and effective treatment strategies.\cite{wang2019machine, xu2023multimodal,zhang2024prototypical, zhou2023cross}. 
A robust FM is essential for improving the precision of survival risk prediction, ultimately leading to better patient outcomes.
To evaluate the performance of various FMs in survival analysis, we conducted experiments on 15 datasets.
Following the methodologies of previous works \cite{wiegrebe2024deep, xu2023multimodal, zhou2023cross}, we adopted the Concordance Index (C-Index) as the evaluation metric to compare the performance of different FMs. 

Across the 15 survival analysis tasks, the GPFM achieved an impressive average ranking score of 2.1, ensuring the best or second-best performance in 13 tasks (Fig. \ref{fig:survival}a, d-f).
In comparison, the second-best performing model, UNI, attained an average ranking score of 4.6, achieving top-2 performance in only 4 tasks (Fig. \ref{fig:survival}a, d-f). 
Furthermore, when evaluated using the widely recognized C-Index metric, the GPFM emerged as the top performer, achieving an average C-Index of 0.665 (Fig. \ref{fig:survival}b). 
This result represents a statistically significant improvement of 3.4\% over UNI (\textit{P} < 0.001), further demonstrating the superior generalization capability of GPFM for survival analysis tasks. 
To further validate the generalization of FMs, we conducted additional validation studies, including one external validation for head and neck cancer (TCGA-HNSC) and one internal validation for lung adenocarcinoma (TCGA-LUAD).
In the head and neck cancer survival prediction task, UNI achieved the best performance in both the TCGA-HNSC and HANCOCK cohorts, while our method ranked as the second-best performer (Fig. \ref{fig:survival}c). However, in the lung adenocarcinoma task, GPFM demonstrated a 10.6\% improvement in the CPTAC-LUAD cohort (Extended Data Fig. \ref{fig:extra_roi_results}h) compared with UNI.

It is noteworthy that survival analysis tasks are inherently more challenging than WSI classification, and no single model has been able to dominate these tasks (Fig. \ref{fig:overall_fig}e). 
The experimental results from both WSI classification and survival analysis highlight the limited generalization capability of existing FMs. This limitation is likely attributable to the data distribution of their training sets and the pretraining methods they employ.
While existing FMs exhibit limited generalization, they demonstrate exceptional performance on specific types of tasks. By leveraging their individual strengths, it is possible to construct a more powerful and versatile model. This is precisely what we have achieved in this study: we propose a unified distillation framework to distill the capabilities of existing models—particularly in tasks where they excel—into the GPFM, thereby significantly enhancing its generalization ability.
\subsection{ROI Classification}
\begin{figure*}[]
    \centering
    \includegraphics[width=1\linewidth]{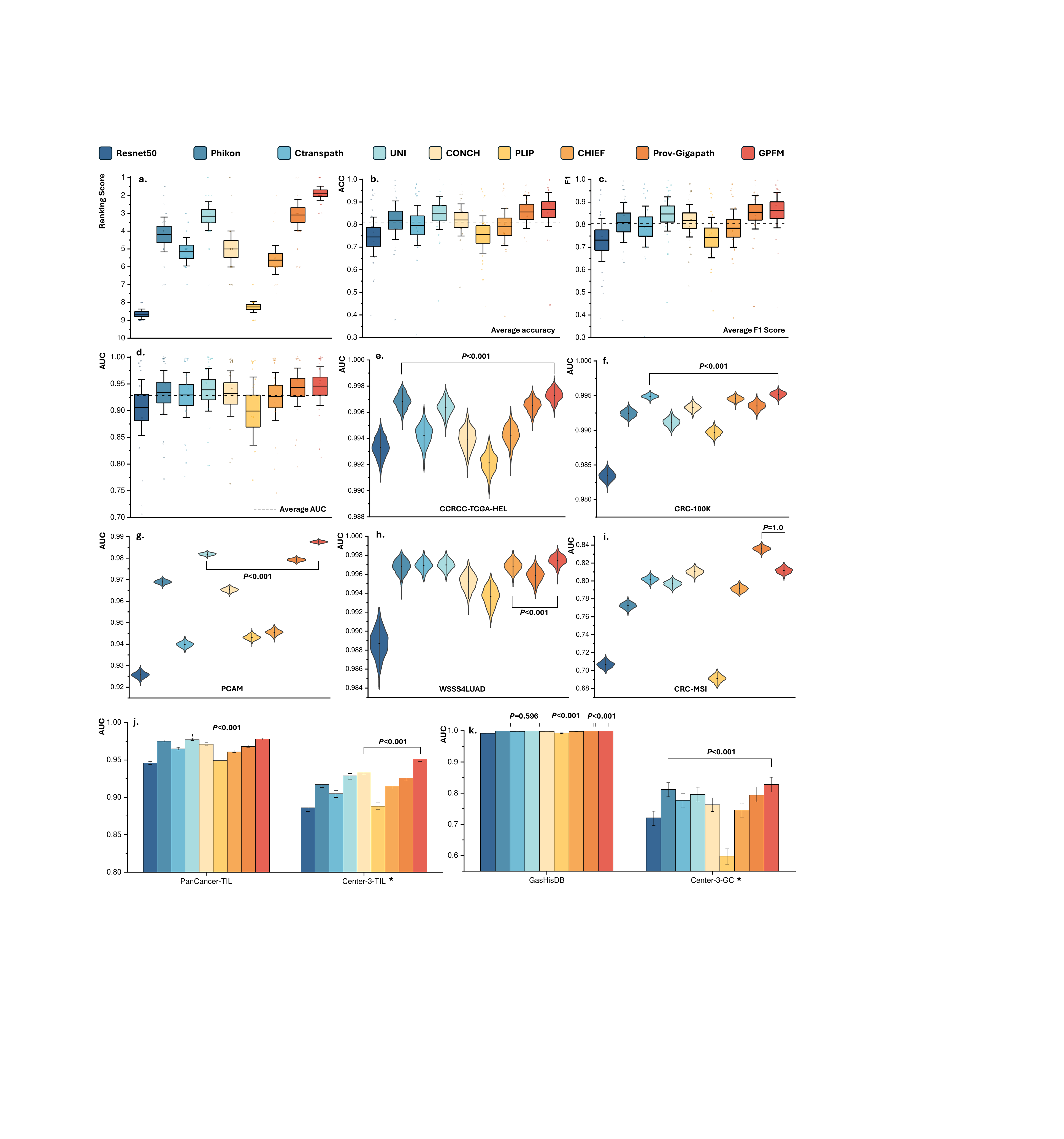}
    \caption{\textbf{Performance of FMs on Tissue Classification Tasks.} 
    \textbf{a.} Average ranking order of FMs based on AUC across 16 tasks. 
    \textbf{b-d.} Average balanced accuracy (ACC), and weighted F1 score (F1), and AUC of FMs across 16 tasks. The center line represents mean and the box limits represents the standard error.
\textbf{e-i.} AUC of FMs across 5 tissue classification tasks. The Wilcoxon signed-rank one-side test is adopted to detect significant difference. Then center black line in the violin plot represents the mean AUC.
\textbf{j.} Tumor infiltrating lymphocytes classification based on the PanCancer-TIL (internal) and Center-3-TIL data (external).
\textbf{k.} Gastric cancer tissue classificaiton with GasHisDB (internal) and Center-3-GC data (external).
In all subfigures, the error bars indicate 95\% CI.
More results are presented in Extended Data \textbf{Fig.} \ref{fig:extra_roi_results}.
}
    \label{fig:patch_cls}
\end{figure*}
The performance of WSI classification is influenced by both the feature extractor (i.e., FM) and the multiple instance learning (MIL) method.
Unlike WSI classification, Region-of-Interest (ROI) classification tasks allow for a direct assessment of the FMs' feature representation capabilities, independent of MIL methods. 
To this end, we employed a linear probe approach, as outlined in \cite{oquab2023dinov2}, to evaluate the FMs. Our assessment spanned 16 ROI classification tasks, encompassing 13 internal and 3 external validation datasets. Comprehensive findings from these evaluations are cataloged in Extended Data Tables \ref{lab:linear:avg}-\ref{lab:linear:GasHisDB}.

GPFM emerged as the top performer across all 16 ROI classification tasks, securing the best ranking score of 1.88, significantly outperforming the second-ranked model, Prov-Gigapath, which scored 3.09 (Fig. \ref{fig:patch_cls}a). In terms of conventional metrics, GPFM achieved the highest average AUC of 0.946 (+0.2\% over Prov-Gigapath, P<0.001; Fig. \ref{fig:patch_cls}d), the best weighted F1 score of 0.865 (+0.9\%, P<0.001; Fig. \ref{fig:patch_cls}c), and the highest balanced accuracy of 0.866 (+1\%, P<0.001;Fig. \ref{fig:patch_cls}b).
GPFM exhibited outstanding performance in several tasks, including the detection of metastatic tissue in breast cancer (Fig. \ref{fig:patch_cls}g), tissue type classification in lung cancer (Fig. \ref{fig:patch_cls}h), the classification of tumor-infiltrating lymphocytes (TILs) (Fig. \ref{fig:patch_cls}j), and the classification of gastric cancer tissues (Fig. \ref{fig:patch_cls}k). 
In relatively simpler ROI classification tasks, GPFM shared the top rank with other FMs.
For instance, in pancancer tissue classification (Extended Data Fig. \ref{fig:extra_roi_results}f), breast tumor classification (Extended Data Fig. \ref{fig:extra_roi_results}b), colorectal cancer tissue classification (Fig. \ref{fig:patch_cls}f), and kidney tissue classification (Fig. \ref{fig:patch_cls}e), GPFM achieved performance on par with other leading FMs.
In tasks where GPFM did not achieve the top performance, it consistently ranked as the second-best method (Extended Data Fig. \ref{fig:extra_roi_results}a, d, e; Fig. \ref{fig:patch_cls}i) or the third-best method (Extended Data Fig. \ref{fig:extra_roi_results}c). This consistent high ranking across diverse tasks contributed to GPFM's overall superior performance.
In addition, the average ranking scores (Fig. \ref{fig:patch_cls}a) of UNI and Prov-Gigapath are close, with ranking scores of 3.2 and 3.1, respectively. This indicates that no single existing model dominates ROI classification tasks. In contrast, by integrating knowledge from all adopted expert models, the unified knowledge distillation enables GPFM to surpass the performance of individual models, achieving a significantly lower average ranking score of 1.88, outperforming the next-best model by more than one point. 
This underscores GPFM's strength as a highly generalizable FM.

Furthermore, to evaluate the robustness of GPFM in handling images with varying resolutions, we visualized the heatmap of attention scores between the [patch] tokens and [CLS] tokens of the ViT transformer (Extended Data Fig. \ref{fig:extra_roi_results}g). Across four resolutions—224$\times$224, 448$\times$448, 896$\times$896, and 1344$\times$1344—we observed consistent attention patterns, highlighting GPFM's robustness in adapting to different image resolutions.
\subsection{Pathological Image Retrieval }
Image retrieval techniques could match the new patient pathology images to a curated database of previously diagnosed cases, providing pathologists with a novel tool to enhance diagnostic accuracy.
Through visual inspection and comparison of similar historical cases, pathologists can leverage image search functionality to enhance their diagnostic decision-making. 
In this study, we employ the CRC-100K dataset \cite{CRC100K} for conducting pathological image retrieval tasks. 

\begin{figure*}[]
    \centering
    \includegraphics[width=1\linewidth]{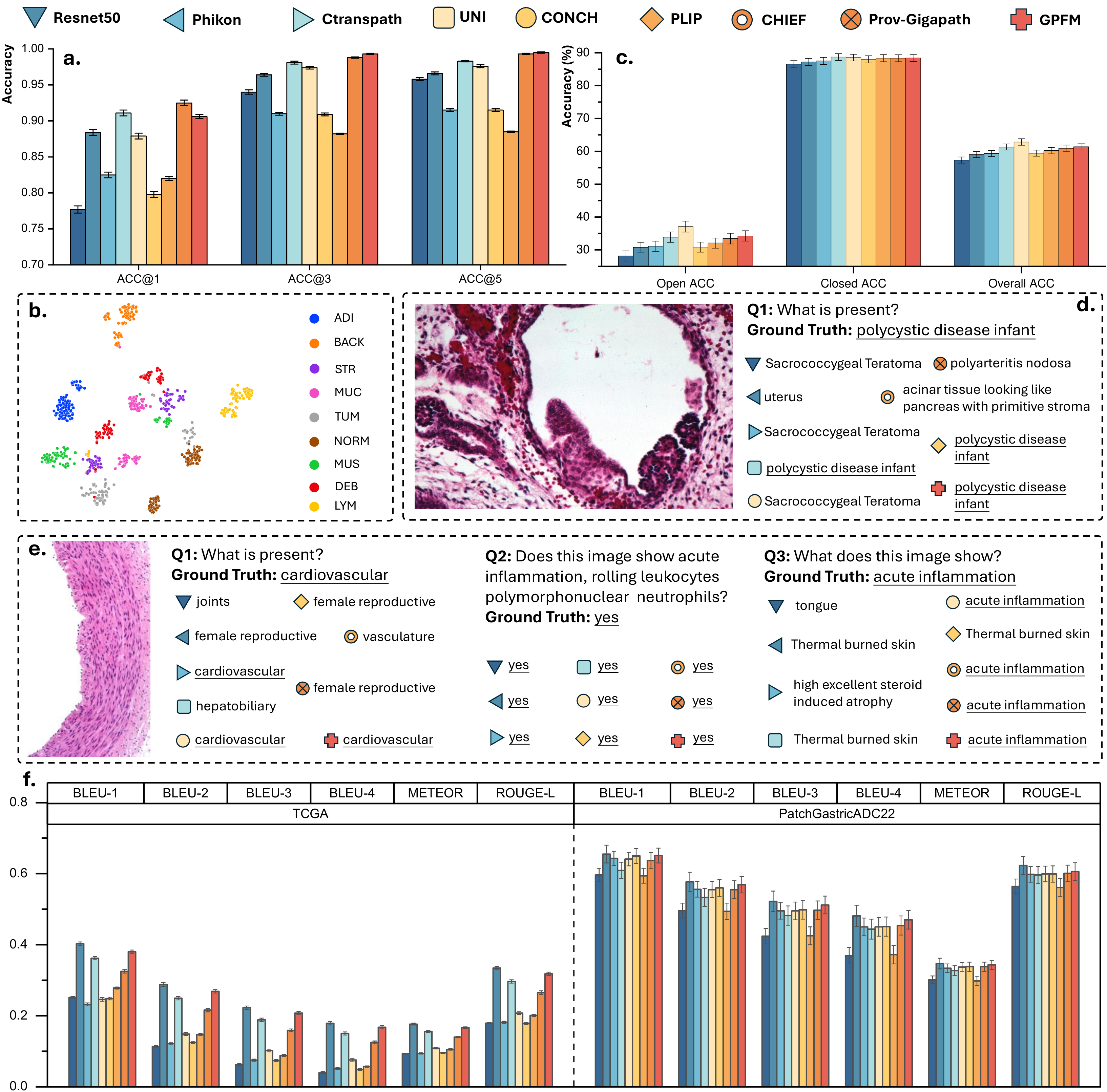}
    \caption{\textbf{Overview of Pathology Tissue Retrieval, VQA, and Report Generation.} \textbf{a.} The top-1, top-3, and top-5 accuracy of different FMs on pathology tissue retrieval tasks. 
    \textbf{b.} The distribution of features extracted by GPFM. For each class, 100 samples from the test set were used, and a total of 900 samples were subjected to t-SNE dimensionality reduction to 2D. 
    \textbf{c.} The performance of VQA on PathVQA dataset, measured by open-ended accuracy, closed-ended accuracy, and overall accuracy, for different FMs. 
\textbf{d.} An example of open-ended questions along with the answers generated by various FMs.
\textbf{e.}  Three example questions and the answers generated by FMs related to the query image.
\textbf{f.} The performance of WSI report generation on TCGA and PatchGastricAD22 datasets. The models are measured by six different language quality metrics, \textit{i.e.}, BLEU-1, BLEU-2, BLEU-3, BLEU-4, METEOR, and ROUGE-L. 
In all subfigures, the error bars indicate standard deviation.
}
    \label{fig:roi_retrieval_vqa}
\end{figure*}

The experimental results (Fig. \ref{fig:roi_retrieval_vqa}a, Extended Data Table \ref{lab:ret:crc-100k}) show that the GPFM model achieved the second-best Top-1 accuracy with a value of 0.906 (-1.9\%, Prov-Gigapath). 
However, GPFM outperforms other models in terms of Top-3 and Top-5 accuracy, achieving values of 0.993 (+0.5\%, Prov-Gigapath) and 0.995 (+0.2\%, Prov-Gigapath), respectively.
To further explore the clustering effect and feature representation ability, we utilized t-Distributed Stochastic Neighbor Embedding (t-SNE) \cite{tsne} to project the features extracted by GPFM into a 2D embedding space. 
The categories are well clustered, further illustrating that the features are highly discriminative (Fig. \ref{fig:roi_retrieval_vqa}b).
We also visualized the feature distribution of other FMs (Extended Data Fig. \ref{fig:roi_retrieval}). 
The features extracted by the GPFM are clustered more tightly and the query image is also located within the candidate cluster, indicating a better clustering effect. 
This observation suggests that the GPFM has superior feature representation capabilities in capturing the intrinsic patterns and structures present in the data.
\subsection{Pathological Images VQA }
VQA is an exciting field of artificial intelligence that aims to enable machines to answer questions about visual content. 
In the domain of pathology, VQA systems can be particularly powerful, allowing clinicians and researchers to quickly and accurately extract relevant information from medical images.

For the patch-level VQA task, our model achieved the second-best performance, with results only slightly lower than those of CONCH (Fig. \ref{fig:roi_retrieval_vqa}c, Extended Data Table \ref{tab:vqa}). It is important to note that CONCH is a vision-language FM trained on millions of image-text pairs, which inherently provides it with an advantage in VQA tasks. Despite this, our results highlight the substantial potential of our approach compared to other pure vision FMs.
To further illustrate the capabilities of our model, we visualized the query images, questions, and answers generated by different FMs (Fig. \ref{fig:roi_retrieval_vqa}d and \ref{fig:roi_retrieval_vqa}e). As demonstrated in the figures, both GPFM and CONCH consistently produced more reliable and accurate answers compared to the other models.

Moreover, in the WSI-level VQA task \cite{chen2025wsi}, our model achieved the best or second-best performance across 6 out of 7 metrics, demonstrating performance comparable to the slide-level FM CHIEF (Extended Data Fig. \ref{fig:vqa-extra} and Table \ref{lab:WSI-VQA}). These results, combined with the patch-level findings, underscore the effectiveness of unified knowledge distillation. Specifically, the knowledge acquired by CONCH from millions of image-text pairs can be successfully distilled into GPFM without requiring access to the original image-text pair data.
The strong performance of GPFM highlights the potential of leveraging textual knowledge indirectly, without the need for direct utilization of text data, thereby offering a promising direction for future research in VQA tasks.
\subsection{Pathology Report Generation}
Pathology reports are essential components of the healthcare system, providing critical information to clinicians and patients about the diagnosis, prognosis, and treatment of various medical conditions. 
These reports summarize the findings from pathological examinations, such as biopsies, cytology samples, and surgical specimens, and play a vital role in guiding clinical decision-making.
Traditionally, pathology reports are written manually by pathologists and their teams, a time-consuming and labor-intensive process. 
Recent advancements in natural language processing (NLP) and machine learning have enabled the development of automated pathology report generation systems, which can dramatically improve the efficiency and consistency of this critical task \cite{guo2024histgen, guevara2023caption, Chen2023WsiCaptionMI}. 
To assess the effectiveness of FMs in this domain, we evaluated their performance on the TCGA WSI-report dataset, curated by Guo et al. \cite{guo2024histgen}, and the PatchGastricADC22 \cite{tsuneki2022inference} dataset. 

The experimental results demonstrate that Phikon achieved the best performance across all six metrics, while GPFM achieved comparable performance and ranked as the second-best model on both tasks (Fig. \ref{fig:roi_retrieval_vqa}f, Extended Data Table \ref{lab:WSI-report} and \ref{lab:PatchGastricADC}).
It is quite surprising to observe that vision FMs (e.g., Phikon and GPFM) performed much better in this task than vision-language FMs such as CONCH and PLIP. 
This performance gap may be attributed to PLIP and CONCH's training paradigm, which relied solely on short descriptions or captions of pathological images without access to global contextual information. Consequently, these text-image pairs proved less effective for comprehensive report generation compared to their original VQA task applications. The examples of generated reports shown in Extended Data Fig. \ref{fig:extra_report_1} and \ref{fig:extra_report_2} certify this assumption.

To further validate these findings, we conducted stratified report generation analyses by stratifying the TCGA WSI-report dataset by major cancer types, \textit{i.e.}, breast, lung, and kidney cancers, for independent evaluation. Results (Extended Data Table \ref{lab:WSI-report-stratify} and Fig. \ref{fig:score_criteria}a-c) reveal that Phikon keeps its superiority in breast and lung cancer report generation, yet is slightly outperformed by our GPFM in kidney cancer report generation. To leverage the complementary strengths of existing FMs, the proposed unified knowledge distillation approach can distill the capabilities of Phikon in report generation into the GPFM. This synergistic integration allows us to combine the respective strengths of these FMs, leading to the development of a more generalizable model. To further assess clinical relevance, an experienced pathologist evaluated the diagnostic reports using a four-tier scoring system (Extended Data Fig. \ref{fig:score_criteria}d). 
The blinded human-based evaluation results demonstrate GPFM's superior performance, achieving the highest average scores across breast, lung, and kidney cancer reports (Extended Data Table \ref{lab:WSI-report-human} and Fig. \ref{fig:score_criteria}a-c). 
These expert-validated results underscore the potential of our unified knowledge distillation approach to generate clinically meaningful reports that align with pathologists' diagnostic standards, marking a significant step toward the practical application of AI in pathology workflow automation.

\subsection{The Effectiveness of Expert Knowledge Distillation}
In the self-supervised learning framework proposed in this study, we introduced a unified knowledge distillation model to facilitate the transfer of knowledge from off-the-shelf FMs to GPFM during the pretraining stage. To assess the effectiveness of this module, we conducted an experiment where we removed the Expert Knowledge Distillation module, resulting in a modified self-supervised learning framework known as DINOv2 \cite{oquab2023dinov2}. We trained both DINOv2 and GPFM on the same dataset and evaluated their performance in tissue classification tasks. 
The experimental results clearly demonstrate the positive impact of Expert Knowledge Distillation on the performance of the models across 12 tasks (Extended Data Fig. \ref{fig:albation} and Table \ref{lab:linear:ablation}).
The experimental results demonstrated significant improvements not only in the performance of individual tasks but also in the overall average performance, with substantial enhancements observed across all three evaluation metrics.
The AUC increased by 0.6\%, the weighted F1 score improved by 1.8\%, and the balanced accuracy showed an increase of 1.8\%.
These findings provide strong evidence for the effectiveness of transferring knowledge from off-the-shelf pathology FMs through the proposed knowledge distillation learning framework. 
However, even with the distillation, GPFM still can not beat vanilla DINOv2 in all tasks such as Chaoyang and BreakHis, illustrating that there is still room for improving the distillation strategy.


\section{Discussion}\label{sec12}
In this study, we construct the most comprehensive benchmark for CPath tasks to date, to the best of our knowledge. Additionally, we introduce GPFM, a generalizable FM designed for a broad spectrum of CPath tasks.
To enhance the model's versatility, we propose a unified knowledge distillation pretraining framework, which effectively consolidates expertise from a variety of existing models. 
This innovative approach ensures that GPFM can adapt and excel across different CPath tasks.
To further maximize the diversity of data used for pretraining, we gathered 190 million images sourced from 56 sources, spanning 34 major tissue types. 
This rich dataset, combined with our advanced pretraining methodology, empowers GPFM to surpass current FMs in performance across 72 CPath tasks. 
Unlike other models that demonstrate proficiency in narrow domains—such as UNI \cite{uni_paper}, which specializes in WSI classification, and Phikon \cite{filiot2023scaling}, which excels in report generation—GPFM showcases exceptional generalization, outperforming its counterparts across a wide array of CPath challenges by combining the strengthens of expert models.
 
Recently, several vision-language \cite{lu2024visual, huang2023visual} and pure vision \cite{uni_paper, filiot2023scaling, ctranspath_paper, xu2024whole} pathology FMs have been developed.
However, the overall performance of these existing FMs is unclear due to the absence of a comprehensive benchmark.
Our analysis reveals that no single existing model consistently exhibits the best performance. 
This is likely because each FM is trained using distinct datasets and pretraining strategies, leading to model-specific advantages for particular domains and datasets.
The root of a model's generalization ability lies in the diversity of the training data. 
Unfortunately, gathering extremely large-scale diverse datasets, especially for sensitive medical data, is very difficult due to security and privacy concerns. 
Therefore, it is almost impossible to access and utilize all the data used to develop the existing FM.
Although accessing the original private training data is limited, the pretrained models themselves are available. Since the knowledge of the pretrained models is derived from the training data, we can indirectly leverage this knowledge by using a unified knowledge distillation framework. 
It provides a feasible method to integrate knowledge from a large number of existing models under the premise of limited data and protecting data privacy, which has better feasibility and scalability in clinical practice. 
The significantly greater generalization ability of GPFM compared to existing FMs, suggests that transferring knowledge from one existing model to another may be a more viable path to further advancing pathology FMs in the future, especially given the challenges of assembling large-scale diverse medical datasets.

This study also has some limitations.
We recognize that current off-the-shelf FMs still exhibit potential in specific tasks, such as Phikon for report generation using TCGA data.
This illustrates that the proposed unified knowledge distillation approach is not perfect and has room for improvement.
Future research should concentrate on developing sophisticated methodologies to effectively distill and incorporate expert knowledge into one model, maximizing their potential across a broader spectrum of tasks. 
For example, further expanding the model's parameter size to enhance its adaptability, facilitating a more comprehensive assimilation of knowledge from diverse FMs. 
Additionally, the current GPFM is an unimodal FM, which limits its ability to effectively handle cross-modal tasks such as VQA.
Given the prevalence of multimodal data in pathology, encompassing WSIs, reports, and genomic data, the development of a multi-modal pathology FM is more attractive. 
Such a model would be more adept at integrating heterogeneous information, offering a more holistic understanding of patient data and enhancing diagnostic accuracy.

\section{Methods}\label{sec11}

\subsection{FM Pretraining}
CPath has emerged as a groundbreaking field that synergizes the power of AI with the expertise of pathologists, revolutionizing the practice of diagnosing and analyzing diseases. At the core of this transformative discipline lies the FM, which serves as the backbone for a wide range of applications in pathology.
While there exist some readily available FMs such as Ctranspath (pretrained on 32K TCGA slides) \cite{ctranspath_paper} and UNI \cite{uni_paper} (pretrained on 100K private slides), the utilization of public data remains incomplete, and the evaluation of these models in CPath tasks is inadequate. 
The limited diversity of primary sites in the pretraining slides also restricts the adaptability of current FMs for public CPath benchmarks.
To facilitate the advancement of CPath, we meticulously curated a comprehensive dataset comprising 56 histopathology datasets, encompassing a wide spectrum of 34 distinct tissue types for pretraining and downstream task evaluation (Extended Data Table \ref{tab:primariy_site}).
Leveraging this large-scale dataset, we developed a self-supervised learning approach with unified knowledge distillation to construct a FM that surpasses existing models.\\\\
\textbf{Dataset Preparation.} To boost the performance of FMs, diverse datasets with various tissues are necessary.
We have collected over 33 datasets as depicted in Extended Data Table \ref{tab:data_links} (from row 1 to row 33).
To process WSIs, we employed the OpenSlide \cite{openslide} and CLAM toolkit \cite{CLAM} to find all non-overlapping 512$\times$512 patches at level 0 that contain tissues.
It is worth noting that we did not scale the patches to a uniform resolution, opting instead to use the original resolution of each WSI. This approach was implemented to increase the robustness of the FMs to varying resolutions.
For datasets that only contain ROI images, we extracted non-overlapping 512$\times$512 patches as well.
Upon processing all 33 datasets, we obtained a comprehensive dataset, as presented in Extended Data Table \ref{tab:dataset_distributation}. 
The pretraining data consists of 72,280 WSIs and a total of 190,212,668 patches.\\\\
\textbf{Pretraining with Self and Expert Knowledge Distillation.} In CPath, current FMs typically rely on state-of-the-art self-supervised pretraining (SSL) methods, such as DINOv2 \cite{oquab2023dinov2} and iBOT \cite{iBOT}. These methods are applied directly to either private or public datasets. 
For instance, Phikon \cite{filiot2023scaling} is constructed based on 6,093 TCGA slides using iBOT, while UNI is built upon approximately 100,000 private and public slides using DINOv2. 
Due to larger training dataset and more powerful SSL methods, UNI outperforms Phikon in various tasks. However, UNI still lags behind other FMs in tasks related to text analysis and survival analysis due to its pretraining strategy and limited coverage of primary sites.
To address the limitations of current FMs and further enhance their performance, we propose a novel pretraining strategy involving Unified Knowledge Distillation.
The framework of the proposed pretraining method is  similar to DINOv2, we employ teacher-student networks with masking image modeling (MIM) loss \cite{MAE} and DINO (self-distillation) \cite{DINO, iBOT} loss to optimize the student network (Fig. \ref{fig:main}c). 
Specifically, given an input image $x$, we obtain two augmented views, $u$ and $v$. 
Random masking is then applied to both $u$ and $v$, resulting in masked views, $\hat{u}$ and $\hat{v}$.
For the MIM objective, the student network takes $\hat{u}$ and $\hat{v}$ as inputs and aims to predict the masked tokens. With the DINO objective, we first crop $n$ additional local views, $w_i$, and extract encoded class ([CLS]) tokens using the student network. Next, we obtain the [CLS] tokens of the global views ($u$ and $v$) using the teacher network. Finally, we compute the cross-entropy loss between the local views and global views' [CLS] tokens.
However, this strategy fails to leverage the knowledge from existing vision FMs, such as UNI and vision-language FMs like CONCH \cite{lu2024visual}, which restricts their applicability to different tissue types. 
To facilitate the transfer of knowledge from established pathology FMs, we propose an Expert Knowledge Distillation module aimed at distilling knowledge into the student network \cite{hinton2015distilling, amradio}.
To maximize the generalizability of the pretrained model, it is crucial to balance the performance and  diversity of expert  models. We evaluated several existing models across six different tasks, selecting those that excelled in classification (UNI), report generation (Phikon), and visual question answering (CONCH) as expert models (see Fig. \ref{fig:main}c). 
The [CLS] token, which represents the overall information of a patch for downstream tasks, serves as a critical component in our approach. 
If the [CLS] token of our model aligns well with those of the expert models, it indicates that our model can effectively assimilate the knowledge from selected experts. 
Similarly, the [PATCH] token also contains rich information. 
For example, some methods use mean pooling to perform downstream tasks \cite{virchow2}.
Therefore, aligning the [PATCH] token can further improve the effect of knowledge transfer.
To achieve above alignments, we use the student network to encode the global views $u$ and $v$ and extract the [CLS] and [PATCH] tokens. 
Additionally, we employ the adopted experts to obtain their [CLS] and [PATCH] tokens, respectively. 
For aligning the class tokens, we utilize cosine similarity. As for the patch token alignment, we employ both cosine similarity and smooth L1 distance. The pseudo-code for this process is outlined in Algorithm \ref{alg:distill}. 
The hyperparameters used in the pretraining phase are provided in Extended Data Table \ref{tab:hyperparameter}.
Once the student network is updated, we adopt the Exponential Moving Average (EMA) to update the teacher network (GPFM).

\textbf{Baselines.} To evaluate the performance of our FM, GPFM, we conducted a comprehensive evaluation by comparing it with other vision FMs, namely Ctranspath \cite{ctranspath_paper}, Phikon \cite{filiot2023scaling}, and UNI \cite{uni_paper}, slide-level FM CHIEF \cite{wang2024pathology} and Prov-Gigapath \cite{xu2024whole}, as well as visual-language FMs PLIP \cite{huang2023visual} and CONCH \cite{lu2024visual}. 
As a baseline, we also compared these FMs with a ResNet50 \cite{resnet_paper} pretrained on the ImageNet dataset \cite{imagenet-dataset}.
The model configurations and training details for all these models are presented in\textbf{ }Extended Data Table \ref{tab:sota_config}. 
For all downstream tasks, it should be emphasized that feature extraction was consistently performed on images resized to 224×224 resolution, except where explicitly stated otherwise in the experimental protocol. \\\\
\subsection{WSI Classification}
In CPath, WSI classification typically employs multiple instance learning (MIL) as the underlying methodology. 
The MIL approach involves the following steps: (1) Non-overlapping tissue patches are cropped from the original WSI, and features are extracted using a feature extractor. (2) A feature aggregator is applied to integrate the patch-level features into a slide-level feature, enabling classification.
To preprocess the WSIs, we utilize the pipeline described in the CLAM toolkit \cite{CLAM}. 
Specifically, we employ the default segmentation configuration of CLAM to extract patches with 512$\times$512 pixels at level 0 for all slides. 
Slides with a limited number of patches are discarded. 
Once all patches are extracted, we resize them to 224$\times$224 pixels. We then utilize FMs to extract features from the resized patches and save these features for subsequent MIL analysis. 
There are several MIL methods available, such as Attention-Based Multiple Instance Learning (ABMIL) \cite{abmil} and TransMIL \cite{TransMIL}. 
After evaluating the performance of different FMs across various WSI classification tasks, we found that ABMIL consistently achieves the best results, which aligns with the findings from previous studies \cite{uni_paper, vorontsov2023virchow}. Therefore, we adopt ABMIL to evaluate the performance of different FMs in our experiments. The architecture and training details of ABMIL are presented in Extended Data Table \ref{tab:abmil_config}.
For CHIEF \cite{wang2024pathology} and Prov-Gigapath \cite{xu2024whole} models, we use their pretrained slide-level FM to perform classification.

To evaluate the performance of the MIL model, we assess the balanced accuracy, weighted F1 score, and AUC, which consider the class imbalance present in the dataset. Our experiments encompass 36 pathology WSI classification tasks, including 20 internal and 16 external validation datasets.
The results of our experiments are presented in Extended Data Tables \ref{lab:wsi:avg_cls}-\ref{lab:WSI:hancock}. \\\\
\textbf{NSCLC Subtyping on TCGA, CPTAC and Center-1 Cohorts (2 classes).} 
To perform subtyping of non-small cell lung cancer (NSCLC), we utilized data from the TCGA \cite{tcga_dataset}, CPTAC \cite{cptac_data}, and Center-1. 
The TCGA cohort comprises 541 lung adenocarcinoma (LUAD) and 512 lung squamous cell carcinoma (LUSC) samples. The data is label-stratified in a ratio of 7:1:2, resulting in 738 slides for training, 105 slides for validation, and 210 slides for testing.
For the CPTAC cohort, there are 1,077 LUSC slides and 1,136 LUAD slides. Similarly, this cohort is label-stratified in a 7:1:2 ratio, yielding 1,549 slides for training, 222 slides for validation, and 442 slides for testing. Additionally, we included 180 LUAD slides and 30 LUSC slides from Center-1 for external validation.
We directly predicted the subtype of the slides using the model trained on the TCGA cohort. The experimental results are presented in Extended Data Table \ref{lab:wsi:tcga-nsclc}.
\\\\
\textbf{Lung Cancer Metastatic  Detection and Primary Site Prediction (2 classes and 6 classes).}
For metastatic detection, we utilized 1,198 WSIs from the Center-1, comprising 705 patients, including 391 primary cases and 314 metastatic cases. To predict the primary site of metastatic cancer, we curated a dataset with six distinct classes: LUAD (391 cases), breast (55 cases), colon (186 cases), kidney (25 cases), liver (34 cases), and carcinoma of unknown primary (CUP, 14 cases). For both tasks, the data were stratified into training, validation, and test sets at a ratio of 7:1:2.
Additionally, we incorporated an external validation cohort consisting of 530 WSIs (431 cases) from Center-2. For the metastatic detection task, the Center-2 cohort included 238 primary cases and 193 metastatic cases. For the primary site prediction task, the Center-2 cohort comprised 238 LUAD cases, 50 breast cases, 96 colon cases, 30 kidney cases, 10 liver cases, and 7 CUP cases.
To facilitate distinction between the datasets, we designated the Center-1 cohort as Center-1-LMD and the Center-2 cohort as Center-2-LMD. The experimental results are presented in Extended Data Table \ref{lab:wsi:lung_osp}.
\\\\
\textbf{RCC Subtyping (3 classes) on TCGA and Center-3 Cohorts.} This task contains kidney renal papillary cell carcinoma (KIRP), kidney chromophobe (KICH) and kidney renal clear cell carcinoma (KIRC) WSIs from TCGA database \cite{tcga_dataset}. 
After preprocessing, 3 KIRP slides without sufficient foreground are excluded, resulting in 297 KIRP slides, 121 KICH slides, and 519 KIRC slides for further analysis. 
For training and evaluation, we label-stratified the TCGA-RCC cohort into 7:1:2 train-validation-test (656:94:187 slides). 
Additionally, we adopted 28 KICH slides, 30 KIRC slides, and 30 KIRP slides from Center-3 (Center-3-RCC) as the external cohort.
The experimental results are reported in Extended Data Table \ref{lab:wsi:tcga-rcc}.
\\\\
\textbf{CAMELYON for Breast Metastasis Detection (2 classes).} This dataset consists of a total of 899 slides, sourced from the Cancer Metastases in Lymph Nodes Challenge 2016 (CAMELYON16, 399 slides)  \cite{c16_dataset} and the CAMELYON17 (500 slides) \cite{c17_dataset}. These slides are divided into two classes: \textbf{normal} and \textbf{metastasis}, with a distribution of 557 slides classified as normal and 341 slides classified as metastasis. After image preprocessing, a corrupted normal slide is removed,  resulting in a total of 898 WSIs. For training and evaluation, we employed a label-stratified train-validation-test split, with a ratio of 7:1:2. This resulted in 630 slides for training, 91 slides for validation, and 180 slides for testing. The experimental result is shown in Extended Data Table \ref{lab:wsi:camelyon}. 
\\\\
\textbf{Lobular and Ductal Carcinoma Subtyping on TCGA and Center-3 Cohorts (2 classes).} 
We utilized the TCGA-BRCA dataset \cite{tcga_dataset} and slides from the Center-3 for both internal and external experiments. The TCGA-BRCA dataset contains 787 slides of invasive ductal carcinoma (IDC) and 198 slides of invasive lobular carcinoma (ILC). For training and evaluation, the dataset was stratified by labels into training, validation, and testing folds in a ratio of 7:1:2, resulting in 689 slides for training, 99 slides for validation, and 197 slides for testing.
We also adopted BRCA slides (Center-3-LD) from Center-3 to conducte external validation. 
This dataset comprises 84 ILC slides and 299 IDC slides.
The subtyping results are presented in Extended Data Table \ref{lab:wsi:tcga-brca}.
\\\\
\textbf{BRACS for Breast Carcinoma Subtyping (3 classes \& 7 classes).} This dataset involves 547 breast carcinoma H\&E slides obtained from 187 patients \cite{bracs_dataset}. 
To ensure the quality of the dataset,  slides that do not meet the criteria for tumor proportion are excluded, resulting a total of 545 slides for analysis. The dataset is derived from the Breast Carcinoma Subtyping (BRCA) task, which encompasses both coarse-grained (Benign Tumors, Atypical Tumors, and Malignant Tumors) and fine-grained (Normal, Pathological Benign, Usual Ductal hyperplasis, Flat Epithelial Atypia, Atypical Ductal Hyperplasia, Ductal Carcinoma In Situ, and Invasive Carcinoma) subtyping tasks. 
For training and evaluation, a label-stratified train-validation-test split is employed, maintaining a ratio of 7:1:2 based on the fine-grained classes. This partitioning results in 382 slides for training, 54 slides for validation, and 109 slides for testing. 
Additionally, we also adopted 84 normal slides and 383 abnormal slides from Center-3 to perform external validation (Center-3-BRCA).
The coarse-grained and fine-grained classification results are presented in Extended Data Table \ref{lab:wsi:bracs-3} and \ref{lab:wsi:bracs-7}, respectively.
\\\\
\textbf{PANDA for Prostate Cancer Grade Assessment (6 classes).} This dataset is designed for prostate cancer grade assessment and consists of a total of 10,616 core needle biopsies sourced from the \textbf{P}rostate c\textbf{AN}cer gra\textbf{D}e \textbf{A}ssessment (PANDA) challenge \cite{panda_dataset}.
After preprocessing, slides without sufficient foreground are excluded, resulting in 10,212 slides available for further analysis. 
The dataset includes the following subtypes: Background or Unknown (2,724 slides), Stroma (2,602 slides), Healthy Epithelium (1,321 slides), Cancerous Epithelium - Gleason 3 (1,205 slides), Cancerous Epithelium - Gleason 4 (1,187 slides), and Cancerous Epithelium - Gleason 5 (1,163 slides).
For training and evaluation, the train-validation-test cohort is label-stratified in a ratio of 7:1:2, resulting in 7,143 slides for training, 1,019 slides for validation, and 2,040 slides for testing. The experimental results are reported in Extended Data Table \ref{lab:wsi:panda}.
\\\\
\textbf{TCGA-LUAD for Lung Adenocarcinoma TP53 Gene Mutation Prediction (2 classes).} The LUAD TP53 gene mutation prediction task consists of 469 FFPE H\&E-stained WSIs of lung adenocarcinoma sourced from the TCGA database, along with their TP53 gene mutation annotations. The slides without reported TP53 mutation status are excluded from the dataset. WSIs used in this task are classified into 2 classes, namely TP53 Mutant (248 slides), and TP53 Wildtype (221 slides). For training and evaluation, we label-stratified the WSIs into a training-validation-test cohort with a ratio of 7:1:2, including 345 slides for training, 41 slides for validation, and 83 slides for testing. The experimental results for TCGA-LUAD TP53 gene mutation prediction could be found in Extended Data Table \ref{lab:wsi:tcga-luad-tp53}.
\\\\
\textbf{The mutation Status of IDH in Glioma  (2 classes).} 
To predict the IDH mutational status in gliomas, we utilized data from TCGA-GBM and TCGA-LGG, comprising a total of 979 slides, including 722 positive slides and 257 negative slides. For model training and evaluation, the dataset was divided into training, validation, and test sets in a label-stratified ratio of 7:1:2. Additionally, to validate the robustness of our model, we incorporated an external validation set consisting of 852 slides (322 positives and 530 negatives) from EBRAINS \cite{ebrains_data}. The detailed experimental results for this task are presented in Extended Data Table \ref{lab:TCGA-IDH1}.
\\\\
\textbf{Ovarian Cancer Subtyping (5 classes) on UBC-OCEAN and Center-3 Cohorts.} 
To perform overian cancer classification, we adopted UBC-OCEAN dataset. This dataset is a collection of 538 slides obtained from the \textbf{UBC} \textbf{O}varian \textbf{C}ancer subtyp\textbf{E} cl\textbf{A}ssification and outlier detectio\textbf{N} (UBC-OCEAN) competition \cite{ubc_ocean_paper1, ubc_ocean_paper2}. 
The main objective of this competition is to accurately classify ovarian cancer subtypes into five distinct categories.
After image preprocessing, the slides without sufficient foregrounds are excluded to reduce data noise, resulting in a total of 527 slides for further analysis. The subtypes of the dataset contains Clear Cell (CC, 98 slides), Endometrioid (EC, 122 slides),  High-Grade Serous Carcinoma (HGSC, 221 slides), Low-Grade Serous Carcinoma (LGSC, 43 slides) , and Mucinous Carcinoma (MC, 43 slides). 
For training and evaluation, we label-stratified into train-validation-test folds into a ratio of 7:1:2 (369:52:104 slides).
In addition, we also adopted 100 CC, 100 HGSC, 38 LGSC, 97 EC and 35 MC slides from Center-3 as the external validation cohort (Center-3-Ovary).
The experimental results are presented in Extended Data Table \ref{lab:wsi:ubc-ocean}. 
\\\\
\textbf{Brain Tumor Subtyping (3 classes).}
To conduct brain tumor subtyping, we utilized a dataset of 1,276 slides from TCGA-GBM and TCGA-LGG, comprising 217 oligodendroglioma slides, 164 anaplastic astrocytoma slides, and 895 glioblastoma slides. For model training and evaluation, the dataset was label-stratified and divided into training, validation, and test sets with 839, 200, and 237 slides, respectively. Additionally, we incorporated an external validation set of 732 slides from the EBRAINS Digital Tumor Atlas \cite{ebrains_data}, which includes 84 oligodendroglioma slides, 89 anaplastic astrocytoma slides, and 559 glioblastoma slides. The experimental results for this task are detailed in Extended Data Table \ref{lab:EBRAINS}.
\\\\
\textbf{Lesion grade Classification of Colon Cancer.}
To perform lesion grade classification in colon cancer, we utilized the IMP-CRS-2024 dataset \cite{Oliveira2021,Neto2022,Neto2024} for experiments.
 This dataset comprises 847 non-neoplastic slides, 2,847 low-grade lesion slides, and 1,638 high-grade lesion slides. 
We adhered to the official dataset splits, using 3,300 slides from CRS2 for training, 1,132 slides from CRS1 for validation, and 900 slides from CRS\_Test for testing. Additionally, we incorporated an external validation set from Center-3, referred to as Center-3-Colon-WSI, which includes 100 non-neoplastic slides, 121 low-grade lesion slides, and 76 high-grade lesion slides. The experimental results for this task are detailed in Extended Data Table \ref{lab:WSI:colon}.
\\\\
\textbf{Head\&Neck Cancer Primary Site Prediction and TNM analysis}
We employed the HANCOCK dataset \cite{hancock} to predict the primary site of head and neck tumors and to determine the T stage of the tumors.
For primary site prediction, we utilized 708 slides, including 80 hypopharynx slides, 182 larynx slides, 317 oropharynx slides, and 129 oral cavity slides. The dataset was label-stratified and divided into 495 WSIs for training, 68 WSIs for validation, and 145 WSIs for testing.
For the TNM analysis task, we used 705 slides from the HANCOCK dataset to predict the tumor stage (T stage). This dataset comprises 259 T1 slides, 256 T2 slides, 123 T3 slides, and 67 T4 slides. The dataset was partitioned into training, validation, and testing sets with 496, 67, and 142 slides, respectively.
The experimental results for both tasks are presented in Extended Data Table \ref{lab:WSI:hancock}.
 \\\\
\textbf{ Lauren Subtyping of Gastric Cancer.}
We utilized the TCGA-STAD dataset to conduct Lauren classification. The TCGA-STAD cohort comprises 81 diffuse-type, 125 mixed-type, and 184 intestinal-type WSIs.
For model training and evaluation, we divided the dataset into training, validation, and test sets in a stratified 7:1:2 ratio based on labels. Furthermore, we incorporated 141 WSIs from the Center-5 and 319 WSIs from Center-4 as external validation cohorts. The Center-5 cohort consists of 77 diffuse-type, 33 mixed-type, and 31 intestinal-type WSIs, while the Center-4 cohort includes 143 diffuse-type, 86 mixed-type, and 90 intestinal-type WSIs. We detail the results of these three datasets for this task in Extended Data Table \ref{lab:WSI:lauren}.
\\\\
 \textbf{Vascular Invasion Detection in Gastric Cancer.}
To detect vascular invasion in gastric cancer, we utilized a dataset comprising 396 WSIs from Center-1, referred to as the Center-1-Vascular dataset. This dataset includes 197 positive cases and 168 negative cases. For the purpose of model training and evaluation, the data was partitioned into training, validation, and test sets in a ratio of 7:1:2. Additionally, we incorporated two external validation sets: 230 WSIs (140 positive and 90 negative) from Center-5 and 319 WSIs (122 positive and 197 negative) from Center-4. The experiment results of all three datasets of this task are shown in Extended Data Table \ref{lab:WSI:vascular}.
 \\\\
 \textbf{Perineural Invasion Detection in Gastric Cancer.}
To detect perineural invasion in gastric cancer, we utilized a dataset consisting of 397 WSIs obtained from Center-1. 
This dataset includes 255 positive cases and 141 negative cases. For model training and evaluation, the data was divided into training, validation, and test sets in a ratio of 7:1:2. 
Furthermore,  we incorporated two additional external validation sets: 232 WSIs (156 positive and 76 negative) from Center-5 and 319 WSIs (112 positive and 207 negative) from Center-4. See Extended Data Table \ref{lab:WSI:perineural} for experimental results.

\subsection{Survival Analysis}
Survival analysis has traditionally been employed to analyze time-to-event data in cancer studies, focusing on events such as disease progression or patient survival. 
When applied to WSIs, survival analysis offers new opportunities for studying various aspects of tissue behavior and predicting patient outcomes \cite{wiegrebe2024deep, chen2021whole}. 
By integrating survival analysis with WSIs, researchers can investigate the correlation between specific morphological features and patient outcomes. 
In our study, we adopt ABMIL \cite{abmil} for survival analysis with Negative Log-Likelihood (NLL) loss \cite{nll_surv_loss}, following a similar model architecture and training configuration as WSI classification reported in Extended Data Table \ref{tab:abmil_config}.
For CHIEF and Prov-Gigapath models, we use their pretrained slide-level FM to perform classification.

To evaluate the effectiveness of different FMs in survival analysis, we employ a train:test split of 8:2 setting and utilize the C-index metric to assess performance.
We conduct survival analysis on 14 TCGA datasets, including breast cancer (BRCA),  bladder cancer (BLCA), kidney renal clear cell carcinoma (KIRC), kidney renal papillary cell carcinoma (KIRP), lung adenocarcinoma (LUAD), stomach adenocarcinoma (STAD), lung squamous cell carcinoma (LUSC), colon adenocarcinoma (COAD), rectum adenocarcinoma (READ), glioblastoma multiforme (GBM), low-grade glioma (LGG), skin cutaneous melanoma (SKCM), cervical squamous cell carcinoma (CESC), and head-neck squamous cell carcinoma (HNSC).
Additionally, we performed external validation on the HANCOCK dataset. 
The number of slides for each dataset is reported in the Extended Data Table \ref{tab:survival_data_info}.
To ensure robust and consistent results, we maintain uniform censorship (survival status information) between the training and testing datasets.
To address the challenge of imbalanced survival times, we employ a stratified approach. 
Specifically, we sort the cases based on survival time and divide them into four equally sized bins. 
We assign the label of the bin to all cases within it.
As a result, we label-stratify the train-test cohort into an 8:2 ratio.
The experimental results are presented in Extended Data Table \ref{tab:avg_survival}-\ref{tab:survival2-2}. 

\subsection{ROI Classification}
For patch-level tissue classification tasks, we evaluate the transfer performance and representation ability of different FMs using a linear probe, inspired by the approach employed in DINOv2. \cite{oquab2023dinov2, balestriero2023cookbook}.
Initially, we extract features from the images using the pretrained FMs. 
Subsequently, we employ a linear layer for performing classification. To optimize the model, we utilize AdamW \cite{adamw} with an initial learning rate of 5e-4 and weight decay of 1e-5. Additionally, we incorporate a cosine annealing scheduler to update the learning rate during training \cite{cosineannellr}. 
In order to obtain the best model, we set the maximum number of epochs to 3000 and implemented early stopping with patience of 100 epochs. For ensuring fair comparison, we maintain a consistent batch size of 256 across all methods.

To evaluate the performance of patch-level tissue classification, we consider the impact of class imbalance in the dataset and assess the metrics of balanced accuracy, weighted F1 score, and AUC. These metrics provide comprehensive insights into the classification performance, accounting for both accuracy and the ability to handle imbalanced class distributions.
Specifically, we compare the FMs across 16 tasks.
For all experiments in this section, we estimate the model performance using non-parametric bootstrapping with 1,000 bootstrap replicates. 
We employ Torchmetrics \cite{torchmetrics} for bootstrapping sampling and obtain the mean and standard deviation of the metrics. 
The experimental results are presented in Extended Data Table \ref{lab:linear:crc-100k} to Extended Data Table \ref{lab:linear:GasHisDB}. 
Furthermore, we report the average performance of the patch-level tissue classification results across 12 tasks in Table \ref{lab:linear:avg}, demonstrating the superior performance of GPFM.

\textbf{CRC-100K for Colorectal Cancer (CRC) Tissue Classification (9 classes).} This dataset consists of NCT-CRC-HE-100K and CRC-VAL-HE-7K \cite{CRC100K}. The NCT-CRC-HE-100K comprises 100,000 non-overlapping 224$\times$224 patches obtained from 86 human cancer tissue slides stained with H\&E. These tissue slides were sourced from the NCT biobank (National Center for Tumor Diseases) and the UMM pathology archive (University Medical Center Mannheim). Concurrently, CRC-VAL-HE-7K consists of 7,180 224$\times$224 images extracted from 50 patients diagnosed with colorectal adenocarcinoma. 
The subtypes of this dataset contains: Adipose (ADI, 11,745 ROIs), Background (BACK, 11,413 ROIs), Debris (DEB, 11,851 ROIs), Lymphocytes (LYM, 12,191 ROIs), Mucus (MUC, 9,931 ROIs), Smooth muscle (MUS, 14,128 ROIs), Normal colon mucosa (NORM, 9,504 ROIs), Cancer-associated stroma (STR, 10,867 ROIs), Colorectal adenocarcinoma epithelium (TUM, 15,550 ROIs). 
For training and evaluation, we use the official train-test split(100,000: 7,180). 
The experimental results are reported in Extended DataTable \ref{lab:linear:crc-100k}.
\\\\
\textbf{CCRCC-TCGA-HEL for CCRCC Tissue Classification (4 classes).} This dataset \cite{brummer2023computational} comprises a total of 52,713 regions of interest (ROI) images, each with dimensions of 300$\times$300 pixels. 
The dataset encompasses six distinct categories, namely: renal cancer (cancer,  13,057 ROIs), normal renal tissue (normal, 8,652 ROIs), stromal tissue (stroma, 5,460 ROIs), red blood cells (blood, 996 ROIs), empty background (empty, 16,026 ROIs), and other textures, including necrotic, torn, and adipose tissue (other, 8,522 ROIs). 
The image tiles were selected at random from two sources: the TCGA-KIRC WSIs and the Helsinki datasets.
For training and evaluation, we focused on four specific categories: cancer, stroma, normal, and blood. 
This decision was made due to the potential ambiguities associated with the $"other"$ category and the lack of meaningful information conveyed by the $"empty"$ category. 
We randomly shuffle the samples and set the train-test split as a 22530:5635 ratio. The experimental results are shown in Extended Data Table \ref{lab:linear:ccrcc-tcga_hel}.
\\\\
\textbf{BACH for Breast Cancer Tissue Classification (4 classes).} The dataset \cite{aresta2019bach} is used for the breast cancer subtyping task and consists of 400 images with dimensions of 2048$\times$1536 pixels. The dataset is labeled into four classes: Normal (100 ROIs), Benign (100 ROIs), In situ carcinoma (100 ROIs), and Invasive carcinoma (100 ROIs).
For training and evaluation, all ROIs are resized to $224\times224$ pixels and we label-stratified the train-test with a ratio of 8:2 (320: 80 ROIs). 
The experimental results are summarized in Extended Data Table \ref{lab:linear:bach}.\\\\
\textbf{BreakHis for Breast Cancer Image Classification (2 classes).} This dataset \cite{spanhol2015dataset} is collected for breast cancer histopathological image classification containing two main groups: benign tumors (2,480 ROIs) and malignant tumors (5,429 ROIs). 
The ROIs in this dataset have 4 different magnifications (40$\times$, 100$\times$, 200$\times$, and 400$\times$). 
For training and evaluation, we resized all images to 224$\times$224 pixels to ensure consistency and label-stratified the train-test with a ratio of 8:2 (6,327:1,582 ROIs).
The experimental results are presented in Extended Data Table \ref{lab:linear:bach}.
\\\\
\textbf{UniToPatho for CRC Polyp Classification (6 classes).} This dataset is a meticulously annotated dataset comprising 9,536 H\&E stained patches extracted from 292 WSIs \cite{barbano2021unitopatho}. 
The primary objective of this dataset is to facilitate the training of deep neural networks for the classification of colorectal polyps and the grading of adenomas.  
The annotations include 6 classes: Normal tissue (950 ROIs), Hyperplastic Polyp (545 ROIs),  Tubular Adenoma with High-Grade dysplasia (454 ROIs),  Tubular Adenoma with Low-Grade dysplasia (3,618 ROIs), Tubulo-Villous Adenoma with High-Grade dysplasia (916 ROIs), and Tubulo-Villous Adenoma with Low-Grade dysplasia (2,186 ROIs). 
For training and evaluation, we use the official train-test split (6,270:2,399 ROIs).  
The experimental result is shown in Extended Data Table \ref{lab:linear:unitopatho}.
\\\\\textbf{CRC-MSI for MSI Screening  (2 classes).} This dataset consists of 51,918 512$\times$512 histological images of colorectal cancer obtained from the TCGA database \cite{kather2019deep}. 
In addition to the visual data, information regarding the Microsatellite Instability (MSI) status of each patient was obtained. Patients were classified into two categories: those with MSI-H (high MSI) and those with either MSI-L (low MSI) or MSS (Microsatellite Stable), collectively referred to as NonMSIH. 
For training and evaluation, we use the official train-test split (19,557:32,361 ROIs). 
The experimental result is shown in Extended Data Table \ref{lab:linear:CRC-MSI}.
\\\\\textbf{PanCancer-TCGA for Tissue Classification (32 classes).} This dataset comprises 271,170 images with dimensions of 256 × 256 pixels \cite{komura2022universal}. The images were extracted from 8,736 histopathology WSIs obtained from the TCGA database. These images represent various cancer types and are annotated with following 32 classes: Head and Neck Squamous Cell Carcinoma (11,790 ROIs),  Bladder Urothelial Carcinoma (9,990 ROIs), Uterine Carcinosarcoma (2,120 ROIs), Colon Adenocarcinoma (8,150 ROIs), Lymphoid Neoplasm Diffuse Large B-cell Lymphoma (8,40 ROIs), Lung Squamous Cell Carcinoma (16,560 ROIs), Brain Lower Grade Glioma (23,530 ROIs), Esophageal Carcinoma (3,380 ROIs), Pheochromocytoma And Paraganglioma (1,350 ROIs), Sarcoma (13,480 ROIs), Glioblastoma Multiforme (23,740 ROIs), Adrenocortical Carcinoma (4,980 ROIs), Uterine Corpus Endometrial Carcinoma (12,480 ROIs), Prostate Adenocarcinoma (9,810 ROIs), Breast Invasive Carcinoma (23,690 ROIs), Stomach Adenocarcinoma (9,670 ROIs), Pancreatic Adenocarcinoma (4,090 ROIs), Skin Cutaneous Melanoma (10,060 ROIs), Ovarian Serous Cystadenocarcinoma (2,520 ROIs), Thymoma (3,600 ROIs), Lung Adenocarcinoma (16,460 ROIs), Kidney Renal Papillary Cell Carcinoma (6,790 ROIs), Testicular Germ Cell Tumors (6,010 ROIs), Kidney Renal Clear Cell Carcinoma (11,650 ROIs), Rectum Adenocarcinoma (1,880 ROIs), Cholangiocarcinoma (900 ROIs), Cervical Squamous Cell Carcinoma And Endocervical Adenocarcinoma (6,270 ROIs), Thyroid Carcinoma (11,360 ROIs), Mesothelioma (2,090 ROIs), Uveal Melanoma (1,640 ROIs), Liver Hepatocellular Carcinoma (8,370 ROIs), Kidney Chromophobe (2,460 ROIs). 
For training and evaluation, the train-test split is set to 21,736:54,342 ROIs. 
The experimental results are summarized in Extended Data Table \ref{lab:linear:pancancer-tcga} indicating that GPFM outperforms other models across all three metrics.
\\\\
\textbf{TIL classification (2 classes).}We use PanCancer-TIL dataset \cite{abousamra2022deep, saltz2018spatial} for tumor infiltrating lymphocyte (TIL) classification. It includes 304,097 images with a size of 100$\times$100 pixels at 0.5 micrometers per pixel. 
The images are labeled with the following two classes: TIL-positive (if there are at least two TILs present in the image, 54,910 ROIs) and TIL-negative (249,187 ROIs). 
For training and evaluation, we use the official train-val-test split (209,221:38,601:56,275 ROIs). 
To ensure consistency, we resize all images to 256$\times$256 pixels. We employ the validation set to select the best model and subsequently evaluate its performance on the test set. 
Additionally, we also adopted the data from Center-3 to conduct external validation.
The TIL-negative samples (8,361 ROIs) were obtained from healthy lymph nodes of pan-cancer type, and TIL-positive samples (10,131 ROIs) were obtained from the marked cancerous areas on the lymph nodes with metastasis.
The experimental results are presented in Extended Data Table \ref{lab:linear:pancancer-til}. 
\\\\
\textbf{ESCA for Esophageal Carcinoma Subtyping (11 classes).} This dataset \cite{tolkach2023artificial} comprises 367,229 images with size of 256$\times$256 pixels. 
These patches were obtained from 320 H\&E WSIs of esophageal adenocarcinoma and adenocarcinoma of the esophagogastric junction, specifically, 22 slides from University Hospital Cologne (UKK), 62 slides from Landesklinikum Wiener Neustadt (WNS), 22 slides from  TCGA, and 214 slides from the University Hospital Berlin Charite (CHA). 
These images were annotated and labeled with one of eleven classes: adventitia (71,131 ROIs), lamina propria mucosae (2,173 ROIs), muscularis mucosae (2,951 ROIs), muscularis propria (83,358 ROIs), regression tissue (56,490 ROIs), mucosa gastric (44,416 ROIs), muscosa oesophagus (18,561 ROIs), submucosa (22,117 ROIs), submucosal glands (1,516 ROIs), tumor (63,863 ROIs), and ulceration (753 ROIs).
For training and evaluation, we adopted CHA dataset, containing 178,187 ROIs, as the training set, and we combined the UKK, WNS, and TCGA datasets as a single testing cohort consisting of 189,142 ROIs.
In our experiment, all images were resized to 224 × 224 pixels to ensure consistency, the experimental result is shown in Extended Data Table \ref{lab:linear:esca}.
\\\\
\textbf{PCAM for Metastatic Tissue Classification  (2 classes).} This dataset consists of 327,680 color images (96 $\times$ 96 pixels) extracted from CAMELYON16 \cite{Veeling2018-qh, c16_dataset}. Each image is annotated with a binary label indicating the presence of metastatic tissue.  
For training and evaluation, we adopt the official train-validation-test split (262,144: 32768:32768 ROIs) and resize all images to 224$\times$224 in our experiment. 
The experimental results are presented in Extended Data Table \ref{tab:linear:pcam}. 
\\\\
\textbf{WSSS4LUAD for Lung Adenocarcinoma Tissue Classification (3 classes).} This dataset \cite{han2022wsss4luad, Han2022Multilayer} was collected from Guangdong Provincial People's Hospital (GDPH) and TCGA. 
It consists of 10,091 images with the following three common and meaningful tissue types: tumor epithelial tissue (6,579 ROIs), tumor-associated stroma tissue (1,680 ROIs), and normal tissue (1,832 ROIs). 
It is worth noting that, in WSSS4LUAD dataset, one image may belong to several categories. 
To avoid ambiguity, we only choose one label for each image based on the order of diagnosability (i.e., from tumor epithelial tissue to normal tissue).
For training and evaluation, all images were resized to 224$\times$224 pixels and we label-stratified the train-test with a ratio of 8:2 (8,072:2019 ROIs).
The experimental results are presented in Extended Data Table \ref{lab:linear:wsss4luad}. 
\\\\
\\\\
\textbf{Chaoyang for Colon Tissue Classification (4 classes).} This dataset \cite{zhu2021hard} contains colon patches from Chaoyang hospital including 1,816 normal ROIs, 1,163 serrated ROIs, 2,244 adenocarcinoma ROIs, and 937 adenoma ROIs. 
For training and evaluation, we resize all patches to 224$\times$224 pixels and use official train-test split (4,021: 2,139 ROIs). 
Additionally, we adopted 9,214 normal ROIs, and 11,854 adenoma ROIs from Center-3 for external validation. 
The experimental results are presented in Extended Data Table \ref{lab:linear:chaoyang}. 
\\\\\

\textbf{GasHisDB for Gastric Tissue Classification (2 classes)}. The dataset consists of a
total of  13,124 160$\times$160 abnormal images, and 20,160 normal images.
For training and evaluation, we resize all patches to 224$\times$224 pixels and label-stratified the train-test with a ratio of 8:2 (26,627: 6,657 ROIs).
Additionally, we adopted the 709 normal tissues and 1,828 abnormal tissues from Center-3 to perform external validation. Results can be found in Extended Data Table \ref{lab:linear:GasHisDB}.

\subsection{Pathological Tissue Retrieval} 
In the linear probe evaluation tasks, we extract semantically-rich features using different FMs and then construct a task-specific classifier.
These features are not only applicable for supervised learning but also prove to be valuable for image-to-image retrieval. The primary goal of this application is to retrieve images that share the same class label as a given query image, thereby facilitating efficient image retrieval.
The CRC-100K dataset comprises 100,000 non-overlapping 224$\times$224 patches extracted from 86 human cancer tissue slides stained with H\&E for training purposes. Additionally, it includes 7,180 images with 224$\times$224 pixels extracted from 50 patients diagnosed with colorectal adenocarcinoma for testing.
The dataset consists of multiple classes, including Adipose (ADI, 11,745 ROIs), Background (BACK, 11,413 ROIs), Debris (DEB, 11,851 ROIs), Lymphocytes (LYM, 12,191 ROIs), Mucus (MUC, 9,931 ROIs), Smooth muscle (MUS, 14,128 ROIs), Normal colon mucosa (NORM, 9,504 ROIs), Cancer-associated stroma (STR, 10,867 ROIs), and Colorectal adenocarcinoma epithelium (TUM, 15,550 ROIs).
For training and evaluation, we utilize the official train-test split, with 100,000 samples for training and 7,180 samples for testing. 

To initiate the pathological tissue image retrieval process, we begin by embedding all images using pretrained FMs. Next, each image in the test set is treated as a query and compared against the images in the training set. To ensure that all features have a comparable impact on the computation of similarity, we independently normalize each feature component to the range [0, 1] \cite{retrieval_norm}. This normalization process involves calculating the mean and variance of the training set features, which are then used to normalize both the training and testing features.

To evaluate the similarity between the query image and candidate images, we employ the L2 distance metric. A lower distance value indicates a higher degree of similarity between the images.
The retrieved images are subsequently ranked based on their similarity scores, and the corresponding class labels are utilized to evaluate the success of the retrieval process. To assess the retrieval performance, we employ evaluation metrics such as Acc@K, where K represents the top K retrieved images (typically 1, 3, and 5). Similar to the patch-level classification evaluation, we estimate the model performance using non-parametric bootstrapping with 1,000 bootstrap replicates. 
Due to the limitation of the number of classes, we primarily focus on the CRC tissue retrieval tasks, and the experimental results are presented in Table \ref{lab:ret:crc-100k}.\\\\
\subsection{Pathology Visual Question Answering}
The objective of this subsection is to evaluate the performance of our proposed pathology  FM in the context of Visual Question Answering (VQA) tasks. To this end, we utilized the PathVQA dataset \cite{he2020pathvqa} and the WSI-VQA dataset \cite{chen2025wsi} as benchmark datasets for our experiments. These datasets provide a comprehensive framework for assessing the model's ability to comprehend and reason about both patch-level and WSI-level visual pathology information, enabling accurate responses to queries related to observed pathological features.\\\\
\textbf{Patch-level VQA on PathVQA dataset.}
To evaluate the effectiveness of FMs in pathology VQA, we utilize the PathVQA dataset \cite{he2020pathvqa}, which is the largest and most widely used dataset in the pathology domain for VQA tasks. The dataset consists of 32,799 image-question-answer triplets, divided into three subsets: a training set (50\%) containing 16,400 triplets used for model training, a validation set (30\%) comprising 9,840 triplets for hyperparameter tuning and overfitting prevention, and a test set (20\%) including 6,560 triplets for final model performance evaluation.
To ensure a rigorous and comparative analysis, we adopt the Multi-modal Unified Medical Captioning (MUMC) method \cite{MUMC}, which currently represents the state-of-the-art approach on the PathVQA dataset. The MUMC method has exhibited superior performance in leveraging the synergies between visual and textual information for medical image understanding tasks.

The VQA model architecture consists of four main components: the image encoder, text encoder, multimodal encoder, and answering decoder. The image encoder is responsible for capturing domain-specific visual features. We employ various pathology FMs as the image encoder. During the fine-tuning process, the weights of the image encoder are kept frozen to preserve the integrity of the pre-trained visual representations and focus on learning task-specific multimodal interactions.
The text encoder is designed to process textual inputs, specifically the questions related to the pathology images. We utilize a 6-layer transformer architecture for the text encoder. It is initialized with the first six layers of a pre-trained BERT model, which has a strong track record in language understanding tasks and has demonstrated excellent performance in several medical and clinical applications.
The multimodal encoder is responsible for fusing visual and textual features. It consists of the last six layers of the pre-trained BERT model and incorporates cross-attention mechanisms at each layer. This integration enables the model to learn robust multimodal interactions, which are crucial for effectively answering questions based on the provided pathology images.
The answering decoder, which comprises a 6-layer transformer, receives the multimodal embeddings generated by the previous components and generates text tokens corresponding to the answers.
During the training stage, we fine-tuned the model for a total of 100 epochs using a batch size of 8. To optimize the model, we employed the AdamW optimizer with an initial learning rate of 2 × 1e-5. Throughout the training process, the learning rate was decayed to 1e-8 to ensure gradual convergence and stability.
To evaluate the performance of the VQA models, we adopt accuracy as the metric, which is consistent with previous research studies \cite{gong2022vqamix, MUMC}. We treat VQA as a generative task by calculating similarities between the generated answers and the candidate list of answers, selecting the answer with the highest score as the final answer.\\\\
\textbf{WSI-level VQA on WSI-VQA dataset.}
The dataset comprises 977 WSIs and 8,671 question-answering pairs, which are divided into three subsets: training, validation, and test. 
Specifically, the training subset consists of 804 WSIs and 7,139 pairs, while the validation subset includes 87 WSIs and 798 pairs. The test subset contains 86 WSIs and 735 pairs. In the close-ended portion of the test subset, the correct answers are distributed as follows: 151 for option A, 107 for B, 86 for C, and 46 for D. 
For the WSI-VQA dataset, we adhere to the implementation framework proposed by Chen et al. \cite{chen2025wsi}, with modifications limited to replacing the visual features.
The experimental result is reported in Table \ref{tab:vqa}.
\\\\
\subsection{Pathology Report Generation}
The task of pathology report generation is inspired by existing works on Chest X-ray and other medical report generation \cite{chen2022cross,chen2020generating,li2022cross}. 
In this task, the report generation model takes a WSI as input and generates the corresponding pathology report. 
Specifically, the input WSI is first processed by FMs to extract an initial representation. This representation is then fed into the encoder-decoder architecture of report generation models to produce the decoded pathology report. 
During this process, the visual encoder further processes the initial representations of WSIs through specific designs \cite{chen2020generating, li2022cross, guo2024histgen} to obtain the optimal WSI features for the report decoding stage. The text decoder of the model then utilizes these features for report generation. A good initial representation of WSI could significantly facilitate both the visual encoding and textual decoding stages. Consequently, the quality of the generated report is directly influenced by the representations provided by the FMs. In this task, we adopt the HistGen model \cite{guo2024histgen} for WSI report generation and set the learning rate to 1e-4, and weight decay to 0.8 per epoch. 
The model is trained for 40 epochs with batch size 1 using features extracted from different FMs. 

To evaluate the report generation performance of FMs, we utilize natural language generation metrics including BLEU \cite{papineni2002bleu}, METEOR \cite{denkowski2011meteor}, and ROUGE-L \cite{lin2004rouge}, in which BLEU is further split into BLEU-1, BLEU-2, BLEU-3, and BLEU-4 for evaluation of different granularity. These metrics provide a robust framework for evaluating machine-generated text, each bringing unique strengths to assess different aspects of text quality. This task is conducted on the TCGA WSI-Report dataset proposed in \cite{guo2024histgen} containing 7,690 WSIs and the paired diagnosis reports in total, and the PatchGastricADC dataset \cite{tsuneki2022inference} which includes 991 pairs of histological descriptions and WSIs of stomach adenocarcinoma endoscopic biopsy specimens. A 7:1:2 train-validation-test split is employed and the experimental results are reported in Extended Data Table \ref{lab:WSI-report} and \ref{lab:PatchGastricADC}. 

To assess the robustness of each FM in report generation, we conducted a stratified analysis of the TCGA WSI-Report dataset based on cancer types, focusing on major organ cancers including breast, lung, and kidney. The stratified evaluation results are presented in Table \ref{lab:WSI-report-stratify}. Additionally, we collaborated with an experienced pathologist to perform a rigorous human evaluation of the reports generated by different models. The evaluation employed a four-tier scoring system (illustrated in Fig. \ref{fig:score_criteria}d), and the scoring distribution and average score of each FM are summarized in Table \ref{lab:WSI-report-human}.
 \subsection{Computing Software and Hardware}
In this project, we utilized PyTorch \cite{paszke2019pytorch} (version 2.1.2 with CUDA 12.1) for both pretraining and evaluating downstream tasks. 
To pretrain the GPFM model, we incorporated established FMs, namely UNI, Phikon, and CONCH, as additional teachers. 
It is worth noting that access to UNI and CONCH requires a prior application submission.
The GPFM model was pretrained using the FullyShardedDataParallel (FSDP) technique on 2$\times$8 80GB NVIDIA H800 GPU nodes. All other data processing and evaluation for downstream tasks were carried out on a server equipped with 8$\times$ NVIDIA RTX 3090 GPUs.
To assess the model's performance, we employed Torchmetrics \cite{torchmetrics} and Scikit-learn \cite{pedregosa2011scikit} for metric evaluation. For WSI processing, we relied on openslide-python (version 1.2.0) \cite{openslide} and the CLAM \cite{CLAM} codebase. Pathology VQA evaluation was conducted using the MUMC \cite{MUMC} codebase. Furthermore, for histology report generation, we utilized the HistGen \cite{guo2024histgen} codebase.
Please refer to \textbf{Extended Data} Table \ref{tab:code_source} for a comprehensive list of the aforementioned models and libraries utilized in this study.
\section{Data availability}
This study incorporates a total of 56 datasets. Out of these, 33 datasets are utilized for pretraining, and a subset of them is also employed for evaluation purposes. The remaining 23 datasets are specifically dedicated to downstream task evaluation. 
For detailed information on the public data used in this project, refer to the Extended Data Table \ref{tab:data_links}.
For the data from Center-1 to Center-5, these datasets are not publicly available due to patient privacy obligations, institutional review board requirements, and data use agreements. 
However, researchers interested in accessing de-identified data may submit a reasonable request directly to the corresponding authors, subject to obtaining the necessary ethical approvals and complying with institutional policies.
The splits of the dataset can be found in our GitHub repository.

\section{Code availability}
The code and weights of the GPFM will be made available upon acceptance.
The code and weights of the GPFM have been made available on GitHub (\url{https://github.com/birkhoffkiki/GPFM/}).
\section{Ethics declarations}
This project has been reviewed and approved by the Human and Artefacts Research Ethics Committee (HAREC) of Hong Kong University of Science and Technology. The protocol number is HREP-2024-0212.

\section{Author contributions}
J.M. conceived the study and designed the experiments. 
J.M., Z.G., and F.Z. collected the data for self-supervised learning and downstream task evaluation. 
J.M. performed model pretraining and conducted patch-level tissue classification tasks. 
Y.L. and X.J. participated in discussions regarding the design of the self-supervised learning framework and were responsible for reproducing the foundation models.
J.M., F.Z., and Y.C. evaluated the weakly-supervised WSI classification task. 
Z.G. performed survival analysis and report generation tasks. 
Y.W. and Y.X. conducted pathological image retrieval and curated the data of WSI-report pairs. 
Z.Z. performed pathological image visual question answering. 
C.J. assisted in result analysis and the creation of visualized attention maps. 
J.M. and Z.G. prepared the manuscript with input from all co-authors. 
R.C.K.C,  A.H., and L.L. provided medical guidance.
L.L., J.L., C.Z., D.L. provided and preprocessed data for some downstream tasks.
S.Z. and F.Y. provided preprocessed data for external validation.
P.F., J.W. offered insightful suggestions for the experimental design and thoughtfully directing the research trajectory.
K.T.C reviewed and refined the draft.  
H.C. supervised the research. 
\section*{Acknowledgments}
This work was supported by the National Natural Science Foundation of China (No. 62202403), 
Hong Kong Innovation and Technology Commission (Project No. PRP/034/22FX and No. ITCPD/17-9),
Research Grants Council of the Hong Kong Special Administrative Region, China (No. R6003-22) and HKUST Frontier Technology Research for Joint Institutes with Industry (No. OKT24EG01).
We thank the support of HKUST SuperPod for providing the GPU platform for foundation model pretraining.

\noindent
\bibliography{sn-article}
\begin{appendices}

\section{Extended Data}\label{secA1}


\begin{algorithm*}
\caption{The PyTorch-like pseudocode of the Expert Knowledge Distillation module.}\label{alg:distill}
\begin{algorithmic}[1]
\Require \texttt{T$_a$}, \texttt{T$_b$}, and \texttt{T$_c$} \textcolor{blue}{\# off-the-shelf foundation models, we used phikon, uni, and conch in this study.}
\Require \texttt{S} \textcolor{blue}{\# student model}
\Require \texttt{v} \textcolor{blue}{\# global views}
\State \texttt{sc}, \texttt{sp =} \texttt{S(v)} \textcolor{blue}{\# [CLS] token and [patch] token encoded by student}
\State \texttt{ac}, \texttt{ap = }\texttt{T$_a$(v)}\textcolor{blue}{\# [CLS] token and [patch] token encoded by T$_a$}
\State \texttt{bc}, \texttt{bp = }\texttt{T$_b$(v)}
\State \texttt{cc}, \texttt{cp = }\texttt{T$_c$(v)}
\State \texttt{d$_{ac}$ = 1-cos(sc,ac)}
\State \texttt{d$_{bc}$ = 1-cos(sc,bc)}
\State \texttt{d$_{cc}$ = 1-cos(sc,cc)}
\State \texttt{d$_c$ = $\alpha$d$_{ac}$ + $\beta$d$_{bc}$ + $\gamma$d$_{cc}$}

\State \texttt{d$_{ap}$ = $\eta$*(1-cos(sp,ap) + $\theta$*SmoothL1(sp,ap)}
\State \texttt{d$_{bp}$ = $\eta$*(1-cos(sp,bp) + $\theta$*SmoothL1(sp,bp)}
\State \texttt{d$_{cp}$ = $\eta$*(1-cos(sp,cp) + $\theta$*SmoothL1(sp,cp)}
\State \texttt{d$_p$ = $\mu$d$_{ap}$ + $\lambda$d$_{bp}$ + $\phi$d$_{cp}$}
\State \texttt{d = d$_c$ + d$_p$}
\end{algorithmic}
\end{algorithm*}

\begin{figure*}[h]
    \centering
    \includegraphics[width=1\linewidth]{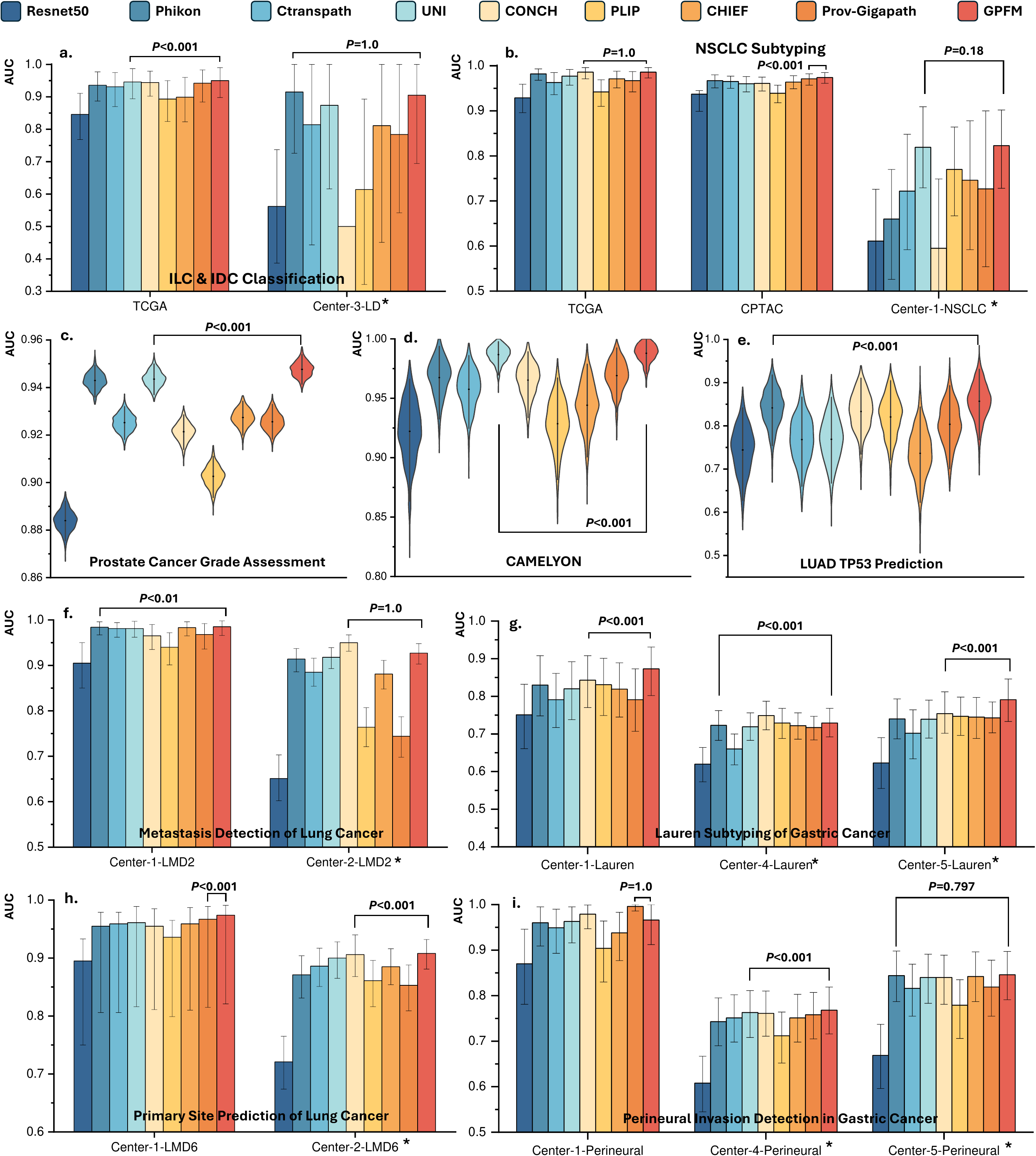}
    \caption{\textbf{Extended Results of WSI Classification.}
\textbf{a.} Performance comparison of foundation models in ILC and IDC classification.
\textbf{b.} NSCLC subtyping performance across models.
\textbf{c-e.} Model performance in prostate cancer grading, breast cancer metastasis detection, and LUAD TP53 mutation prediction, respectively.
\textbf{f-i.} Extended evaluation including lung cancer metastasis detection, gastric cancer Lauren subtyping, lung cancer primary site prediction, and gastric cancer perineural invasion detection.
Violin plots show the distribution of 1,000 bootstrap replicates.
Error bars represent 95\% CI.
External validation cohorts are marked with *.
}
    \label{fig:WSI_ext1}
\end{figure*}
\begin{figure*}
    \centering
    \includegraphics[width=1\linewidth]{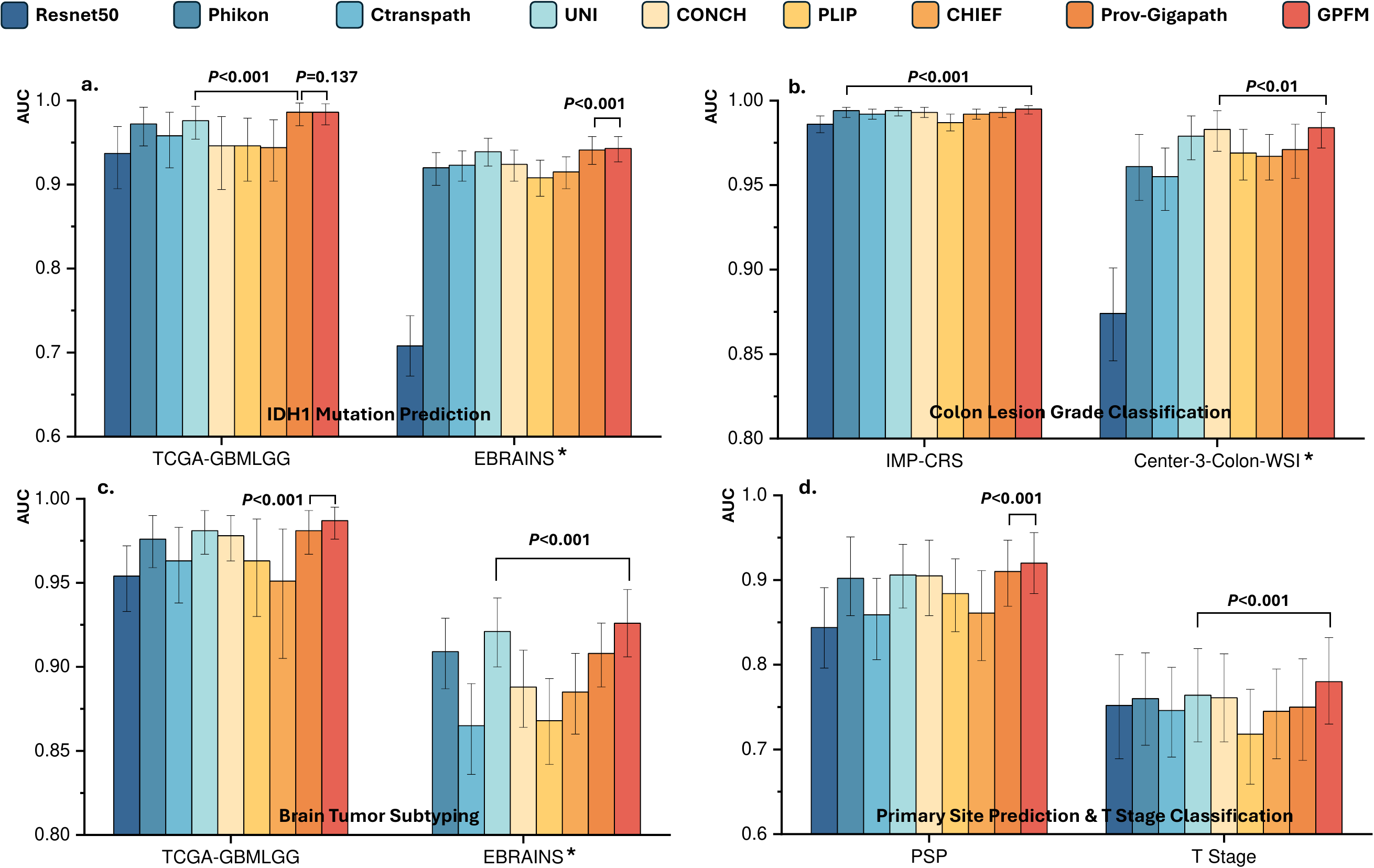}
    \caption{\textbf{Extended Results of WSI Classification.}
\textbf{a.} IDH-1 mutation prediction in brain tumors.
\textbf{b.} Lesion grading in colon cancer.
\textbf{c.} Brain tumor subtyping performance.
\textbf{d.} Dual-task evaluation: primary site prediction and T-stage classification in head \& neck cancer.
Error bars represent 95\% CI.
External validation cohorts are marked with *.
}
    \label{fig:WSI_ext2}
\end{figure*}
\begin{figure*}
    \centering
    \includegraphics[width=1\linewidth]{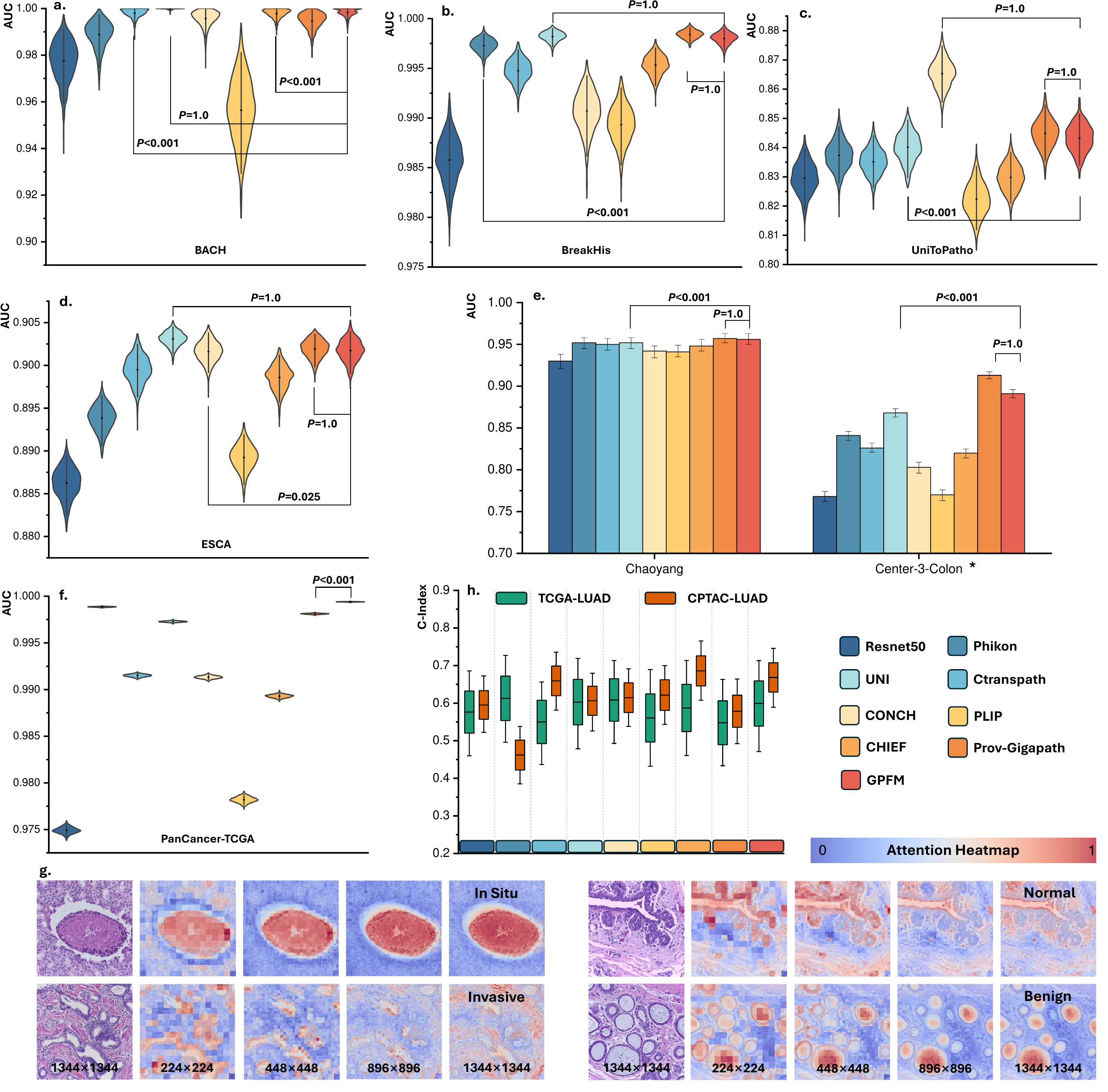}
    \caption{\textbf{Extended Result of ROI Classification Tasks.}
    \textbf{a-d.} The AUC of foundation models on BACH, BreakHis, UniToPatho, and ESCA, respectively.
    \textbf{e.} The colon tissue classification performance. The Chaoyang and Center-3-Colon serve as internal and external, respectively.
    \textbf{f.} The performance of pancancer classificaiton of different foundation models.
    \textbf{g.}Attention heatmap of GPFM across various image resolutions for BRCA subtyping in BACH dataset. The colored squares represent the 14×14 [PATCH] tokens encoded by the GPFM model. The heatmap values indicate the similarity between each [PATCH] token and the [CLS] token generated by the last layer of GPFM, measured using Euclidean distance.  The consistent attention patterns observed across varying image resolutions and tissue types underscore the robust capabilities of the GPFM model.
    \textbf{h.} Results on TCGA-LUAD data and the CPTAC-LUAD cohort. The survival prediction model was trained on the TCGA-LUAD cohort and subsequently tested on the CPTAC-LUAD cohort. 
    The box limits represent the standard error.
    For all subfigures, the error bar indicates the 95\% CI.
    }
    \label{fig:extra_roi_results}
\end{figure*}
\begin{figure*}
    \centering
    \includegraphics[width=1\linewidth]{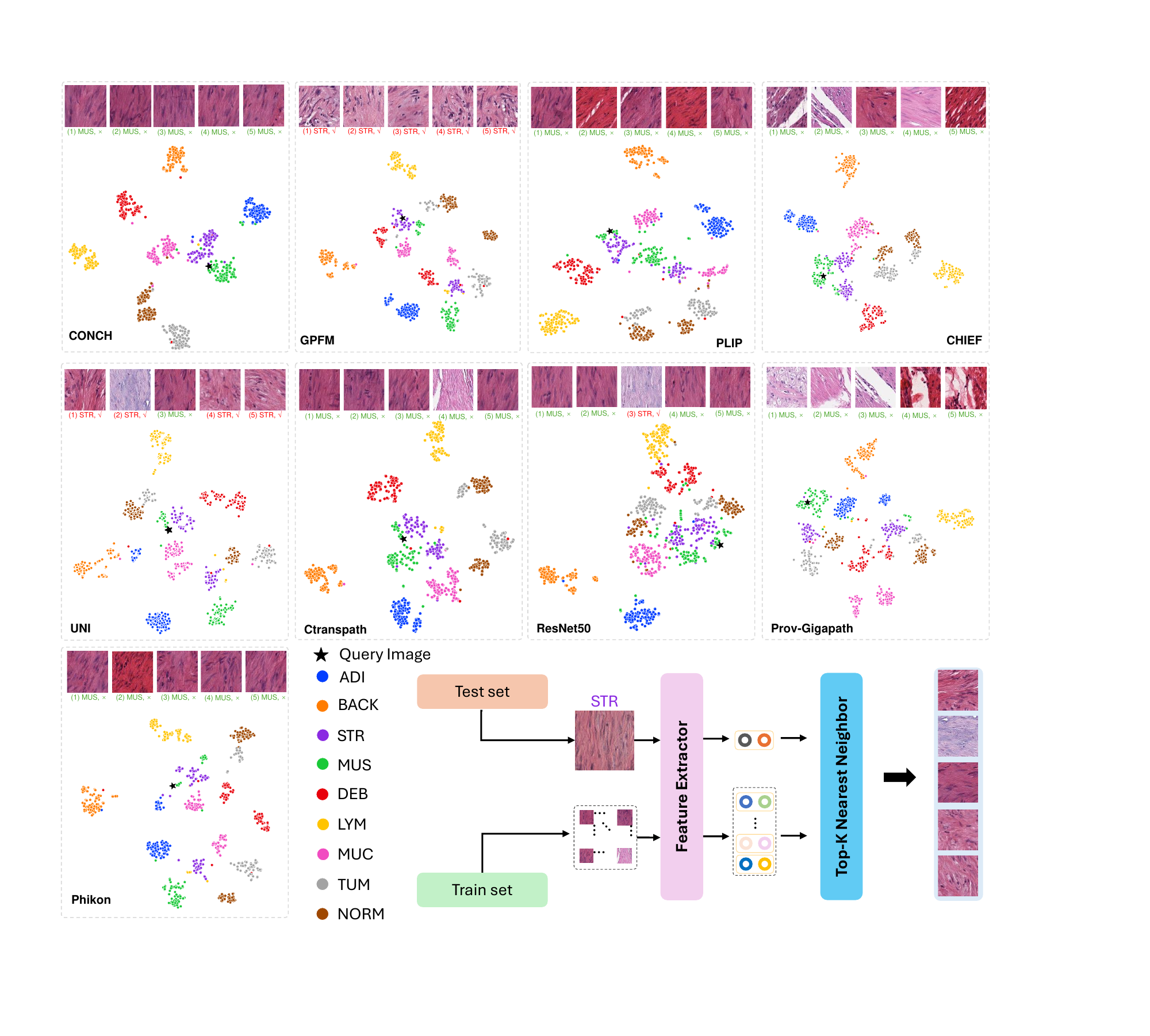}
    \caption{\textbf{Overview of Pathology  ROI Tissue Retrieval.} The central figure illustrates the framework for pathology tissue ROI retrieval.
    The surrounding figures visualize the distribution of features extracted by different models using t-SNE dimensionality reduction to 2D. For each class, 100 samples from the test set were used, and together with the query image, a total of 901 samples were subjected to the t-SNE analysis. The different classes are distinctly colored in the 2D t-SNE plot.
    The retrieved top-5 images for the query are also shown, demonstrating the GPFM's performance on this pathology tissue retrieval task.
}
    \label{fig:roi_retrieval}
\end{figure*}
\begin{figure*}[h]
    \centering
    \includegraphics[width=1.0\linewidth]{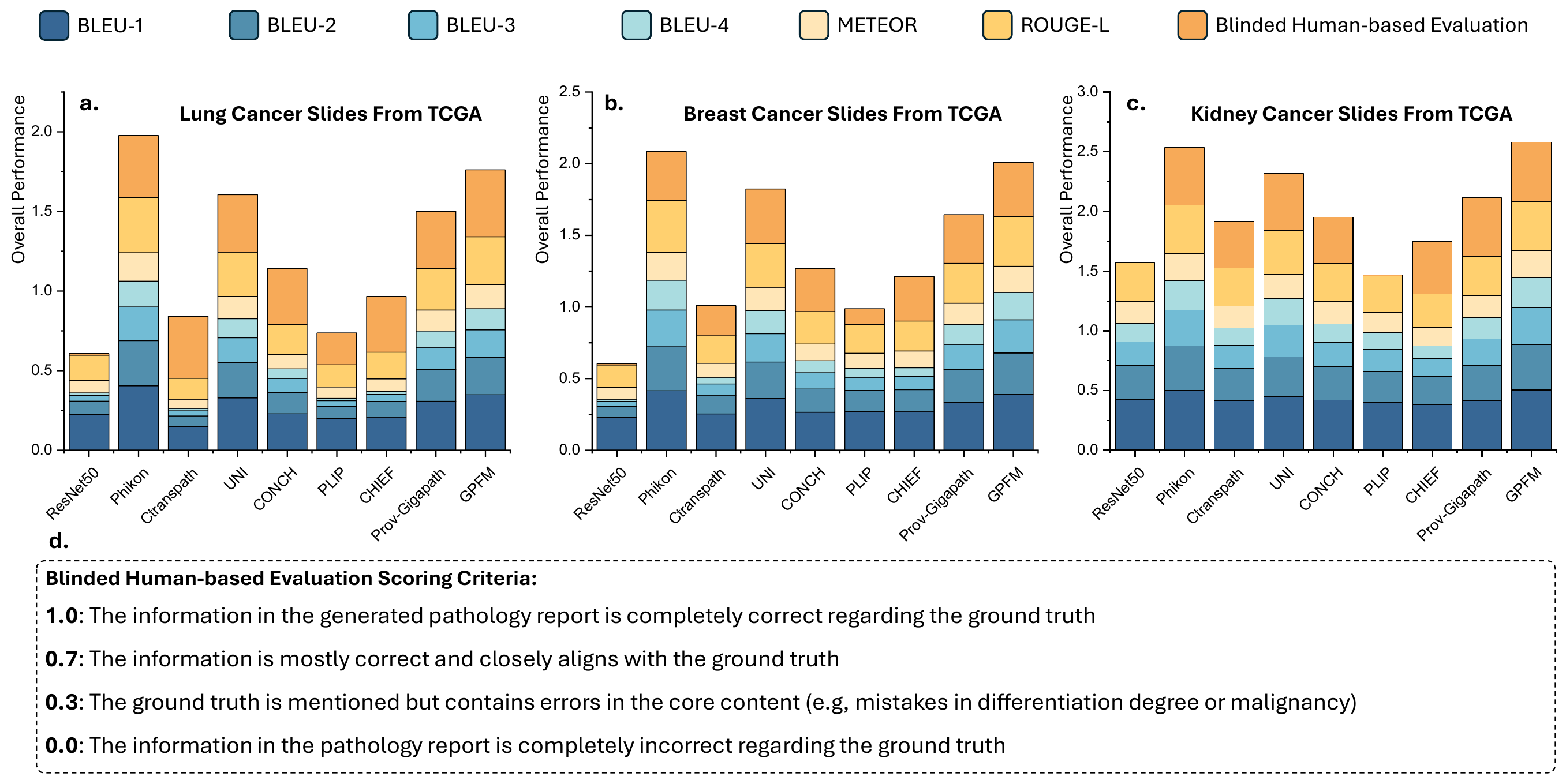}
\caption{\textbf{Evaluation of Report Quality Based on Organ-Specific Analysis.}
\textbf{a-c.} Performance assessment of generated pathology reports for lung cancer, breast cancer, and kidney cancer, respectively.
    \textbf{d.} Scoring criteria for human-based blind evaluation of foundation-model-generated pathology reports. The scoring system ranges from 0.0 to 1.0, where 1.0 indicates complete accuracy with ground truth, 0.7 represents mostly correct information, 0.3 indicates presence of core content errors, and 0.0 denotes completely incorrect information.}
    \label{fig:score_criteria}
\end{figure*}
\begin{figure*}[h]
    \centering
    \includegraphics[width=1\linewidth]{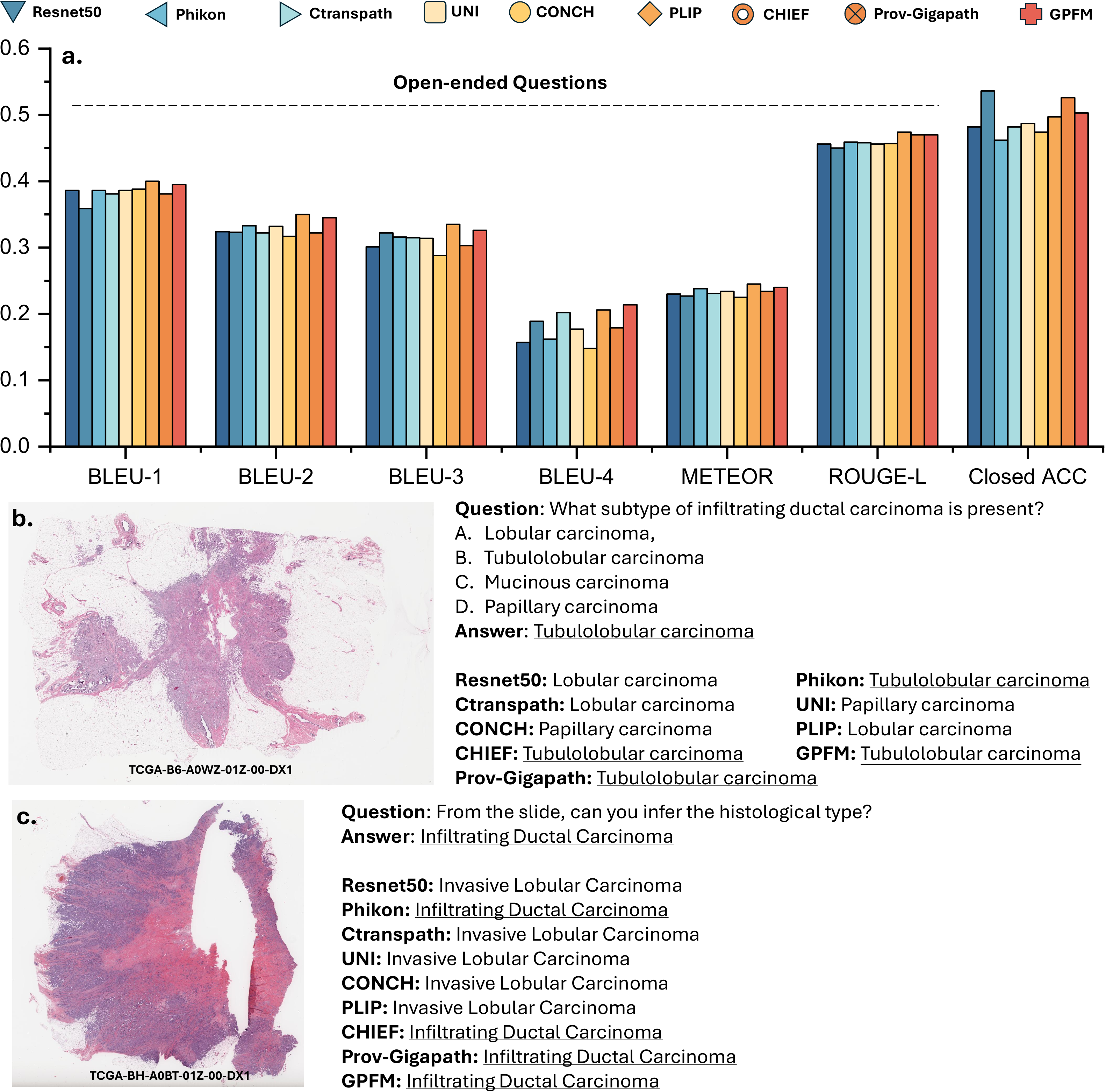}
    \caption{\textbf{VQA results on WSI-VQA dataset.}
    \textbf{a.} Open-ended and close-ended statistical results.
    \textbf{b.} A close-ended question and corresponding answers.
    \textbf{c.}An open-ended question and corresponding answers.
    }
    \label{fig:vqa-extra}
\end{figure*}
\begin{figure*}
    \centering
    \includegraphics[width=1\linewidth]{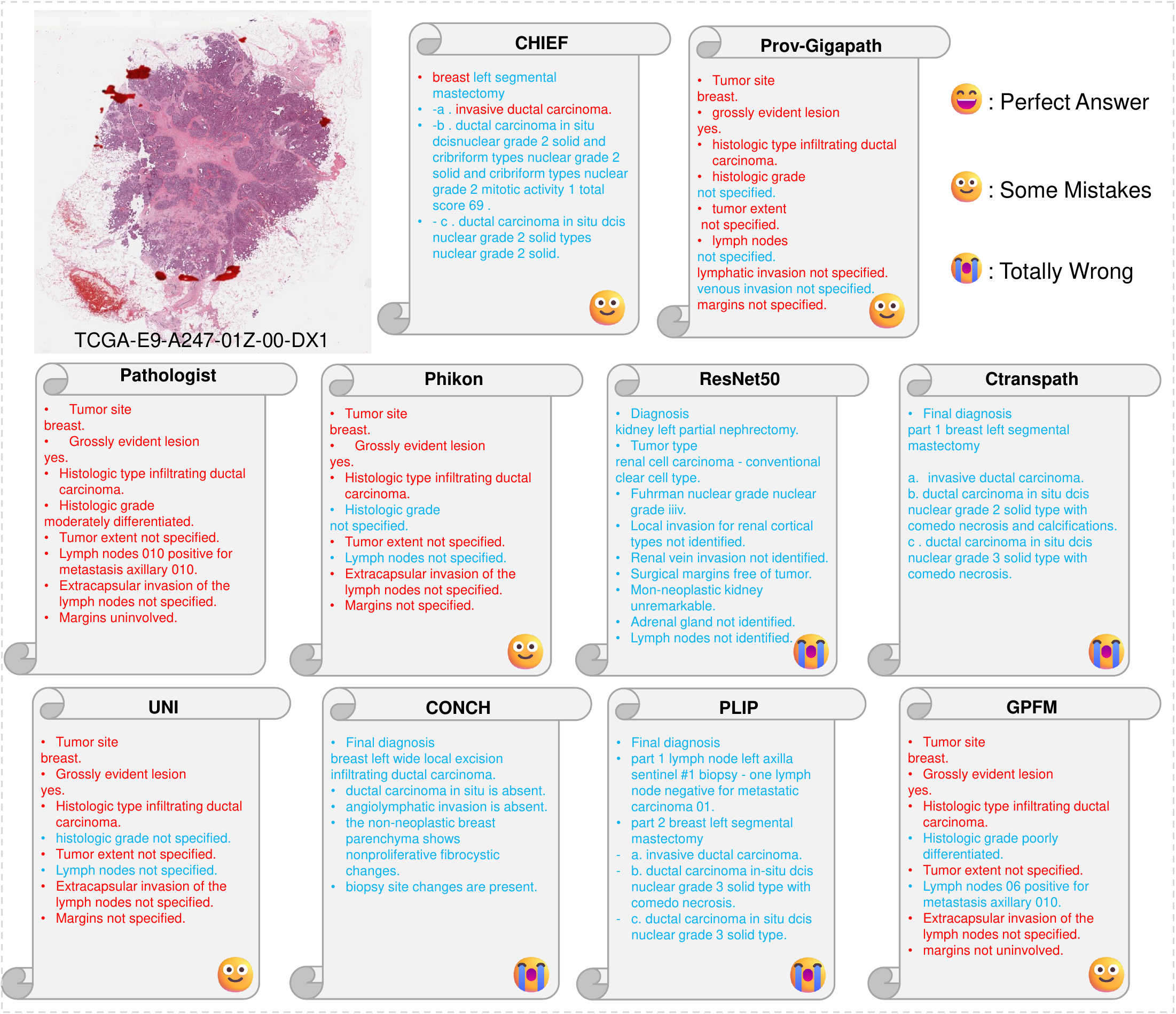}
    \caption{\textbf{Generated Example Reports} The ground truth report is provided by pathologist. The text in red indicates correct predictions, the text in blue represents incorrect predictions. 
    }
    \label{fig:extra_report_1}
\end{figure*}
\begin{figure*}
    \centering
    \includegraphics[width=1\linewidth]{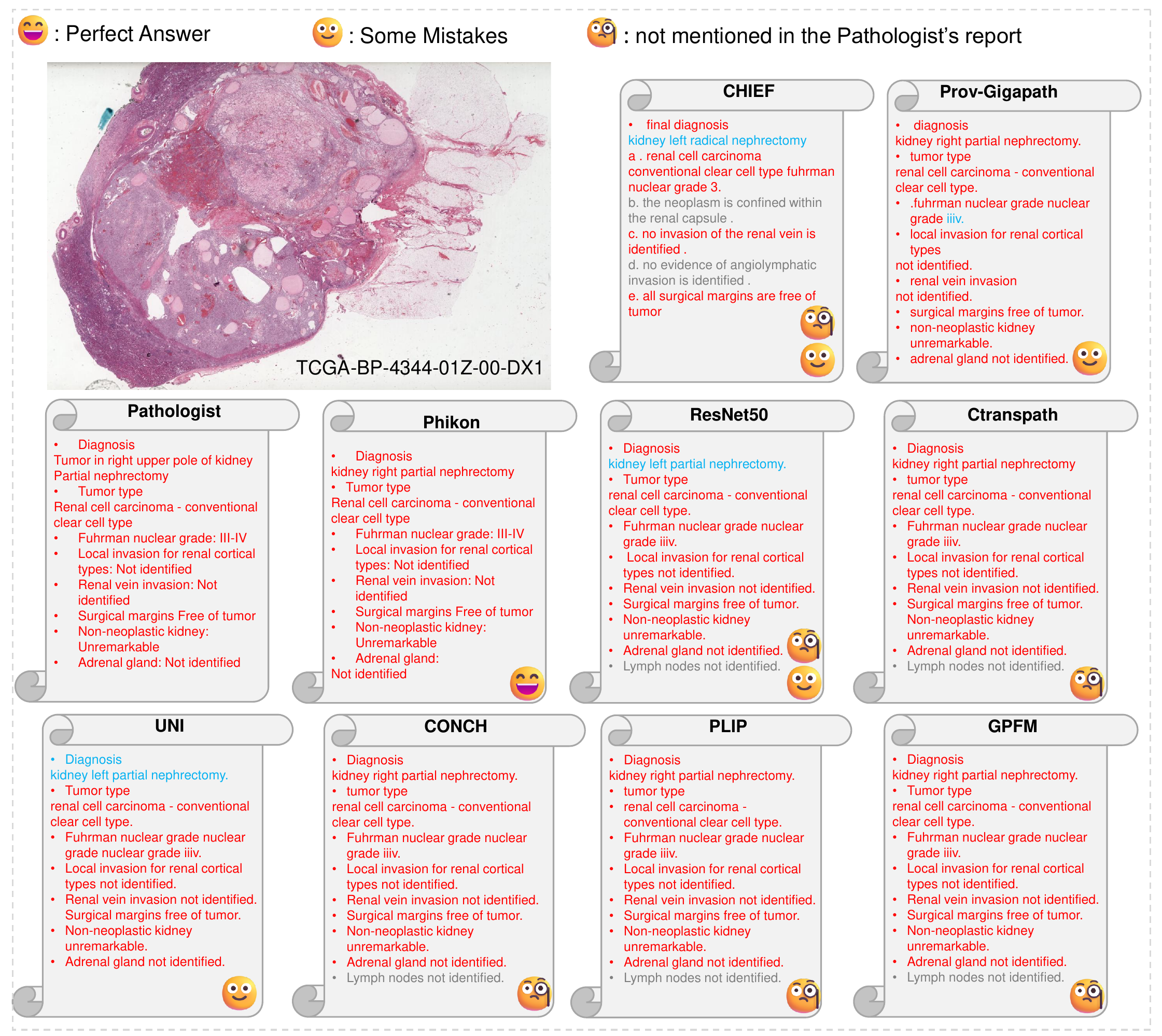}
    \caption{\textbf{Generated Example Reports.} The ground truth report is provided by pathologist. The text in red indicates correct predictions, the text in blue represents incorrect predictions, and the text in gray is the predicted text not mentioned in the pathologist's report.}
    \label{fig:extra_report_2}
\end{figure*}

\begin{figure*}
    \centering
    \includegraphics[width=1\linewidth]{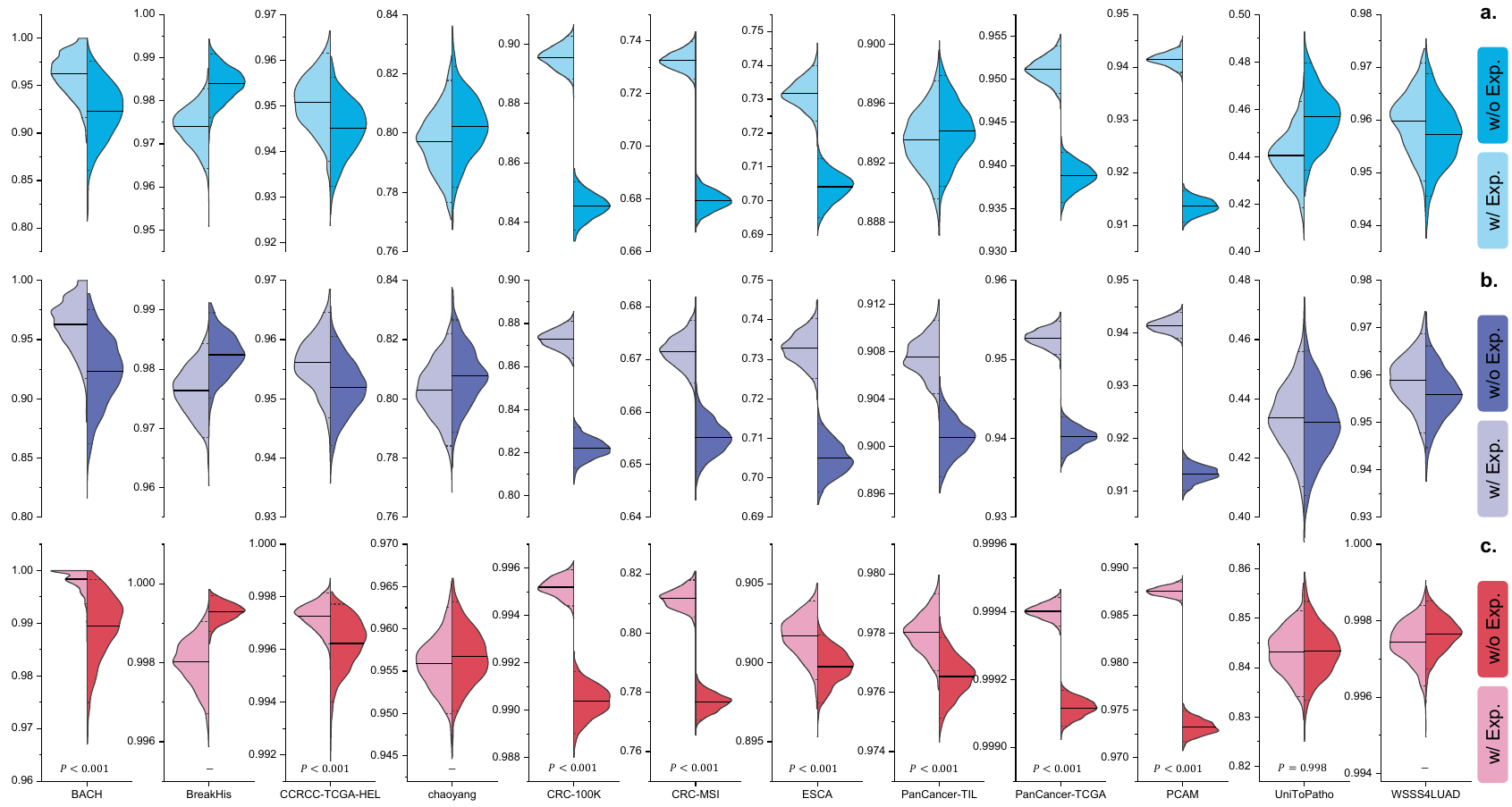}
    \caption{\textbf{The Effectiveness of Expert Knowledge Distillation. } The figure presents the performance difference between GPFM (with Expert Knowledge Distillation, \textit{i.e.}, w/ Exp. in figure) and DINOv2 (without Expert Knowledge Distillation, \textit{i.e.}, w/o Exp. in figure).
    The horizontal black lines indicate the mean AUC.
    If GPFM outperforms DINOv2, the \textit{p}-value is also reported.
    \textbf{a.} The balanced accuracy of the models with and without Expert Knowledge Distillation.
    \textbf{b.} The weighted F1 score of the models with and without Expert Knowledge Distillation.
    \textbf{c.} The AUC of the models with and without Expert Knowledge Distillation.
    The center lines represent mean and the dashed lines indicate the 2.5-th and 97.5-th percentile, respectively.
    Significance testing was conducted using the Wilcoxon signed-rank one-sided test, demonstrating that Expert Knowledge Distillation consistently improves performance across the majority of tasks, highlighting the effectiveness of this technique in enhancing the GPFM.
    }
    \label{fig:albation}
\end{figure*}
\begin{table*}[h]
\centering
\caption{\textbf{Average WSI classification performance of foundation models across 36 tasks.}
The features have been pre-extracted, and the subsequent downstream tasks are conducted using ABMIL.
Best performing model for each metric is \textbf{bolded} and second-best performing model is \underline{underlined}. 
The standard deviation is included.
}\label{lab:wsi:avg_cls}%
\begin{tabular}{llll}
\hline
 & Balanced ACC & Weighted F1 & AUC  \\
 \hline
ResNet50&0.563±0.183&0.518±0.215&0.769±0.135\\
Phikon&0.693±0.191&0.670±0.228&0.868±0.105\\
Ctranspath&0.698±0.183&0.677±0.212&0.855±0.111\\
UNI&\underline{0.721±0.181}&\underline{0.706±0.214}&\underline{0.875±0.105}\\
CONCH&0.695±0.180&0.664±0.214&0.849±0.140\\
PLIP&0.668±0.156&0.639±0.186&0.835±0.109\\
 CHIEF& 0.687±0.171& 0.667±0.202&0.856±0.108\\
 Prov-Gigapath& 0.703±0.173& 0.676±0.209&0.854±0.113\\
GPFM&\pmb{0.752±0.161}&\pmb{0.736±0.179}&\pmb{0.891±0.096}\\
\hline
\end{tabular}
\end{table*}
\begin{table*}[h]
\centering
\caption{\textbf{NSCLC subtyping performance of different foundation models.}
Non-parametric bootstrapping with 1,000 bootstrap replicates is employed for statistical analysis. 
The 95\% CI is included in parentheses.
Best performing model for each metric is \textbf{bolded} and second-best performing model is \underline{underlined}. 
* indicates the external validation and the trained model is from TCGA cohort.
}\label{lab:wsi:tcga-nsclc}%
\begin{tabular}{lllll}
\hline
  &Cohort& Balanced ACC & Weighted F1 & AUC  \\
 \hline
ResNet50 &TCGA-NSCLC&0.845 (0.800-0.893)&0.842 (0.791-0.893)&0.929 (0.896-0.959)\\
Phikon &TCGA-NSCLC&0.888 (0.845-0.928)&0.885 (0.843-0.924)&\underline{0.982 (0.968-0.993)}\\
Ctranspath &TCGA-NSCLC&0.894 (0.850-0.934)&0.895 (0.851-0.933)&0.963 (0.936-0.985)\\
UNI &TCGA-NSCLC&\underline{0.928 (0.891-0.961)}&\underline{0.928 (0.890-0.962)}&0.977 (0.957-0.992)\\
CONCH &TCGA-NSCLC&0.924 (0.888-0.957)&0.924 (0.881-0.957)&\pmb{0.986 (0.971-0.996)}\\
PLIP &TCGA-NSCLC&0.865 (0.821-0.908)&0.865 (0.812-0.909)&0.942 (0.910-0.969)\\
 CHIEF &TCGA-NSCLC& 0.910 (0.874-0.951)& 0.910 (0.866-0.943)&0.971 (0.951-0.988)\\
\hline
\end{tabular}
\end{table*}
\begin{table*}
\caption{\textbf{The lung cancer metastasis detection (2 classes) and primary site prediction (6 classes).}
Non-parametric bootstrapping with 1,000 bootstrap replicates is employed for statistical analysis. 
The 95\% CI is included in parentheses.
Best performing model for each metric is \textbf{bolded} and second-best performing model is \underline{underlined}. cls represents "class number".
* indicates the external validation cohort.
}
\label{lab:wsi:lung_osp}
\centering
\begin{tabular}{llllll}
\hline
   &cls&Cohort& Balanced ACC & Weighted F1 & AUC  \\
 \hline
ResNet50  &2&Center-1-LMD&0.816 (0.752-0.876)&0.819 (0.759-0.880)&0.905 (0.850-0.950)\\
Phikon  &2&Center-1-LMD&0.902 (0.853-0.945)&0.894 (0.844-0.943)&\underline{0.984 (0.967-0.996)}\\
Ctranspath  &2&Center-1-LMD&0.913 (0.863-0.956)&0.908 (0.858-0.957)&0.981 (0.962-0.994)\\
UNI  &2&Center-1-LMD&0.908 (0.860-0.952)&0.901 (0.852-0.950)&0.981 (0.962-0.997)\\
CONCH  &2&Center-1-LMD&0.910 (0.861-0.958)&0.908 (0.852-0.951)&0.965 (0.935-0.990)\\
PLIP  &2&Center-1-LMD&0.838 (0.777-0.896)&0.837 (0.768-0.894)
&0.940 (0.901-0.972)\\
 CHIEF  &2&Center-1-LMD& \underline{0.930 (0.882-0.972)}& \underline{0.935 (0.890-0.971)}&0.983 (0.965-0.996)\\
\hline
\end{tabular}
\end{table*}
\begin{table*}
\centering
\caption{\textbf{RCC subtyping performance of different foundation models.}
Non-parametric bootstrapping with 1,000 bootstrap replicates is employed for statistical analysis. 
The 95\% CI is included in parentheses.
Best performing model for each metric is \textbf{bolded} and second-best performing model is \underline{underlined}. 
* indicates the external validation.
}\label{lab:wsi:tcga-rcc}
\begin{tabular}{lllll}
\hline
  &Cohort& Balanced ACC & Weighted F1 & AUC  \\
 \hline
ResNet50 &TCGA-RCC&0.870 (0.805-0.926)&0.858 (0.797-0.910)&0.963 (0.937-0.983)\\
Phikon &TCGA-RCC&\pmb{0.964 (0.936-0.987)}&\pmb{0.960 (0.937-0.982)}&\underline{0.995 (0.989-0.999)}\\
Ctranspath &TCGA-RCC&0.883 (0.820-0.938)&0.888 (0.833-0.939)&0.981 (0.965-0.992)\\
UNI &TCGA-RCC&0.903 (0.847-0.954)&0.913 (0.862-0.957)&0.982 (0.966-0.995)\\
CONCH &TCGA-RCC&\underline{0.941 (0.891-0.979)}&\underline{0.937 (0.896-0.974)}&0.988 (0.977-0.997)\\
PLIP &TCGA-RCC&0.899 (0.845-0.947)&0.904 (0.853-0.948)&0.980 (0.963-0.993)\\
 CHIEF &TCGA-RCC& 0.889 (0.824-0.945)& 0.900 (0.843-0.945)&0.981 (0.964-0.993)\\
 Prov-Gigapath &TCGA-RCC& 0.903 (838-0.958)& 0.907 (0.850-0.955)&0.986 (0.975-0.996)\\
GPFM &TCGA-RCC&0.925 (0.874-0.967)&0.930 (0.885-0.966)&\pmb{0.996 (0.992-0.999)}\\
\hline
 ResNet50
 &Center-3-RCC*& 0.490 (0.415-0.562)& 0.409 (0.332-0.496)&0.747 (0.662-0.824)\\
 Phikon
 &Center-3-RCC*& 0.433 (0.378-0.495)& 0.337 (0.254-0.422)&0.831 (0.744-0.909)\\
 Ctranspath
 &Center-3-RCC*& 0.735 (0.641-0.817)& 0.735 (0.640-0.821)&0.910 (0.853-0.953)\\
 UNI
 &Center-3-RCC*& 0.713 (0.614-0.799)& 0.717 (0.614-0.808)&\underline{0.921 (0.868-0.959)}\\
 CONCH
 &Center-3-RCC*& 0.560 (0.472-0.647)& 0.521 (0.407-0.618)&0.822 (0.745-0.890)\\
 PLIP
 &Center-3-RCC*& 0.523 (0.453-0.593)& 0.451 (0.365-0.530)&0.818 (0.743-0.883)\\
 CHIEF
 &Center-3-RCC*& \underline{0.757 (0.671-0.841)}& \underline{0.755 (0.658-0.840)}&\underline{0.921 (0.869-0.960)}\\
 Prov-Gigapath
 &Center-3-RCC*& 0.628 (0.532-0.718)& 0.607 (0.503-0.704)&0.751 (0.678-0.817)\\
 GPFM &Center-3-RCC*& \pmb{0.759 (0.667-0.835)}& \pmb{0.756 (0.666-0.843)}&\pmb{0.922 (0.868-0.963)}\\
\hline
\end{tabular}
\end{table*}
\begin{table*}
\caption{\textbf{The breast metastasis detection performance of different foundation models on CAMELYON dataset.}
Non-parametric bootstrapping with 1,000 bootstrap replicates is employed for statistical analysis. 
The 95\% CI is included in parentheses.
Best performing model for each metric is \textbf{bolded} and second-best performing model is \underline{underlined}. 
}
\label{lab:wsi:camelyon}
\centering
\begin{tabular}{llll}
\hline
 & Balanced ACC & Weighted F1 & AUC  \\
 \hline
ResNet50&0.855 (0.797-0.909)&0.857 (0.800-0.910)&0.922 (0.864-0.966)\\
Phikon&0.945 (0.900-0.979)&0.952 (0.918-0.982)&0.967 (0.932-0.993)\\
Ctranspath&0.898 (0.852-0.941)&0.908 (0.860-0.951)&0.957 (0.924-0.986)\\
UNI&\underline{0.963 (0.930-0.992)}&\pmb{0.970 (0.940-0.994)}&\underline{0.987 (0.969-0.998)}\\
CONCH&0.936 (0.896-0.974)&0.945 (0.910-0.977)&0.965 (0.934-0.989)\\
PLIP&0.882 (0.826-0.930)&0.890 (0.840-0.936)&0.929 (0.882-0.967)\\
 CHIEF
& 0.902 (0.858-0.947)& 0.905 (0.856-0.947)&0.944 (0.901-0.979)\\
 Prov-Gigapath
& 0.941 (0.900-0.977)& 0.951 (0.917-0.982)&0.969 (0.939-0.993)\\
GPFM&\pmb{0.964 (0.931-0.991)}&\underline{0.964 (0.932-0.988)}&\pmb{0.988 (0.971-1.000)}\\
\hline
\end{tabular}
\end{table*}

\begin{table*}
\centering
\caption{\textbf{Lobular and ductal carcinoma subtyping performance of different foundation models.}
Non-parametric bootstrapping with 1,000 bootstrap replicates is employed for statistical analysis. 
The 95\% CI is included in parentheses.
Best performing model for each metric is \textbf{bolded} and second-best performing model is \underline{underlined}. 
* indicates the external validation cohort.
}\label{lab:wsi:tcga-brca}%
\begin{tabular}{lllll}
\hline
  &Cohort& Balanced ACC & Weighted F1 & AUC  \\
 \hline
ResNet50 &TCGA-BRCA&0.658 (0.585-0.735)&0.691 (0.596-0.783)&0.846 (0.768-0.911)\\
Phikon &TCGA-BRCA&0.794 (0.718-0.865)&0.835 (0.751-0.901)&0.936 (0.887-0.977)\\
Ctranspath &TCGA-BRCA&0.843 (0.767-0.914)&0.859 (0.790-0.917)&0.931 (0.870-0.975)\\
UNI &TCGA-BRCA&0.869 (0.797-0.929)&\underline{0.879 (0.810-0.932)}&\underline{0.946 (0.894-0.987)}\\
CONCH &TCGA-BRCA&0.835 (0.750-0.905)&0.875 (0.807-0.934)&0.944 (0.902-0.979)\\
PLIP &TCGA-BRCA&0.823 (0.747-0.897)&0.820 (0.750-0.888)&0.893 (0.824-0.950)\\
 CHIEF
 &TCGA-BRCA& 0.790 (0.717-0.866)& 0.829 (0.756-0.896)&0.899 (0.823-0.960)\\
 Prov-Gigapath
 &TCGA-BRCA& \pmb{0.884 (0.821-0.940)}& 0.877 (0.813-0.931)&0.942 (0.884-0.983)\\
GPFM &TCGA-BRCA&\underline{0.881 (0.813-0.947)}&\pmb{0.907 (0.850-0.956)}&\pmb{0.950 (0.898-0.990)}\\
\hline
 ResNet50 
& Center-3-LD*& 0.441 (0.394-0.480)& 0.441 (0.406-0.472)&0.562 (0.387-0.737)\\
 Phikon 
& Center-3-LD*& \underline{0.824 (0.613-1.000)}& \pmb{0.849 (0.597-1.000)}&\pmb{0.915 (0.726-1.000)}\\
 Ctranspath 
& Center-3-LD*& 0.814 (0.606-0.991)& 0.814 (0.578-0.957)&0.814 (0.443-1.000)\\
 UNI 
& Center-3-LD*& 0.750 (0.500-1.000)& 0.819 (0.491-1.000)&0.874 (0.616-1.000)\\
 CONCH 
& Center-3-LD*& 0.500 (0.500-0.500)& 0.472 (0.447-0.491)&0.500 (0.500-0.500)\\
 PLIP 
& Center-3-LD*& 0.637 (0.412-0.827)& 0.502 (0.374-0.635)&0.614 (0.321-0.893)\\
 CHIEF
& Center-3-LD*& 0.500 (0.500-0.500)& 0.472 (0.447-0.491)&0.811 (0.451-1.000)\\
  
Prov-Gigapath
& Center-3-LD*& 0.652 (0.404-0.898)& 0.472 (0.441-0.736)&0.784 (0.542-1.000)\\
  
GPFM& Center-3-LD*& \pmb{0.887 (0.686-0.991)}& \underline{0.837 (0.648-0.966)}&\underline{0.905 (0.694-1.000)}\\
\hline
\end{tabular}
\end{table*}
\begin{table*}
\caption{\textbf{Coarse-grained breast carcinoma subtyping performance of different foundation.}
Non-parametric bootstrapping with 1,000 bootstrap replicates is employed for statistical analysis. 
The 95\% CI is included in parentheses.
Best performing model for each metric is \textbf{bolded} and second-best performing model is \underline{underlined}. 
* indicates the external validation.
}\label{lab:wsi:bracs-3}
\centering
\begin{tabular}{lllll}
\hline
  &Cohort& Balanced ACC & Weighted F1 & AUC  \\
 \hline
ResNet50 &BRACS&0.568 (0.515-0.615)&0.522 (0.463-0.571)&0.835 (0.776-0.892)\\
Phikon &BRACS&0.707 (0.621-0.797)&0.701 (0.602-0.800)&0.898 (0.852-0.942)\\
Ctranspath &BRACS&0.674 (0.592-0.757)&0.664 (0.559-0.754)&0.908 (0.871-0.946)\\
UNI &BRACS&\underline{0.746 (0.660-0.840)}&\underline{0.738 (0.640-0.824)}&0.913 (0.865-0.956)\\
CONCH &BRACS&0.677 (0.606-0.752)&0.668 (0.575-0.771)&0.923 (0.883-0.958)\\
PLIP &BRACS&0.679 (0.596-0.773)&0.676 (0.579-0.782)&0.866 (0.805-0.917)\\
 CHIEF &BRACS& 0.717 (0.635-0.804)& 0.723 (0.617-0.814)&0.922 (0.879-0.958)\\
 Prov-Gigapath &BRACS& 0.720 (0.620-0.810)& 0.715 (0.613-0.802)&\underline{0.927 (0.885-0.961)}\\
GPFM &BRACS&\pmb{0.749 (0.660-0.834)}&\pmb{0.758 (0.658-0.841)}&\pmb{0.936 (0.896-0.965)}\\
\hline
 ResNet50
 &Center-3-BRCA*& 0.618 (0.563-0.669)& 0.521 (0.472-0.564)&0.624 (0.555-0.688)\\
 Phikon
 &Center-3-BRCA*& \underline{0.636 (0.584-0.691)}& 0.539 (0.493-0.589)&0.659 (0.607-0.713)\\
 Ctranspath
 &Center-3-BRCA*& 0.583 (0.521-0.645)& 0.543 (0.493-0.596)&0.670 (0.617-0.720)\\
 UNI
 &Center-3-BRCA*& 0.556 (0.493-0.616)& 0.504 (0.454-0.552)&0.614 (0.559-0.666)\\
 CONCH
 &Center-3-BRCA*& 0.479 (0.450-0.503)& 0.178 (0.146-0.208)&0.500 (0.498-0.501)\\
 PLIP
 &Center-3-BRCA*& 0.578 (0.522-0.634)& \underline{0.566 (0.514-0.618)}&\underline{0.681 (0.627-0.736)}\\
 CHIEF
 &Center-3-BRCA*& 0.568 (0.506-0.624)& 0.546 (0.495-0.597)&0.670 (0.615-0.723)\\
 Prov-Gigapath
 &Center-3-BRCA*& 0.587 (0.530-0.642)& \pmb{0.574 (0.521-0.629)}&0.678 (0.625-0.732)\\
 GPFM &Center-3-BRCA*& \pmb{0.675 (0.625-0.720)}& 0.543 (0.490-0.594)&\pmb{0.694 (0.641-0.747)}\\
\hline

\end{tabular}

\end{table*}
\begin{table*}
\caption{\textbf{Fine-grained breast carcinoma subtyping performance of different foundation models on BRACS dataset.}
Non-parametric bootstrapping with 1,000 bootstrap replicates is employed for statistical analysis. 
The 95\% CI is included in parentheses.
Best performing model for each metric is \textbf{bolded} and second-best performing model is \underline{underlined}. 
}\label{lab:wsi:bracs-7}
\centering
\begin{tabular}{llll}
\hline
 & Balanced ACC & Weighted F1 & AUC  \\
 \hline
ResNet50&0.309 (0.266-0.357)&0.250 (0.181-0.320)&0.772 (0.719-0.818)\\
Phikon&0.363 (0.322-0.406)&0.293 (0.251-0.332)&0.818 (0.768-0.866)\\
Ctranspath&\pmb{0.530 (0.450-0.626)}&\pmb{0.520 (0.407-0.615)}&\underline{0.857 (0.811-0.896)}\\
UNI&0.433 (0.356-0.511)&0.411 (0.325-0.490)&0.855 (0.811-0.893)\\
CONCH&0.424 (0.352-0.505)&0.367 (0.287-0.439)&0.841 (0.797-0.884)\\
PLIP&0.420 (0.342-0.511)&0.414 (0.324-0.493)&0.814 (0.763-0.864)\\
 CHIEF& 0.445 (0.357-0.541)& \underline{0.445 (0.332-0.547)}&0.854 (0.810-0.897)\\
 Prov-Gigapath& \underline{0.463 (0.380-0.549)}& 0.433 (0.346-0.505)&0.808 (0.753-0.861)\\
GPFM&0.437 (0.360-0.514)&0.408 (0.326-0.493)&\pmb{0.871 (0.829-0.904)}\\
\hline
\end{tabular}
\end{table*}
\begin{table*}
\centering
\caption{\textbf{Prostate cancer grade assessment performance of different foundation models on PANDA dataset.}
Non-parametric bootstrapping with 1,000 bootstrap replicates is employed for statistical analysis. 
The 95\% CI is included in parentheses.
Best performing model for each metric is \textbf{bolded} and second-best performing model is \underline{underlined}. 
}\label{lab:wsi:panda}%
\begin{tabular}{llll}
\hline
 & Balanced ACC & Weighted F1 & AUC  \\
 \hline
ResNet50&0.531 (0.510-0.552)&0.531 (0.508-0.553)&0.884 (0.875-0.892)\\
Phikon&\underline{0.731 (0.709-0.750)}&\underline{0.735 (0.715-0.755)}&0.943 (0.936-0.949)\\
Ctranspath&0.649 (0.627-0.670)&0.651 (0.629-0.671)&0.925 (0.918-0.932)\\
UNI&0.728 (0.707-0.749)&0.734 (0.712-0.753)&\underline{0.944 (0.937-0.950)}\\
CONCH&0.656 (0.635-0.678)&0.657 (0.637-0.679)&0.921 (0.914-0.929)\\
PLIP&0.607 (0.583-0.628)&0.612 (0.591-0.635)&0.903 (0.894-0.911)\\
 CHIEF
& 0.665 (0.643-0.688)& 0.667 (0.643-0.689)&0.927 (0.920-0.934)\\
 Prov-Gigapath
& 0.674 (0.653-0.697)& 0.676 (0.653-0.699)&0.926 (0.918-0.933)\\
GPFM&\pmb{0.740 (0.720-0.760)}&\pmb{0.742 (0.722-0.762)}&\pmb{0.948 (0.941-0.954)}\\
\hline
\end{tabular}
\end{table*}
\begin{table*}
\centering
\caption{\textbf{Lung adenocarcinoma TP53 gene mutation prediction performance of different foundation models on TCGA-LUAD dataset.}
5-fold cross validation is employed for statistical analysis. 
The 95\% CI is included in parentheses.
Best performing model for each metric is \textbf{bolded} and second-best performing model is \underline{underlined}. 
}\label{lab:wsi:tcga-luad-tp53}%
\begin{tabular}{llll}
\hline
 & Balanced ACC & Weighted F1 & AUC \\
 \hline
ResNet50& 0.675 (0.549-0.708)& 0.609 (0.493-0.714) & 0.742 (0.629-0.842)\\
Phikon& \underline{0.783 (0.704-0.874)}& \underline{0.782 (0.687-0.867)}& \underline{0.841 (0.754-0.918)}\\
Ctranspath&0.711 (0.621-0.810) & 0.710 (0.601-0.806)& 0.770 (0.660-0.867)\\
UNI& 0.639 (0.553-0.749)& 0.638 (0.530-0.746)& 0.766 (0.667-0.867)\\
CONCH& 0.735 (0.618-0.820)& 0.730 (0.629-0.818)& 0.836 (0.734-0.911)\\
PLIP& 0.759 (0.643-0.818)& 0.739 (0.629-0.832)& 0.821 (0.721-0.905)\\
 CHIEF& 0.683 (0.586-0.773)& 0.682 (0.570-0.783)&0.736 (0.622-0.844)\\
 Prov-Gigapath& 0.627 (0.537-0.730)& 0.616 (0.505-0.731)&0.804 (0.700-0.894)\\
GPFM& \pmb{0.795 (0.707-0.878)}& \pmb{0.794 (0.694-0.878)}& \pmb{0.855 (0.767-0.931)}\\
\hline
\end{tabular}
\end{table*}
\begin{table*}
\centering
\caption{\textbf{WSI-level  IDH1 gene mutation prediction performance of different foundation models}.
Non-parametric bootstrapping with 1,000 bootstrap replicates is employed for statistical analysis. 
The 95\% CI is included in parentheses.
Best performing model for each metric is \textbf{bolded} and second-best performing model is \underline{underlined}. 
* indicates the external validation.
}\label{lab:TCGA-IDH1}%
\begin{tabular}{lllll}
\hline
  &Cohort& Balanced ACC & Weighted F1 & AUC  \\
 \hline
ResNet50 &TCGA-GBMLGG&0.842 (0.781-0.900)&0.849 (0.788-0.904)&0.937 (0.895-0.969)\\
Phikon &TCGA-GBMLGG&0.885 (0.828-0.938)& 0.900 (0.846-0.947)&0.972 (0.946-0.992)\\
Ctranspath &TCGA-GBMLGG&0.860 (0.804-0.911)&0.838 (0.774-0.896)&0.958 (0.920-0.986)\\
UNI &TCGA-GBMLGG&0.917 (0.867-0.957)&0.906 (0.853-0.951)&\underline{0.976 (0.954-0.993)}\\
CONCH &TCGA-GBMLGG&0.856 (0.795-0.915)&0.869 (0.808-0.921)&0.946 (0.894-0.981)\\
PLIP &TCGA-GBMLGG&0.888 (0.829-0.938)&0.889 (0.834-0.939)&0.946 (0.904-0.979)\\
CHIEF&TCGA-GBMLGG& 0.886 (0.833-0.935)& 0.883 (0.826-0.936)&0.944 (0.904-0.977)\\
Prov-Gigapath &TCGA-GBMLGG& \underline{0.928 (0.884-0.963)}& \underline{0.920 (0.870-0.963)}&\pmb{0.986 (0.970-0.997)}\\
GPFM &TCGA-GBMLGG&\pmb{0.936 (0.891-0.974)}&\pmb{0.934 (0.888-0.971)}&\pmb{0.986 (0.971-0.996)}\\
\hline
ResNet50 & EBRAINS*& 0.531 (0.517-0.547)& 0.455 (0.428-0.485)&0.708 (0.672-0.744)\\
Phikon & EBRAINS*& 0.749 (0.724-0.771)& 0.697 (0.667-0.726)&0.920 (0.899-0.938)\\
Ctranspath & EBRAINS*& 0.854 (0.829-0.877)& 0.851 (0.826-0.875)&0.923 (0.904-0.940)\\
UNI & EBRAINS*& \pmb{0.882 (0.860-0.905)}& \pmb{0.875 (0.851-0.896)}&0.939 (0.922-0.955)\\
CONCH & EBRAINS*& 0.795 (0.772-0.820)& 0.758 (0.726-0.786)&0.924 (0.904-0.941)\\
PLIP & EBRAINS*& 0.800 (0.777-0.823)& 0.765 (0.738-0.792)&0.908 (0.886-0.929)\\
CHIEF& EBRAINS*& 0.810 (0.784-0.835)& 0.779 (0.750-0.805)&0.915 (0.895-0.933)\\
Prov-Gigapath& EBRAINS*& 0.838 (0.815-0.861)& 0.811 (0.785-0.839)&\underline{0.941 (0.924-0.957)}\\
 GPFM& EBRAINS*& \underline{0.875 (0.852-0.896)}& \underline{0.863 (0.838-0.886)}&\pmb{0.943 (0.927-0.957)}\\
 \hline
\end{tabular}
\end{table*}
\begin{table*}
\centering
\caption{\textbf{Ovarian cancer subtyping performance of different foundation models.}
Non-parametric bootstrapping with 1,000 bootstrap replicates is employed for statistical analysis. 
The 95\% CI is included in parentheses.
Best performing model for each metric is \textbf{bolded} and second-best performing model is \underline{underlined}. 
* indicates the external validation.
}\label{lab:wsi:ubc-ocean}
\begin{tabular}{lllll}
\hline
  &Cohort& Balanced ACC & Weighted F1 & AUC  \\
 \hline
ResNet50 &UBC-OCEAN&0.487 (0.417-0.557)&0.480 (0.385-0.563)&0.903 (0.866-0.936)\\
Phikon &UBC-OCEAN&0.731 (0.654-0.820)&0.751 (0.654-0.828)&0.970 (0.947-0.988)\\
Ctranspath &UBC-OCEAN&\underline{0.830 (0.744-0.906)}&\underline{0.820 (0.727-0.891)}&0.973 (0.956-0.989)\\
UNI &UBC-OCEAN&0.737 (0.661-0.814)&0.740 (0.634-0.826)&\underline{0.975 (0.956-0.990)}\\
CONCH &UBC-OCEAN&\pmb{0.841 (0.748-0.918)}&\pmb{0.830 (0.728-0.905)}&0.972 (0.952-0.989)\\
PLIP &UBC-OCEAN&0.670 (0.596-0.748)&0.665 (0.572-0.754)&0.944 (0.912-0.972)\\
 CHIEF &UBC-OCEAN& 0.782 (0.684-0.875)& 0.801 (0.689-0.882)&0.971 (0.949-0.988)\\
 Prov-Gigapath &UBC-OCEAN& 0.764 (0.671-0.857)& 0.788 (0.690-0.869)&0.951 (0.915-0.981)\\
GPFM &UBC-OCEAN&0.809 (0.717-0.888)&0.810 (0.701-0.888)&\pmb{0.984 (0.969-0.994)}\\
\hline
 ResNet50 
& Center-3-Ovary*& 0.259 (0.225-0.295)& 0.212 (0.179-0.252)&0.637 (0.603-0.674)\\
 Phikon 
& Center-3-Ovary*& 0.250 (0.215-0.285)& 0.215 (0.178-0.252)&0.684 (0.657-0.709)\\
 Ctranspath 
& Center-3-Ovary*& 0.239 (0.206-0.275)& 0.242 (0.204-0.280)&0.584 (0.557-0.612)\\
 UNI 
& Center-3-Ovary*& 0.267 (0.232-0.306)& 0.300 (0.260-0.342)&\underline{0.685 (0.654-0.715)}\\
 CONCH 
& Center-3-Ovary*& 0.279 (0.233-0.328)& \pmb{0.323 (0.263-0.376)}&0.593 (0.565-0.624)\\
 PLIP 
& Center-3-Ovary*& \pmb{0.351 (0.302-0.404)}& \underline{0.320 (0.278-0.357)}&0.627 (0.592-0.659)\\
 CHIEF 
& Center-3-Ovary*& 0.299 (0.257-0.341)& 0.248 (0.209-0.292)&0.673 (0.643-0.703)\\
 Prov-Gigapath 
& Center-3-Ovary*& \underline{0.338 (0.294-0.380)}& 0.311 (0.270-0.350)&0.660 (0.626-0.692)\\
 GPFM & Center-3-Ovary*& 0.276 (0.237-0.320)& 0.308 (0.275-0.342)&\pmb{0.687 (0.661-0.712)}\\
 \hline
\end{tabular}
\end{table*}
\begin{table*}
\centering
\caption{\textbf{Brain tumor subtyping performance of different foundation models.}
Non-parametric bootstrapping with 1,000 bootstrap replicates is employed for statistical analysis. 
The 95\% CI is included in parentheses.
Best performing model for each metric is \textbf{bolded} and second-best performing model is \underline{underlined}. 
* indicates the external validation.
}\label{lab:EBRAINS}%
\begin{tabular}{lllll}
\hline
  &Cohort& Balanced ACC & Weighted F1 & AUC  \\
 \hline
ResNet50 &TCGA-GBMLGG&0.736 (0.651-0.818)&0.732 (0.640-0.811)&0.954 (0.933-0.972)\\
Phikon &TCGA-GBMLGG&0.790 (0.706-0.865)&0.803 (0.715-0.871)&0.976 (0.959-0.990)\\
Ctranspath &TCGA-GBMLGG&0.729 (0.643-0.814)
&0.740 (0.635-0.818)&0.963 (0.938-0.983)\\
UNI &TCGA-GBMLGG&0.795 (0.708-0.880)
&\underline{0.823 (0.738-0.895)}&\underline{0.981 (0.967-0.993)}\\
CONCH &TCGA-GBMLGG&\underline{0.813 (0.723-0.886)}&0.815 (0.726-0.889)
&0.978 (0.963-0.990)\\
PLIP &TCGA-GBMLGG&0.791 (0.707-0.871)&0.806 (0.716-0.873)&0.963 (0.930-0.988)\\
CHIEF & TCGA-GBMLGG& 0.711 (0.620-0.798)& 0.718 (0.627-0.798)&0.951 (0.905-0.982)\\
Prov-Gigapath & TCGA-GBMLGG& \pmb{0.816 (0.727-0.901)}& \pmb{0.825 (0.738-0.895)}&\underline{0.981 (0.967-0.993)}\\
GPFM &TCGA-GBMLGG&0.782 (0.693-0.867)&0.804 (0.706-0.882)&\pmb{0.987 (0.976-0.995)}\\
\hline
ResNet50 & EBRAINS*& 0.333 (0.333-0.333)& 0.289 (0.282-0.295)&0.554 (0.506-0.600)\\
Phikon & EBRAINS*& 0.784 (0.743-0.827)& \underline{0.726 (0.690-0.763)}&0.909 (0.887-0.929)\\
Ctranspath & EBRAINS*& 0.758 (0.718-0.795)& 0.622 (0.585-0.658)&0.865 (0.836-0.890)\\
UNI & EBRAINS*& \underline{0.806 (0.764-0.847)}& \pmb{0.767 (0.727-0.804)}&\underline{0.921 (0.900-0.941)}\\
CONCH & EBRAINS*& 0.764 (0.729-0.801)& 0.642 (0.608-0.675)&0.888 (0.864-0.910)\\
PLIP & EBRAINS*& 0.687 (0.643-0.732)& 0.682 (0.634-0.724)&0.868 (0.842-0.893)\\
CHIEF & EBRAINS*& 0.755 (0.718-0.794)& 0.656 (0.617-0.694)&0.885 (0.860-0.908)\\
Prov-Gigapath & EBRAINS*& 0.776 (0.736-0.814)& 0.691 (0.654-0.727)&0.908 (0.888-0.926)\\
 GPFM & EBRAINS*& \pmb{0.809 (0.770-0.846)}& 0.713 (0.678-0.746)&\pmb{0.926 (0.906-0.946)}\\
 \hline
\end{tabular}
\end{table*}
\begin{table*}
\centering
\caption{\textbf{Lesion grade classification for colon cancer.}
Non-parametric bootstrapping with 1,000 bootstrap replicates is employed for statistical analysis. 
The 95\% CI is included in parentheses.
Best performing model for each metric is \textbf{bolded} and second-best performing model is \underline{underlined}. 
* indicates the external validation.
}\label{lab:WSI:colon}%
\begin{tabular}{lllll}
\hline
  &Cohort& Balanced ACC & Weighted F1 & AUC  \\
 \hline
ResNet50 &IMP-CRS&0.922 (0.902-0.941)&0.922 (0.904-0.939)&0.986 (0.981-0.991)\\
Phikon &IMP-CRS&\pmb{0.946 (0.931-0.960)}&\pmb{0.952 (0.938-0.964)}&\underline{0.994 (0.990-0.996)}\\
Ctranspath &IMP-CRS&0.941 (0.925-0.957)&\underline{0.947 (0.933-0.960)}&0.992 (0.989-0.995)\\
UNI &IMP-CRS&0.922 (0.902-0.941)&0.938 (0.923-0.953)&\underline{0.994 (0.991-0.996)}\\
CONCH &IMP-CRS&0.933 (0.915-0.951)&0.943 (0.928-0.958)&0.993 (0.990-0.996)\\
PLIP &IMP-CRS&0.899 (0.878-0.920)&0.915 (0.895-0.934)&0.987 (0.982-0.992)\\
CHIEF & IMP-CRS& 0.942 (0.925-0.957)& 0.945 (0.931-0.958)&0.992 (0.989-0.995)\\
Prov-Gigapath & IMP-CRS& \underline{0.944 (0.930-0.957)}& 0.934 (0.918-0.948)&0.993 (0.990-0.996)\\
GPFM &IMP-CRS&0.916 (0.895-0.934)&0.932 (0.914-0.947)&\pmb{0.995 (0.992-0.997)}\\
\hline
ResNet50 & Center-3-Colon-WSI*& 0.584 (0.541-0.629)& 0.563 (0.504-0.621)&0.874 (0.846-0.901)\\
Phikon & Center-3-Colon-WSI*& 0.869 (0.832-0.909)& 0.879 (0.842-0.914)&0.961 (0.941-0.980)\\
Ctranspath & Center-3-Colon-WSI*& 0.827 (0.784-0.865)& 0.839 (0.797-0.876)&0.955 (0.935-0.972)\\
UNI & Center-3-Colon-WSI*& \underline{0.917 (0.884-0.948)}& \underline{0.921 (0.890-0.951)}&0.979 (0.965-0.991)\\
CONCH & Center-3-Colon-WSI*& 0.899 (0.858-0.935)& 0.909 (0.873-0.940)&\underline{0.983 (0.970-0.994)}\\
PLIP & Center-3-Colon-WSI*& 0.776 (0.734-0.816)& 0.788 (0.739-0.837)&0.969 (0.953-0.983)\\
CHIEF & Center-3-Colon-WSI*& 0.835 (0.793-0.875)& 0.848 (0.802-0.886)&0.967 (0.953-0.980)\\
 Prov-Gigapath & Center-3-Colon-WSI*& 0.900 (0.865-0.930)& 0.904 (0.868-0.934)&0.971 (0.954-0.986)\\
 GPFM & Center-3-Colon-WSI*& \pmb{0.925 (0.895-0.957)}& \pmb{0.937 (0.908-0.965)}&\pmb{0.984 (0.972-0.993)}\\
 \hline
\end{tabular}
\end{table*}
\begin{table*}
\centering
\caption{\textbf{Primary site prediction (PSP) and T stage classification for head \& neck cancers on HANCOCK dataset.}
Non-parametric bootstrapping with 1,000 bootstrap replicates is employed for statistical analysis. 
The 95\% CI is included in parentheses.
Best performing model for each metric is \textbf{bolded} and second-best performing model is \underline{underlined}. 
}\label{lab:WSI:hancock}%
\begin{tabular}{lllll}
\hline
  &Task& Balanced ACC & Weighted F1 & AUC  \\
 \hline
ResNet50 &PSP&0.555 (0.492-0.615)&0.519 (0.451-0.575)&0.844 (0.796-0.891)\\
Phikon & PSP& \underline{0.727 (0.648-0.800)}& 0.713 (0.640-0.784)&0.902 (0.858-0.951)\\
Ctranspath & PSP& 0.674 (0.612-0.738)& 0.655 (0.586-0.733)&0.859 (0.806-0.902)\\
UNI & PSP& 0.722 (0.664-0.783)& \underline{0.725 (0.654-0.797)}&0.906 (0.867-0.942)\\
CONCH & PSP& 0.698 (0.626-0.767)& 0.696 (0.605-0.775)&0.905 (0.858-0.947)\\
PLIP & PSP& 0.672 (0.605-0.741)& 0.673 (0.592-0.757)&0.884 (0.839-0.925)\\
CHIEF & PSP& 0.632 (0.553-0.717)& 0.659 (0.570-0.744)&0.861 (0.805-0.911)\\
Prov-Gigapath & PSP& 0.667 (0.620-0.712)& 0.634 (0.582-0.676)&\underline{0.910 (0.869-0.947)}\\
GPFM & PSP& \pmb{0.754 (0.704-0.810)}& \pmb{0.736 (0.664-0.805)}&\pmb{0.920 (0.884-0.956)}\\
 \hline
ResNet50 & T Stage& 0.473 (0.381-0.564)& \underline{0.487 (0.386-0.578)}&0.752 (0.689-0.812)\\
 Phikon& T Stage& 0.452 (0.362-0.544)& 0.456 (0.366-0.541)&0.760 (0.705-0.814)\\
Ctranspath & T Stage& 0.405 (0.321-0.491)& 0.402 (0.314-0.484)&0.746 (0.691-0.797)\\
 UNI & T Stage& 0.453 (0.381-0.540)& 0.424 (0.329-0.516)&\underline{0.764 (0.709-0.819)}\\
CONCH & T Stage& 0.418 (0.327-0.502)& 0.393 (0.311-0.466)&0.761 (0.709-0.813)\\
PLIP & T Stage& 0.438 (0.342-0.527)& 0.418 (0.325-0.501)&0.718 (0.659-0.771)\\
CHIEF & T Stage& 0.433 (0.362-0.502)& 0.415 (0.331-0.493)&0.745 (0.689-0.795)\\
Prov-Gigapath & T Stage& \underline{0.487 (0.395-0.575)}& 0.453 (0.362-0.544)&0.750 (0.687-0.807)\\
GPFM & T Stage& \pmb{0.513 (0.425-0.607)}& \pmb{0.515 (0.409-0.602)}&\pmb{0.780 (0.730-0.832)}\\
 \hline
\end{tabular}
\end{table*}
\begin{table*}
\centering
\caption{\textbf{Lauren subtyping of gastric cancer.}
Non-parametric bootstrapping with 1,000 bootstrap replicates is employed for statistical analysis. 
The 95\% CI is included in parentheses. 
Best performing model for each metric is \textbf{bolded} and second-best performing model is \underline{underlined}. 
* indicates the external validation.
}\label{lab:WSI:lauren}%
\begin{tabular}{lllll}
\hline
  &Cohort& Balanced ACC & Weighted F1 & AUC  \\
 \hline
ResNet50 &TCGA-STAD&0.320 (0.282-0.358)&0.227 (0.175-0.291)&0.751 (0.661-0.832)\\
Phikon & TCGA-STAD& 0.557 (0.449-0.666)& 0.572 (0.443-0.688)&0.830 (0.748-0.908)\\
Ctranspath & TCGA-STAD& 0.373 (0.316-0.437)& 0.311 (0.226-0.390)&0.791 (0.717-0.861)\\
UNI & TCGA-STAD& 0.575 (0.461-0.693)& 0.594 (0.469-0.706)&0.820 (0.738-0.892)\\
CONCH & TCGA-STAD& 0.614 (0.501-0.715)& 0.619 (0.500-0.720)&\underline{0.843 (0.770-0.908)}\\
PLIP & TCGA-STAD& \underline{0.641 (0.521-0.750)}& \underline{0.635 (0.521-0.739)}&0.831 (0.749-0.901)\\
CHIEF & TCGA-STAD& 0.595 (0.475-0.711)& 0.596 (0.473-0.703)&0.819 (0.745-0.889)\\
\hline
\end{tabular}
\end{table*}
\begin{table*}
\centering
\caption{\textbf{Vascular invasion detection of gastric cancer.}
Non-parametric bootstrapping with 1,000 bootstrap replicates is employed for statistical analysis. 
The 95\% CI is included in parentheses. 
Best performing model for each metric is \textbf{bolded} and second-best performing model is \underline{underlined}. 
* indicates the external validation.
}\label{lab:WSI:vascular}%
\begin{tabular}{lllll}
\hline
  &Cohort& Balanced ACC & Weighted F1 & AUC  \\
 \hline
ResNet50 &Center-1-GC&0.600 (0.492-0.710)&0.594 (0.486-0.707)&0.625 (0.490-0.754)\\
Phikon & Center-1-GC& 0.700 (0.603-0.795)& 0.697 (0.589-0.799)&0.760 (0.653-0.862)\\
Ctranspath & Center-1-GC& 0.725 (0.628-0.816)& 0.721 (0.612-0.807)&\underline{0.798 (0.693-0.889)}\\
UNI & Center-1-GC& \pmb{0.750 (0.647-0.844)}& \pmb{0.749 (0.644-0.837)}&0.787 (0.680-0.881)\\
CONCH & Center-1-GC& 0.675 (0.580-0.768)& 0.665 (0.559-0.770)&0.764 (0.664-0.866)\\
PLIP & Center-1-GC& 0.675 (0.582-0.768)& 0.665 (0.557-0.762)&0.756 (0.640-0.859)\\
CHIEF & Center-1-GC& 0.637 (0.530-0.740)& 0.636 (0.525-0.735)&0.712 (0.589-0.813)\\
\hline
\end{tabular}
\end{table*}
\begin{table*}
\centering
\caption{\textbf{Perineural invasion detection in gastric cancer.}
Non-parametric bootstrapping with 1,000 bootstrap replicates is employed for statistical analysis. 
The 95\% CI is included in parentheses. 
Best performing model for each metric is \textbf{bolded} and second-best performing model is \underline{underlined}. * indicates the external validation.
}\label{lab:WSI:perineural}%
\begin{tabular}{lllll}
\hline
  &Cohort& Balanced ACC & Weighted F1 & AUC  \\
 \hline
 ResNet50 &Center-1-GC&0.516 (0.467-0.578)&0.453 (0.365-0.561)&0.870 (0.781-0.946)\\
Phikon & Center-1-GC& 0.887 (0.812-0.947)& 0.868 (0.782-0.937)&0.960 (0.909-0.995)\\
Ctranspath & Center-1-GC& \underline{0.915 (0.853-0.971)}& 0.895 (0.821-0.958)&0.949 (0.893-0.990)\\
UNI & Center-1-GC& 0.900 (0.823-0.963)& 0.903 (0.832-0.963)&0.963 (0.916-0.995)\\
CONCH & Center-1-GC& 0.907 (0.840-0.963)& 0.893 (0.813-0.959)&\underline{0.979 (0.947-0.999)}\\
PLIP & Center-1-GC& 0.824 (0.739-0.907)& 0.813 (0.716-0.899)&0.904 (0.830-0.964)\\
CHIEF & Center-1-GC& 0.872 (0.782-0.953)& 0.875 (0.788-0.947)&0.938 (0.877-0.983)\\
\hline
\end{tabular}
\end{table*}
\begin{table*}
    \centering
\caption{\textbf{Average C-Index of Foundation Models Across 15 Survival Analysis Tasks.}
The best-performing and second-best-performing models are highlighted in \textbf{bold} and \underline{underlined}, respectively.}
\label{tab:avg_survival}
    \begin{tabular}{ll}
    \hline
    Models&C-Index \\
    \hline
 ResNet50& 0.619±0.080\\
 Phikon& 0.636±0.088\\
 Ctranspath& 0.631±0.071\\
 UNI& \underline{0.643±0.079}\\
 CONCH& 0.637±0.077\\
 PLIP& 0.623±0.073\\
 CHIEF &   0.638±0.070\\
 Prov-Gigapath &  0.642±0.072\\
 GPFM& \pmb{0.665±0.071}\\
 \hline
\end{tabular}
\end{table*}

\begin{table*}
\centering
\caption{ \textbf{Performance of Survival Analysis on TCGA-BRCA, TCGA-BLCA, TCGA-KIRC, and TCGA-KIRP Datasets.}
The 95\% CI is included in parentheses.
The best and second-best performed models are \textbf{bolded} and \underline{underlined}.}
\label{tab:survival1}

\begin{tabular}{l l l l l}     
\hline
& \textbf{TCGA-BRCA}& \textbf{TCGA-BLCA}& \textbf{TCGA-KIRC} &\textbf{TCGA-KIRP}\\
\hline
ResNet50 &0.569 (0.451-0.699)& 0.576 (0.450-0.677) & 0.731 (0.642-0.820)& 0.784 (0.472-0.966)\\
Phikon & \underline{0.730 (0.631-0.820)} &\pmb{0.664 (0.548-0.763)} &\underline{0.764 (0.673-0.848)} & 0.697 (0.385-0.923)\\
Ctranspath &0.658 (0.535-0.776)&0.550 (0.416-0.680)&0.726 (0.628-0.813)&0.788 (0.595-0.979)\\
UNI &0.613 (0.446-0.757)&0.564 (0.449-0.680)&0.755 (0.646-0.851)&\pmb{0.863 (0.659-1.000)}\\
CONCH &0.666 (0.538-0.795)&0.620 (0.519-0.720)&0.760 (0.675-0.835)&0.743 (0.462-0.919)\\
PLIP &0.601 (0.471-0.739)&0.555 (0.446-0.664)&0.739 (0.622-0.827)&0.784 (0.574-0.959)\\
CHIEF &0.632 (0.483-0.763) &0.601 (0.486-0.711) &0.717 (0.615-0.816) &0.793 (0.559-0.965)\\
Prov-Gigapath & 0.655 (0.514-0.798) & 0.627 (0.501-0.739) &0.733 (0.639-0.824) &0.772 (0.468-0.959)\\
GPFM & \pmb{0.739 (0.643-0.837)} & \underline{0.633 (0.523-0.732)} &\pmb{0.774 (0.694-0.854)} &\underline{0.797 (0.531-0.991)}\\
\hline
\end{tabular}
\end{table*}
\begin{table*}
\centering
\caption{ \textbf{Performance of Survival Analysis on TCGA-STAD, TCGA-CESC, TCGA-LUAD, and CPTAC-LUAD Datasets.}
The 95\% CI is included in parentheses.
Models trained on the TCGA-LUAD dataset were directly applied to the CPTAC-LUAD dataset for testing.
The best and second-best performed models are \textbf{bolded} and \underline{underlined}.}
\label{tab:survival1-2}
\begin{tabular}{l l l ll}     
\hline
& \textbf{TCGA-STAD}
&\textbf{TCGA-CESC} 
&\textbf{TCGA-LUAD}&\textbf{CPTAC-LUAD}\\
\hline
ResNet50& 0.580 (0.459-0.679)&0.703 (0.525-0.872)&0.578 (0.464-0.685)&0.596 (0.521-0.667)\\
Phikon&0.601 (0.505-0.700)&\pmb{0.768 (0.655-0.866)}&\pmb{0.614 (0.498-0.727)}&0.461 (0.384-0.541)\\
Ctranspath&\underline{0.632 (0.540-0.720)}&0.683 (0.507-0.825)&0.552 (0.439-0.664)&0.658 (0.578-0.733)\\
UNI&0.595 (0.481-0.705)&0.683 (0.505-0.841)&\underline{0.605 (0.468-0.728)}&0.605 (0.536-0.684)\\
CONCH&0.624 (0.524-0.726)&0.681 (0.509-0.826)&0.603 (0.474-0.720)&0.615 (0.541-0.691)\\
PLIP&0.573 (0.467-0.667)&0.655 (0.489-0.814)&0.560 (0.433-0.684)&0.623 (0.547-0.702)\\
CHIEF &0.628 (0.516-0.730) &0.672 (0.528-0.810) & 0.588 (0.472-0.705)&\pmb{0.687 (0.611-0.763)}\\
Prov-Gigapath &0.572 (0.462-0.676) &\underline{0.731 (0.610-0.846)} & 0.549 (0.426-0.670)&0.580 (0.494-0.670)\\
GPFM&\pmb{0.636 (0.534-0.733)}&0.683 (0.491-0.849)&0.599 (0.479-0.707)&\underline{0.669 (0.595-0.747)}\\
\hline
\end{tabular}
\end{table*}
\begin{table*}
\centering
\caption{\textbf{Performance of Survival Analysis on TCGA-COADREAD, TCGA-GBM, TCGA-LGG, and TCGA-LUSC datasets}.
The 95\% CI is included in parentheses.
The best and second-best performed model are \textbf{bolded} and \underline{underlined}.}
\label{tab:survival2-1}
\begin{tabular}{lllll}
\hline
& \textbf{TCGA-COADREAD}& \textbf{TCGA-GBM}&\textbf{TCGA-LGG} &\textbf{TCGA-LUSC}
\\
\hline
ResNet50&0.632 (0.496-0.767)&0.533 (0.434-0.629) & 0.723 (0.566-0.860)  &0.573 (0.462-0.692)
\\
Phikon&0.660 (0.534-0.762)&0.534 (0.452-0.613) & 0.710 (0.555-0.850)  &0.575 (0.462-0.677)
\\
Ctranspath&0.656 (0.522-0.784)&0.565 (0.475-0.647) & 0.693 (0.493-0.868)  &0.579 (0.468-0.682)
\\
UNI&0.662 (0.543-0.779)&0.580 (0.498-0.659) &0.687 (0.527-0.830)  &0.572 (0.456-0.693)
\\
CONCH&0.593 (0.434-0.766)&0.559 (0.474-0.651) & \pmb{0.771 (0.635-0.893)} &0.531 (0.412-0.637)
\\
PLIP&0.673 (0.540-0.805)&0.539 (0.447-0.629) & 0.684 (0.538-0.816)  &0.570 (0.476-0.670)
\\
CHIEF &\pmb{0.682 (0.557-0.803)}&0.516 (0.422-0.605) & 0.694 (0.551-0.824) &0.572 (0.462-0.701)
\\
Prov-Gigapath &0.675 (0.537-0.802) &\underline{0.581 (0.496-0.663)} &0.728 (0.599-0.845)  &\pmb{0.599 (0.502-0.696)}
\\
GPFM&\underline{0.678 (0.552-0.788)}&\pmb{0.590 (0.500-0.676)} & \underline{0.731 (0.562-0.872)} &\underline{0.581 (0.458-0.692)}\\
\hline
\end{tabular}
\end{table*}

\begin{table*}
\centering
\caption{\textbf{Performance of Survival Analysis on TCGA-HNSC, HANCOCK (external validation), and TCGA-SKCM datasets}. 
The 95\% CI is included in parentheses.
The best and second-best performed model are \textbf{bolded} and \underline{underlined}.}
\label{tab:survival2-2}
\begin{tabular}{llll}
\hline
& \textbf{TCGA-HNSC}&\textbf{HANCOCK}&\textbf{TCGA-SKCM}\\
\hline
ResNet50&0.558 (0.467-0.650)& 0.494 (0.452-0.537)& 0.661 (0.543-0.762)\\
Phikon&0.638 (0.541-0.733)& 0.515 (0.475-0.558)&0.604 (0.491-0.707) \\
Ctranspath&0.584 (0.481-0.690)& 0.528 (0.484-0.572)&0.616 (0.507-0.707) \\
UNI&\pmb{0.663 (0.567-0.751)}& \pmb{0.560 (0.517-0.601)}&0.639 (0.536-0.736) \\
CONCH&0.625 (0.523-0.725)& 0.502 (0.462-0.542)&\pmb{0.669 (0.585-0.749)} \\
PLIP&0.608 (0.519-0.694)& 0.522 (0.481-0.566)& 0.663 (0.567-0.755)\\
CHIEF &0.639 (0.543-0.716) & 0.531 (0.487-0.572)& 0.625 (0.527-0.716)\\
Prov-Gigapath &0.651 (0.564-0.739) & 0.533 (0.491-0.572)& 0.643 (0.528-0.749)\\
GPFM&\underline{0.661 (0.567-0.759)}& \underline{0.535 (0.497-0.579)}& \underline{0.667 (0.547-0.777)}\\
\hline
\end{tabular}
\end{table*}
\begin{table*}
\centering
\caption{\textbf{Average Tissue Classification Performance of Foundation Models across 16 Patch-level Tissue tasks.}
The best-performing and second-best-performing models are highlighted in \textbf{bold} and \underline{underlined}, respectively.
}\label{lab:linear:avg}%
\begin{tabular}{llll}
\hline
 & Balanced ACC & Weighted F1 & AUC  \\
 \hline
ResNet50&0.745±0.160&0.732±0.174&0.906±0.096\\
Phikon&0.820±0.155&0.810±0.162&0.934±0.075\\
Ctranspath&0.797±0.161&0.792±0.165&0.929±0.076\\
UNI&0.851±0.133&0.848±0.137&0.939±0.072\\
CONCH&0.820±0.130&0.818±0.131&0.932±0.077\\
PLIP&0.756±0.150&0.743±0.164&0.899±0.116\\
 CHIEF& 0.790±0.150& 0.785±0.154&0.926±0.082\\
 Prov-Gigapath& \underline{0.856±0.132}& \underline{0.856±0.135}&\underline{0.944±0.065}\\
GPFM&\pmb{0.866±0.136}&\pmb{0.865±0.142}&\pmb{0.946±0.066}\\
\hline
\end{tabular}
\end{table*}
\begin{table*}
\centering
\caption{\textbf{CRC tissue classification performance of different foundation models on CRC-100K dataset. }Non-parametric bootstrapping with 1,000 bootstrap replicates is employed for statistical analysis. 
The 95\% CI is included in parentheses.
Best performing model for each metric is \textbf{bolded} and second-best performing model is \underline{underlined}. }\label{lab:linear:crc-100k}%
\begin{tabular}{llll}
\hline
 & Balanced ACC & Weighted F1 & AUC  \\
 \hline
ResNet50&0.792 (0.782-0.802)&0.775 (0.765-0.785)&0.983 (0.982-0.985)\\
Phikon&0.867 (0.859-0.875)&0.842 (0.833-0.850)&0.992 (0.991-0.993)\\
Ctranspath&0.853 (0.844-0.861)&0.833 (0.825-0.843)&\pmb{0.995 (0.994-0.996)}\\
UNI&0.879 (0.872-0.886)&0.849 (0.841-0.858)&0.991 (0.990-0.992)\\
CONCH&0.855 (0.847-0.863)&0.824 (0.815-0.833)&0.993 (0.992-0.994)\\
PLIP&0.804 (0.796-0.813)&0.764 (0.755-0.772)&0.990 (0.989-0.992)\\
 CHIEF
& 0.802 (0.795-0.810)& 0.749 (0.741-0.758)&\pmb{0.995 (0.994-0.995)}\\
 Prov-Gigapath
& \pmb{0.940 (0.934-0.947)}& \pmb{0.935 (0.928-0.941)}&\underline{0.994 (0.992-0.995)}\\
GPFM&0.896 (0.888-0.902)&0.872 (0.865-0.881)&\pmb{0.995 (0.994-0.996)}\\
\hline
\end{tabular}
\end{table*}
\begin{table*}
\centering
\caption{\textbf{CCRCC tissue classification performance of different foundation models on CCRCC-TCGA-HEL dataset.} Non-parametric bootstrapping with 1,000 bootstrap replicates is employed for statistical analysis.  
The 95\% CI is included in parentheses.
Best performing model for each metric is \textbf{bolded} and second-best performing model is \underline{underlined}. 
}\label{lab:linear:ccrcc-tcga_hel}%
\begin{tabular}{llll}
\hline
 & Balanced ACC & Weighted F1 & AUC  \\
 \hline
ResNet50&0.930 (0.919-0.942)&0.934 (0.925-0.944)&0.993 (0.991-0.995)\\
Phikon&\underline{0.949 (0.936-0.960)}&\underline{0.955 (0.946-0.963)}&\pmb{0.997 (0.996-0.998)}\\
Ctranspath&0.936 (0.923-0.948)&0.938 (0.926-0.946)&0.994 (0.992-0.996)\\
UNI&0.946 (0.932-0.956)&0.950 (0.941-0.959)&\underline{0.996 (0.995-0.997)}\\
CONCH&0.934 (0.920-0.946)&0.939 (0.929-0.949)&0.994 (0.992-0.995)\\
PLIP&0.920 (0.905-0.932)&0.919 (0.909-0.929)&0.992 (0.991-0.994)\\
 CHIEF
& 0.933 (0.921-0.944)& 0.935 (0.924-0.944)&0.994 (0.993-0.995)\\
 Prov-Gigapath
& 0.946 (0.935-0.957)& 0.948 (0.938-0.957)&\pmb{0.997 (0.995-0.997)}\\
GPFM&\pmb{0.953 (0.939-0.962)}&\pmb{0.956 (0.947-0.964)}&\pmb{0.997 (0.994-0.998)}\\
\hline
\end{tabular}
\end{table*}
\begin{table*}
\centering
\caption{\textbf{Breast cancer tissue classification performance of different foundation models on BACH  and BreakHis dataset. }
Non-parametric bootstrapping with 1,000 bootstrap replicates is employed for statistical analysis. 
The 95\% CI is included in parentheses.
Best performing model for each metric is \textbf{bolded} and second-best performing model is \underline{underlined}. 
}\label{lab:linear:bach}%
\begin{tabular}{lllll}
\hline
  &Cohort& Balanced ACC & Weighted F1 & AUC  \\
 \hline
ResNet50 &BACH&0.865 (0.788-0.932)&0.856 (0.776-0.928)&0.977 (0.958-0.992)\\
Phikon &BACH&0.918 (0.845-0.971)&0.915 (0.842-0.965)&0.988 (0.975-0.998)\\
Ctranspath &BACH&0.927 (0.865-0.975)&0.919 (0.861-0.965)&\underline{0.998 (0.993-1.000)}\\
UNI &BACH&\underline{0.960 (0.915-1.000)}&\pmb{0.966 (0.911-1.000)}&\pmb{1.000 (0.999-1.000)}\\
CONCH &BACH&0.934 (0.879-0.981)&0.933 (0.885-0.986)&0.996 (0.988-1.000)\\
PLIP &BACH&0.799 (0.714-0.871)&0.791 (0.698-0.880)&0.959 (0.926-0.981)\\
 CHIEF &BACH& 0.925 (0.865-0.975)& 0.924 (0.859-0.975)&\underline{0.998 (0.993-1.000)}\\
 Prov-Gigapath &BACH& 0.925 (0.866-0.975)& 0.924 (0.859-0.974)&0.995 (0.986-1.000)\\
GPFM &BACH&\pmb{0.963 (0.919-1.000)}&\underline{0.965 (0.915-1.000)}&\underline{0.998 (0.994-1.000)}\\
\hline
ResNet50 &BreakHis&0.937 (0.923-0.950)&0.938 (0.925-0.951)&0.986 (0.981-0.990)\\
Phikon &BreakHis&0.973 (0.964-0.981)&0.973 (0.965-0.982)&\underline{0.997 (0.996-0.998)}\\
Ctranspath &BreakHis&0.962 (0.952-0.972)&0.961 (0.951-0.971)&0.995 (0.992-0.997)\\
UNI &BreakHis&\pmb{0.977 (0.967-0.984)}&\pmb{0.976 (0.968-0.984)}&\pmb{0.998 (0.997-0.999)}\\
CONCH &BreakHis&0.950 (0.935-0.961)&0.952 (0.941-0.963)&0.991 (0.986-0.994)\\
PLIP &BreakHis&0.943 (0.929-0.954)&0.940 (0.927-0.951)&0.989 (0.986-0.993)\\
 CHIEF &BreakHis& 0.961 (0.950-0.972)& 0.961 (0.950-0.971)&0.995 (0.993-0.997)\\
 Prov-Gigapath &BreakHis& \underline{0.974 (0.966-0.983)}& \underline{0.974 (0.965-0.982)}&\pmb{0.998 (0.997-0.999)}\\
GPFM &BreakHis&\underline{0.974 (0.965-0.984)}&\pmb{0.976 (0.968-0.984)}&\pmb{0.998 (0.997-0.999)}\\
\hline
\end{tabular}
\end{table*}
\begin{table*}
\centering
\caption{\textbf{CRC polyp classification performance of different foundation models on UniToPatho datasets.}
Non-parametric bootstrapping with 1,000 bootstrap replicates is employed for statistical analysis. 
The 95\% CI is included in parentheses.
Best performing model for each metric is \textbf{bolded} and second-best performing model is \underline{underlined}.
}\label{lab:linear:unitopatho}%
\begin{tabular}{llll}
\hline
  & Balanced ACC & Weighted F1 & AUC  \\
\hline
 ResNet50 
&  0.397 (0.376-0.417)&  0.384 (0.359-0.406)& 0.830 (0.819-0.840)\\
 Phikon 
& 0.379 (0.358-0.398)&  0.375 (0.356-0.395)&0.838 (0.828-0.847)\\
 Ctranspath 
&  0.310 (0.289-0.331)&  0.302 (0.285-0.324)& 0.836 (0.828-0.844)\\
 UNI 
& \underline{0.462 (0.443-0.486)}& \underline{0.455 (0.433-0.474)}& 0.840 (0.830-0.850)\\
 CONCH 
&  \pmb{0.522 (0.499-0.550)}& \pmb{0.527 (0.501-0.550)}& \pmb{0.865 (0.855-0.875)}\\
 PLIP & 0.437 (0.413-0.458)& 0.418 (0.395-0.441)& 0.823 (0.812-0.834)\\
 CHIEF
& 0.394 (0.373-0.415)& 0.386 (0.362-0.410)&0.830 (0.821-0.838)\\
  
Prov-Gigapath& 0.442 (0.422-0.462)& 0.437 (0.413-0.461)&\underline{0.845 (0.835-0.855)}\\
  
GPFM& 0.444 (0.420-0.463)&  0.433 (0.412-0.456)& 0.844 (0.834-0.851)\\
 \hline
\end{tabular}
\end{table*}
\begin{table*}
\centering
\caption{\textbf{MSI screening performance of different foundation models on CRC-MSI dataset.}
Non-parametric bootstrapping with 1,000 bootstrap replicates is employed for statistical analysis. 
The 95\% CI is included in parentheses.
Best performing model for each metric is \textbf{bolded} and second-best performing model is \underline{underlined}. 
}\label{lab:linear:CRC-MSI}%
\begin{tabular}{llll}
\hline
 & Balanced ACC & Weighted F1 & AUC  \\
 \hline
ResNet50&0.654 (0.646-0.661)&0.587 (0.581-0.592)&0.706 (0.699-0.714)\\
Phikon&0.695 (0.689-0.703)&0.632 (0.626-0.638)&0.772 (0.766-0.779)\\
Ctranspath&0.728 (0.721-0.734)&0.647 (0.641-0.652)&0.802 (0.796-0.808)\\
UNI&0.719 (0.713-0.727)&0.670 (0.664-0.676)&0.797 (0.790-0.803)\\
CONCH&0.734 (0.727-0.741)&0.669 (0.663-0.675)&0.810 (0.804-0.817)\\
PLIP&0.639 (0.633-0.647)&0.589 (0.583-0.595)&0.691 (0.683-0.698)\\
 CHIEF
& 0.717 (0.710-0.724)& 0.648 (0.642-0.653)&0.791 (0.785-0.798)\\
 Prov-Gigapath
& \pmb{0.740 (0.734-0.746)}& \pmb{0.696 (0.689-0.701)}&\pmb{0.836 (0.830-0.842)}\\
GPFM&\underline{0.733 (0.726-0.740)}&\underline{0.672 (0.666-0.678)}&\underline{0.812 (0.805-0.818)}\\
\hline
\end{tabular}
\end{table*}
\begin{table*}
\centering
\caption{\textbf{Pan-cancer tissue classification performance of different foundation models on PanCancer-TCGA dataset.}
Non-parametric bootstrapping with 1,000 bootstrap replicates is employed for statistical analysis. 
The 95\% CI is included in parentheses.
Best performing model for each metric is \textbf{bolded} and second-best performing model is \underline{underlined}. 
As shown in \textbf{Figure} \ref{fig:patch_cls}\textbf{.j}, the distribution of bootstrapped AUC values is highly centered. As a result, the CI for the AUC is very narrow.
}\label{lab:linear:pancancer-tcga}
\begin{tabular}{llll}
\hline
 & Balanced ACC & Weighted F1 & AUC  \\
 \hline
ResNet50&0.630 (0.625-0.636)&0.640 (0.636-0.646)&0.975 (0.974-0.976)\\
Phikon&\underline{0.924 (0.921-0.928)}&\underline{0.926 (0.923-0.928)}&\pmb{0.999 (0.999-0.999)}\\
Ctranspath&0.785 (0.780-0.790)&0.790 (0.786-0.795)&0.992 (0.991-0.992)\\
UNI&0.885 (0.882-0.889)&0.888 (0.885-0.892)&0.997 (0.997-0.997)\\
CONCH&0.784 (0.779-0.788)&0.789 (0.785-0.794)&0.991 (0.991-0.992)\\
PLIP&0.661 (0.656-0.667)&0.669 (0.664-0.675)&0.978 (0.978-0.979)\\
 CHIEF& 0.762 (0.757-0.767)& 0.765 (0.760-0.770)&0.989 (0.989-0.990)\\
 Prov-Gigapath& 0.909 (0.905-0.912)& 0.912 (0.909-0.915)&0.998 (0.998-0.998)\\
GPFM&\pmb{0.951 (0.949-0.954)}&\pmb{0.953 (0.950-0.955)}&\pmb{0.999 (0.999-0.999)}\\
\hline
\end{tabular}
\end{table*}
\begin{table*}
\centering
\caption{\textbf{TIL classification performance of different foundation models.}
Non-parametric bootstrapping with 1,000 bootstrap replicates is employed for statistical analysis. 
The 95\% CI is included in parentheses.
Best performing model for each metric is \textbf{bolded} and second-best performing model is \underline{underlined}. * indicates the external validation.
}\label{lab:linear:pancancer-til}%
\begin{tabular}{lllll}
\hline
  &Cohort& Balanced ACC & Weighted F1 & AUC  \\
 \hline
ResNet50 &PanCancer-TIL&0.813 (0.809-0.818)&0.843 (0.839-0.847)&0.946 (0.944-0.948)\\
Phikon &PanCancer-TIL&0.893 (0.889-0.896)&0.901 (0.897-0.904)&0.975 (0.974-0.977)\\
Ctranspath &PanCancer-TIL&0.857 (0.852-0.860)&0.880 (0.876-0.883)&0.965 (0.963-0.967)\\
UNI &PanCancer-TIL&\pmb{0.897 (0.893-0.900)}&\underline{0.905 (0.902-0.908)}&\underline{0.977 (0.976-0.979)}\\
CONCH &PanCancer-TIL&0.866 (0.862-0.870)&0.889 (0.885-0.892)&0.971 (0.969-0.973)\\
PLIP &PanCancer-TIL&0.810 (0.805-0.815)&0.843 (0.838-0.847)&0.949 (0.947-0.951)\\
 CHIEF &PanCancer-TIL& 0.845 (0.841-0.849)& 0.873 (0.869-0.876)&0.961 (0.960-0.963)\\
 Prov-Gigapath &PanCancer-TIL& 0.886 (0.883-0.890)& 0.894 (0.891-0.898)&0.968 (0.967-0.970)\\
GPFM &PanCancer-TIL&\underline{0.894 (0.890-0.897)}&\pmb{0.908 (0.904-0.911)}&\pmb{0.978 (0.977-0.979)}\\
\hline
 ResNet50 
& Center-3-TIL*& 0.768 (0.763-0.773)
& 0.749 (0.743-0.755)&0.886 (0.881-0.891)\\
 Phikon 
& Center-3-TIL*& 0.781 (0.776-0.786)
& 0.762 (0.756-0.768)&0.917 (0.913-0.921)\\
 Ctranspath 
& Center-3-TIL*& 0.771 (0.766-0.776)
& 0.750 (0.744-0.757)&0.905 (0.900-0.909)\\
 UNI 
& Center-3-TIL*& \underline{0.917 (0.913-0.921)}& \underline{0.914 (0.910-0.918)}&0.929 (0.924-0.932)\\
 CONCH 
& Center-3-TIL*& 0.915 (0.911-0.919)
& 0.912 (0.908-0.916)&\underline{0.934 (0.930-0.938)}\\
 PLIP 
& Center-3-TIL*& 0.807 (0.802-0.812)
& 0.793 (0.788-0.799)&0.888 (0.884-0.893)\\
 CHIEF 
& Center-3-TIL*& 0.758 (0.753-0.764)
& 0.733 (0.727-0.740)&0.915 (0.911-0.919)\\
 Prov-Gigapath 
& Center-3-TIL*& 0.837 (0.832-0.842)
& 0.826 (0.820-0.831)&0.926 (0.922-0.930)\\
 GPFM & Center-3-TIL*& \pmb{0.942 (0.939-0.946)}& \pmb{0.940 (0.937-0.944)}&\pmb{0.951 (0.948-0.955)}\\
 \hline
\end{tabular}
\end{table*}
\begin{table*}
\centering
\caption{\textbf{ESCA subtyping performance of different foundation models on ESCA  dataset.}
Non-parametric bootstrapping with 1,000 bootstrap replicates is employed for statistical analysis. 
The 95\% CI is included in parentheses.
Best performing model for each metric is \textbf{bolded} and second-best performing model is \underline{underlined}. 
}\label{lab:linear:esca}%
\begin{tabular}{llll}
\hline
 & Balanced ACC & Weighted F1 & AUC  \\
 \hline
ResNet50&0.601 (0.591-0.611)&0.553 (0.544-0.563)&0.886 (0.882-0.889)\\
Phikon&0.668 (0.662-0.676)&0.642 (0.635-0.651)&0.894 (0.890-0.897)\\
Ctranspath&0.642 (0.632-0.651)&0.660 (0.649-0.669)&0.899 (0.896-0.902)\\
UNI&\pmb{0.754 (0.744-0.761)}&\pmb{0.758 (0.749-0.765)}&\pmb{0.903 (0.901-0.904)}\\
CONCH&0.690 (0.682-0.698)&0.700 (0.691-0.707)&\underline{0.902 (0.899-0.904)}\\
PLIP&0.601 (0.593-0.608)&0.552 (0.544-0.559)&0.889 (0.886-0.892)\\
 CHIEF
& 0.609 (0.599-0.620)& 0.628 (0.617-0.637)&0.899 (0.895-0.901)\\
 Prov-Gigapath
& 0.725 (0.717-0.734)& \underline{0.738 (0.729-0.745)}&\underline{0.902 (0.900-0.904)}\\
GPFM&\underline{0.732 (0.724-0.740)}&0.734 (0.725-0.740)&\underline{0.902 (0.899-0.904)}\\
\hline
\end{tabular}
\end{table*}
\begin{table*}
\centering
\caption{\textbf{Metastatic tissue classification performance of different foundation models on PCAM  dataset.}
Non-parametric bootstrapping with 1,000 bootstrap replicates is employed for statistical analysis. 
The 95\% CI is included in parentheses.
Best performing model for each metric is \textbf{bolded} and second-best performing model is \underline{underlined}. 
}\label{tab:linear:pcam}
\begin{tabular}{llll}
\hline
 & Balanced ACC & Weighted F1 & AUC  \\
 \hline
ResNet50&0.837 (0.834-0.841)&0.836 (0.832-0.840)&0.926 (0.923-0.928)\\
Phikon&0.898 (0.894-0.901)&0.897 (0.894-0.900)&0.969 (0.967-0.971)\\
Ctranspath&0.866 (0.862-0.869)&0.866 (0.862-0.869)&0.940 (0.937-0.942)\\
UNI&\underline{0.932 (0.929-0.934)}&\underline{0.931 (0.929-0.934)}&\underline{0.982 (0.981-0.983)}\\
CONCH&0.903 (0.900-0.906)&0.903 (0.900-0.906)&0.965 (0.963-0.967)\\
PLIP&0.859 (0.856-0.863)&0.858 (0.854-0.862)&0.943 (0.941-0.945)\\
 CHIEF& 0.874 (0.871-0.878)& 0.874 (0.870-0.877)&0.946 (0.943-0.948)\\
 Prov-Gigapath& 0.934 (0.931-0.936)& 0.934 (0.931-0.936)&0.979 (0.978-0.980)\\
GPFM&\pmb{0.941 (0.939-0.944)}&\pmb{0.942 (0.939-0.944)}&\pmb{0.988 (0.987-0.989)}\\
\hline
\end{tabular}
\end{table*}
\begin{table*}
\centering
\caption{\textbf{Lung adenocarcinoma tissue classification performance of different foundation models on WSSS4LUAD  dataset.}
Non-parametric bootstrapping with 1,000 bootstrap replicates is employed for statistical analysis. 
The 95\% CI is included in parentheses.
Best performing model for each metric is \textbf{bolded} and second-best performing model is \underline{underlined}. 
}\label{lab:linear:wsss4luad}%
\begin{tabular}{llll}
\hline
 & Balanced ACC & Weighted F1 & AUC  \\
 \hline
ResNet50&0.911 (0.894-0.926)&0.910 (0.897-0.926)&0.989 (0.986-0.992)\\
Phikon&\underline{0.956 (0.944-0.967)}&\underline{0.957 (0.944-0.966)}&\underline{0.997 (0.995-0.998)}\\
Ctranspath&0.947 (0.935-0.960)&0.949 (0.937-0.960)&\underline{0.997 (0.996-0.998)}\\
UNI&0.951 (0.938-0.962)&0.951 (0.940-0.962)&\underline{0.997 (0.996-0.998)}\\
CONCH&0.946 (0.933-0.960)&0.947 (0.935-0.959)&0.995 (0.993-0.997)\\
PLIP&0.927 (0.915-0.945)&0.934 (0.920-0.947)&0.994 (0.992-0.995)\\
 CHIEF& 0.950 (0.937-0.962)& 0.951 (0.939-0.962)&\underline{0.997 (0.996-0.998)}\\
 Prov-Gigapath& 0.943 (0.928-0.955)& 0.941 (0.928-0.953)&0.996 (0.994-0.997)\\
GPFM&\pmb{0.961 (0.949-0.971)}&\pmb{0.959 (0.948-0.969)}&\pmb{0.998 (0.996-0.998)}\\
\hline
\end{tabular}
\end{table*}
\begin{table*}
\centering
\caption{\textbf{Colon tissue classification performance of different foundation models.}
Non-parametric bootstrapping with 1,000 bootstrap replicates is employed for statistical analysis. 
The 95\% CI is included in parentheses.
Best performing model for each metric is \textbf{bolded} and second-best performing model is \underline{underlined}. * indicates the external validation.
}\label{lab:linear:chaoyang}%
\begin{tabular}{lllll}
\hline
  &Cohort& Balanced ACC & Weighted F1 & AUC  \\
 \hline
ResNet50 &Chaoyang&0.725 (0.704-0.746)&0.735 (0.715-0.757)&0.930 (0.921-0.938)\\
Phikon &Chaoyang&0.782 (0.762-0.804)&0.784 (0.763-0.803)&0.952 (0.945-0.958)\\
Ctranspath &Chaoyang&0.772 (0.752-0.793)&0.779 (0.757-0.798)&0.950 (0.943-0.957)\\
UNI &Chaoyang&\underline{0.790 (0.770-0.809)}&0.789 (0.770-0.809)&0.952 (0.945-0.958)\\
CONCH &Chaoyang&0.759 (0.738-0.778)&0.762 (0.743-0.783)&0.942 (0.934-0.948)\\
PLIP &Chaoyang&0.747 (0.724-0.768)&0.755 (0.735-0.775)&0.941 (0.935-0.949)\\
 CHIEF &Chaoyang& 0.765 (0.743-0.785)& 0.772 (0.749-0.792)&0.948 (0.942-0.956)\\
 Prov-Gigapath &Chaoyang& \pmb{0.797 (0.776-0.816)}& \underline{0.799 (0.779-0.818)}&\pmb{0.957 (0.952-0.963)}\\
GPFM &Chaoyang&\pmb{0.797 (0.776-0.817)}&\pmb{0.803 (0.784-0.821)}&\underline{0.956 (0.950-0.963)}\\
\hline
 ResNet50
 &Center-3-Colon*& 0.560 (0.556-0.564)& 0.495 (0.488-0.502)&0.768 (0.762-0.774)\\
 Phikon
 &Center-3-Colon*& 0.684 (0.679-0.689)& 0.678 (0.671-0.684)&0.841 (0.835-0.846)\\
 Ctranspath
 &Center-3-Colon*& 0.700 (0.695-0.706)& 0.701 (0.694-0.707)&0.826 (0.821-0.832)\\
 UNI
 &Center-3-Colon*& 0.724 (0.718-0.729)& 0.724 (0.717-0.730)&0.868 (0.863-0.873)\\
 CONCH
 &Center-3-Colon*& 0.731 (0.725-0.737)& 0.730 (0.724-0.736)&0.803 (0.796-0.809)\\
 PLIP
 &Center-3-Colon*& 0.626 (0.621-0.632)& 0.603 (0.596-0.610)&0.770 (0.763-0.776)\\
 CHIEF
 &Center-3-Colon*& 0.690 (0.684-0.695)& 0.688 (0.682-0.695)&0.820 (0.814-0.825)\\
 Prov-Gigapath
 &Center-3-Colon*& \pmb{0.885 (0.881-0.890)}& \pmb{0.893 (0.889-0.898)}&\pmb{0.913 (0.909-0.917)}\\
 GPFM &Center-3-Colon*& \underline{0.828 (0.823-0.833)}& \underline{0.836 (0.831-0.842)}&\underline{0.891 (0.886-0.896)}\\
 \hline
\end{tabular}
\end{table*}

\begin{table*}
\centering
\caption{\textbf{Gastric tissue classification performance of different foundation models.}
Non-parametric bootstrapping with 1,000 bootstrap replicates is employed for statistical analysis. 
The 95\% CI is included in parentheses.
Best performing model for each metric is \textbf{bolded} and second-best performing model is \underline{underlined}. * indicates the external validation.
}\label{lab:linear:GasHisDB}%
\begin{tabular}{lllll}
\hline
  &Cohort& Balanced ACC & Weighted F1 & AUC  \\
 \hline
ResNet50 &GasHisDB&0.953 (0.947-0.958)&0.954 (0.949-0.959)&0.992 (0.990-0.993)\\
Phikon &GasHisDB&0.995 (0.993-0.997)&0.995 (0.994-0.997)&\pmb{1.000 (1.000-1.000)}\\
Ctranspath &GasHisDB&0.980 (0.976-0.983)&0.980 (0.976-0.983)&\underline{0.998 (0.998-0.999)}\\
UNI &GasHisDB&\underline{0.996 (0.994-0.997)}&\underline{0.996 (0.994-0.997)}&\pmb{1.000 (1.000-1.000)}\\
CONCH &GasHisDB&0.981 (0.978-0.985)&0.981 (0.978-0.985)&\underline{0.998 (0.998-0.999)}\\
PLIP &GasHisDB&0.958 (0.954-0.963)&0.958 (0.953-0.963)&0.993 (0.991-0.994)\\
 CHIEF &GasHisDB& 0.979 (0.976-0.983)& 0.980 (0.976-0.983)&\underline{0.998 (0.997-0.999)}\\
 Prov-Gigapath &GasHisDB& 0.995 (0.994-0.997)& \underline{0.996 (0.994-0.997)}&\pmb{1.000 (1.000-1.000)}\\
GPFM &GasHisDB&\pmb{0.997 (0.996-0.998)}&\pmb{0.997 (0.996-0.998)}&\pmb{1.000 (1.000-1.000)}\\
\hline
ResNet50 &Center-3-GC*&0.551 (0.539-0.563)&0.525 (0.506-0.547)&0.721 (0.696-0.742)\\
Phikon &Center-3-GC*&0.751 (0.730-0.769)&0.729 (0.711-0.746)&\underline{0.812 (0.789-0.834)}\\
Ctranspath &Center-3-GC*&0.708 (0.689-0.727)&0.716 (0.696-0.736)&0.777 (0.753-0.799)\\
UNI &Center-3-GC*&\underline{0.819 (0.800-0.835)}&\underline{0.841 (0.823-0.857)}&0.796 (0.772-0.819)\\
CONCH &Center-3-GC*&0.623 (0.605-0.641)&0.634 (0.614-0.654)&0.763 (0.741-0.785)\\
PLIP &Center-3-GC*&0.555 (0.534-0.576)&0.505 (0.486-0.525)&0.598 (0.572-0.622)\\
 CHIEF &Center-3-GC*& 0.681 (0.660-0.701)& 0.692 (0.673-0.713)&0.746 (0.723-0.768)\\
 Prov-Gigapath &Center-3-GC*& \underline{0.819 (0.800-0.836)}& \underline{0.841 (0.825-0.859)}&0.794 (0.772-0.820)\\
GPFM &Center-3-GC*&\pmb{0.852 (0.837-0.870)}&\pmb{0.886 (0.871-0.900)}&\pmb{0.828 (0.804-0.851)}\\
\hline
\end{tabular}
\end{table*}

\begin{table*}
\centering
\caption{\textbf{CRC Tissue Retrieval Performance on CRC-100K Dataset.}
The table reports the Top-1, Top-3, and Top-5 ACC of different foundation models on the CRC-100K dataset for CRC tissue retrieval. 
Non-parametric bootstrapping with 1,000 bootstrap replicates is used for statistical analysis. 
The 95\% CI is included in parentheses.
The best performing model for each metric is \textbf{bolded} and the second-best performing model is \underline{underlined}.
}\label{lab:ret:crc-100k}%
\begin{tabular}{llll}
\hline
 & ACC@1 & ACC@3 & ACC@5  \\
 \hline
ResNet50&0.777 (0.767-0.787)&0.940 (0.934-0.946)&0.958 (0.954-0.962)\\
Phikon&0.884 (0.876-0.892)&0.964 (0.960-0.968)&0.966 (0.962-0.970)\\
Ctranspath&0.825 (0.817-0.833)&0.910 (0.906-0.914)&0.915 (0.911-0.919)\\
UNI&\underline{0.911 (0.903-0.919)}&0.981 (0.977-0.985)&0.983 (0.981-0.985)\\
CONCH&0.879 (0.871-0.887)&0.974 (0.970-0.978)&0.976 (0.972-0.980)\\
PLIP&0.798 (0.790-0.806)&0.909 (0.905-0.913)&0.915 (0.911-0.919)\\
 CHIEF& 0.820 (0.814-0.826)& 0.882 (0.880-0.884)&0.885 (0.883-0.887)\\
 Prov-Gigapath& \pmb{0.925 (0.917-0.933)}& \underline{0.988 (0.986-0.990)}&\underline{0.993 (0.991-0.995)}\\
GPFM&0.906 (0.900-0.912)&\pmb{0.993 (0.991-0.995)}&\pmb{0.995 (0.993-0.997)}\\
\hline
\end{tabular}
\end{table*}
\begin{table*}
    \centering
\caption{\textbf{VQA performance of different foundation models on PathVQA dataset.}
The open-ended, closed-ended and overall ACC are reported. The best performing model for each metric is \textbf{bolded} and the second-best performing model is \underline{underlined}.
}
\label{tab:vqa}
    \begin{tabular}{l l l l}
    \hline
         &  Open ACC&  Closed ACC& Overall ACC\\
         \hline
         ResNet50&  28.17\%(26.63\%-29.70\%)&  86.52\%(85.43\%-87.61\%)& 57.32\% (56.41\%-58.28\%)\\
         Phikon&  30.78\%(29.28\%-32.29\%)&  87.20\%(86.13\%-88.27\%)& 58.97\% (58.10\%-59.93\%)\\
         Ctranspath&  31.11\%(29.58\%-32.65\%)&  87.51\%(86.44\%-88.58\%)& 59.35\% (58.42\%-60.28\%)\\
         UNI&  33.85\%(32.28\%-35.42\%)&  \pmb{88.69\%(87.64\%-89.74\%)}& 61.28\% (60.39\%-62.23\%)\\
         CONCH&  \pmb{37.08\%(35.40\%-38.77\%)}&  \underline{88.51\%(87.49\%-89.53\%)}& \pmb{62.84\% (61.84\%-63.81\%)}\\
 PLIP& 30.83\%(29.29\%-32.37\%)& 88.02\%(86.94\%-89.09\%)&59.42\% (58.48\%-60.35\%)\\
 CHIEF& 32.11\% (30.49\%-33.60\%)& 88.36\% (87.28\%-89.46\%)&60.23\% (59.26\%-61.18\%)\\
 Prov-Gigapath& 33.46\% (31.80\%-35.04\%)& 88.35\% (87.26\%-89.40\%)&60.91\% (59.88\%-61.90\%)\\
         GPFM&  \underline{34.26\%(32.67\%-35.84\%)}&  88.41\%(87.32\%-89.49\%)& \underline{61.39\% (60.39\%-62.30\%)}\\
         \hline
    \end{tabular}
\end{table*}
\begin{table*}
\centering
\caption{\textbf{Performance of WSI-level VQA on WSI-VQA dataset.} The best performing model for each metric is \textbf{bolded} and the second-best performing model is \underline{underlined}.
CE ACC represents Close-Ended accuracy.
}\label{lab:WSI-VQA}%
\begin{tabular}{llllllll}
\hline
 & BLEU-1 & BLEU-2 & BLEU-3 & BLEU-4 &METEOR &  ROUGE-L  &CE ACC\\
 \hline
ResNet50& 0.386& 0.324& 0.301& 0.157& 0.230&  0.456
&0.482
\\
Phikon& 0.359& 0.323& 0.322& 0.189& 0.227&  0.450&\pmb{0.536}\\
Ctranspath& 0.386& 0.333& 0.316& 0.162& 0.238&  0.459
&0.462
\\
UNI& 0.381& 0.322& 0.315& 0.202& 0.231&  0.458
&0.482
\\
CONCH& 0.386& 0.332& 0.314& 0.177& 0.234&  0.456
&0.487
\\
PLIP& 0.388& 0.317& 0.288& 0.148& 0.225&  0.457
&0.474
\\
CHIEF& \pmb{0.400}& \pmb{0.350}& \pmb{0.335}& \underline{0.206}& \pmb{0.245}&  \pmb{0.474}&0.497
\\
Prov-Gigapath & 0.381& 0.322& 0.303& 0.179& 0.234&  \underline{0.470}&\underline{0.526}\\
GPFM& \underline{0.395}& \underline{0.345}& \underline{0.326}& \pmb{0.214}& \underline{0.240}&  \underline{0.470}&0.503
\\
\hline
\end{tabular}
\end{table*}
\begin{table*}
\centering
\caption{\textbf{Performance of foundation models in WSI report generation on TCGA WSI-Report dataset.} The best performing model for each metric is \textbf{bolded} and the second-best performing model is \underline{underlined}.
}\label{lab:WSI-report}%
\begin{tabular}{lllllll}
\hline
 & BLEU-1 & BLEU-2 & BLEU-3 & BLEU-4 &METEOR &  ROUGE-L \\
 \hline
ResNet50& 0.252±0.003 & 0.113±0.003 & 0.062±0.003 & 0.039±0.003 & 0.093±0.001 & 0.179±0.002\\
Phikon& \pmb{0.404±0.005}& \pmb{0.290±0.005}& \pmb{0.225±0.005}& \pmb{0.181±0.005}& \pmb{0.178±0.003}& \pmb{0.336±0.005}\\
Ctranspath& 0.254±0.004 & 0.131±0.003 & 0.079±0.003 & 0.052±0.003 & 0.097±0.002 & 0.189±0.003\\
UNI& 0.363±0.005 & 0.250±0.005 & 0.189±0.005 & 0.151±0.004 & 0.156±0.003 & 0.298±0.005\\
CONCH& 0.246±0.005 & 0.149±0.004 & 0.104±0.004 & 0.077±0.003 & 0.110±0.002 & 0.208±0.004\\
PLIP& 0.265±0.004 & 0.135±0.003 & 0.080±0.003 & 0.053±0.003 & 0.102±0.002 & 0.188±0.003\\
CHIEF& 0.278±0.003 & 0.147±0.003 & 0.088±0.003 & 0.057±0.002 & 0.105±0.002 & 0.201±0.003 \\
Prov-Gigapath & 0.325±0.005 & 0.216±0.005 & 0.159±0.004 & 0.125±0.004 & 0.140±0.002 & 0.265±0.005 \\
GPFM& \underline{0.384±0.005}& \underline{0.271±0.005}& \underline{0.210±0.005}& \underline{0.169±0.005}& \underline{0.168±0.003}& \underline{0.320±0.005}\\
\hline
\end{tabular}
\end{table*}

\begin{table*}
\centering
\caption{\textbf{Performance of foundation models in WSI report generation on TCGA WSI-Report dataset, split by cancer types. Report generation results on breast, lung, and kidney are reported respectively.} The best performing model for each metric is \textbf{bolded} and the second-best performing model is \underline{underlined}.
}\label{lab:WSI-report-stratify}%
\begin{tabular}{lllllll}
\hline
 & BLEU-1 & BLEU-2 & BLEU-3 & BLEU-4 &METEOR &  ROUGE-L \\
 \hline
 \multicolumn{7}{c}{Breast} \\
 \hline
ResNet50& 0.228±0.007 & 0.079±0.004 & 0.034±0.003 & 0.016±0.002 & 0.081±0.002 & 0.157±0.004\\
Phikon& \pmb{0.416±0.006} & \pmb{0.312±0.005} & \pmb{0.251±0.004} & \pmb{0.208±0.004} & \pmb{0.194±0.003} & \pmb{0.364±0.005}\\
Ctranspath& 0.254±0.007 & 0.131±0.004 & 0.078±0.002 & 0.047±0.002 & 0.097±0.004 & 0.192±0.004\\
UNI& 0.361±0.006 & 0.255±0.005 & 0.198±0.004 & 0.161±0.003 & 0.162±0.004 & 0.306±0.005\\
CONCH& 0.265±0.006 & 0.163±0.005 & 0.113±0.004 & 0.084±0.003 & 0.117±0.003 & 0.226±0.005\\
PLIP& 0.269±0.005 & 0.148±0.003 & 0.093±0.003 & 0.061±0.002 & 0.106±0.003 & 0.201±0.002\\
\hline
\end{tabular}
\end{table*}

\begin{table*}
\centering
\caption{\textbf{Performance of foundation models in WSI report generation on PatchGastricADC22 dataset.} The best performing model for each metric is \textbf{bolded} and the second-best performing model is \underline{underlined}.
}\label{lab:PatchGastricADC}%
\begin{tabular}{lllllll}
\hline
 & BLEU-1 & BLEU-2 & BLEU-3 & BLEU-4 &METEOR &  ROUGE-L \\
 \hline
ResNet50& 0.596±0.019 & 0.496±0.021 & 0.424±0.022 & 0.369±0.023 & 0.301±0.011 & 0.564±0.021\\
Phikon& \pmb{0.655±0.025} & \pmb{0.577±0.027} & \pmb{0.522±0.029} & \pmb{0.481±0.030} & \pmb{0.347±0.015} & \pmb{0.623±0.026} \\
Ctranspath& 0.643±0.020 & 0.556±0.022 & 0.495±0.023 & 0.450±0.025 & 0.334±0.012 &  0.598±0.022\\
UNI& 0.609±0.023 & 0.533±0.025 & 0.482±0.027 & 0.444±0.028 & 0.327±0.014 & 0.596±0.024\\
CONCH& 0.641±0.019 & 0.555±0.023 & 0.495±0.025 & 0.450±0.026 & 0.337±0.013 &  0.599±0.022\\
PLIP& 0.650±0.021 & 0.560±0.024 & 0.498±0.026 & 0.451±0.027 & 0.338±0.013 & 0.599±0.023\\
CHIEF& 0.594±0.021 & 0.494±0.023 & 0.425±0.025 & 0.372±0.026 & 0.298±0.013 & 0.561±0.025 \\
Prov-Gigapath & 0.637±0.022 & 0.555±0.025 & 0.497±0.026 & 0.454±0.027 & 0.338±0.013 & 0.601±0.023 \\
GPFM& \underline{0.651±0.021} & \underline{0.569±0.023} & \underline{0.512±0.025} & \underline{0.470±0.026} & \underline{0.343±0.013} & \underline{0.606±0.025}\\
\hline
\end{tabular}
\end{table*}

\begin{table*}
\centering
\caption{\textbf{Human-based blind evaluation of foundation models in WSI report generation on TCGA WSI-report dataset, where the generated reports of breast, lung, and kidney cancers are used for evaluation.} The number of reports in each score rated by the pathologist is listed and the average score is reported. The best performing model for each metric is \textbf{bolded} and the second-best performing model is \underline{underlined}.
}\label{lab:WSI-report-human}%
\begin{tabular}{llllll}
\hline
 & Score: 0 & Score: 0.3 & Score: 0.7 & Score: 1 & Avg. \\
 \hline
\multicolumn{6}{c}{Breast} \\
\hline
ResNet50& 184 & 4  & 0  & 0  &  0.01   \\
Phikon& 18 & 136 & 34  & 0 &  \underline{0.34}  \\
Ctranspath& 80  & 88  & 20  & 0  &  0.21   \\
UNI& 9 & 134  & 45  & 0  &  \pmb{0.38}   \\
CONCH& 36 & 125  & 27  & 0  & 0.30  \\
PLIP&  118 &  69 & 1  &  0 &  0.11   \\
CHIEF& 25 & 139  & 24 & 0  &  0.31  \\
Prov-Gigapath&  24 & 127  & 37  & 0 &  \underline{0.34} \\
GPFM& 15  & 126  & 47  &  0 &  \pmb{0.38}   \\
\hline
\multicolumn{6}{c}{Lung} \\
\hline
ResNet50& 153  & 3  &  1 & 0  & 0.01  \\
Phikon& 7 & 109  & 41  & 0  &  \underline{0.39} \\
Ctranspath&  19 & 87  & 51  & 0  &  \underline{0.39}  \\
UNI& 13  & 109 & 35 & 0  &  0.36   \\
CONCH& 11  & 120  & 26  & 0  &  0.35  \\
PLIP& 53 & 104 & 0  & 0  &  0.20  \\
CHIEF& 15 & 113 & 29 & 0  &  0.35   \\
Prov-Gigapath& 13 & 111  & 33  & 0 &  0.36   \\
GPFM& 6  & 100 & 51 & 0  &  \pmb{0.42}   \\
\hline
\multicolumn{6}{c}{Kidney} \\
\hline
ResNet50&  175 &  0 & 0  & 0  & 0.00    \\
Phikon& 5  & 86 & 84  & 0  & 0.48  \\
Ctranspath& 32  & 76 & 67 & 0 & 0.39 \\
UNI& 2  & 91  & 82  &  0  &  0.48 \\
CONCH& 11  & 118  & 46 & 0 &  0.39  \\
PLIP& 172  &  3 & 0  & 0  & 0.01  \\
CHIEF&  9 & 100  &  66 & 0  &  0.44  \\
Prov-Gigapath& 5  & 86  & 82 & 2 & \underline{0.49}  \\
GPFM& 2 & 83  & 90  & 0  & \pmb{0.50}   \\
\hline
\end{tabular}
\end{table*}

\begin{table*}
\centering
\caption{\textbf{Performance Comparison of DINOv2 and GPFM Pretraining Methods Across 12 Tasks.}
DINOv2 represents the pretrainined foundation model without using Expert Knowledge Distillation compared with GPFM.
Overall, the Expert Knowledge Distillation module shows an average improvement across balanced ACC, weighted F1 score, and AUC.
}\label{lab:linear:ablation}
\begin{tabular}{lllll}
\hline
  &Method& Balanced ACC & Weighted F1 & AUC  \\
  \hline
\multirow{2}{*}{CRC-100K} & DINOv2 & 0.845& 0.822& 0.990\\
 & GPFM & 0.896(+0.051)& 0.872(+0.050)&0.995(+0.005)\\
 \multirow{2}{*}{WSSS4LUAD}& DINOv2 
& 0.957& 0.956& 0.998
\\
 & GPFM & 0.961(+0.004)& 0.959(+0.003)&0.998(+0.000)\\
 \multirow{2}{*}{PCAM}& DINOv2 
& 0.925& 0.925& 0.976
\\
 & GPFM & 0.941(+0.016)& 0.942(+0.017)&0.988(+0.012)\\
 \multirow{2}{*}{PanCancer-TCGA}& DINOv2 
& 0.939& 0.940& 0.999
\\
 & GPFM & 0.951(+0.012)& 0.953(+0.013)&0.999(+0.000)
 \\
 \multirow{2}{*}{PanCancer-TIL}& DINOv2 
& 0.857& 0.864& 0.963
\\
 & GPFM & 0.894(+0.037)& 0.908(+0.044)&0.978(+0.015)\\
 \multirow{2}{*}{chaoyang}& DINOv2 
& 0.802& 0.808& 0.957
\\
 & GPFM & 0.797(-0.005)& 0.803(-0.005)&0.956(-0.001)\\
 \multirow{2}{*}{CCRCC-TCGA-HEL}& DINOv2 
& 0.945& 0.951& 0.996
\\
 & GPFM & 0.953(+0.008)& 0.956(+0.005)&0.997(+0.001)\\
  \multirow{2}{*}{BreakHis}& DINOv2 
& 0.984& 0.982& 0.999
\\
 & GPFM & 0.974(-0.008)& 0.976(-0.006)&0.998(-0.001)\\
  \multirow{2}{*}{BACH}& DINOv2 
& 0.922& 0.920& 0.990\\
 & GPFM & 0.963(+0.041)& 0.965(+0.045)&0.998(+0.008)\\
  \multirow{2}{*}{UniToPatho}& DINOv2 
& 0.457& 0.431& 0.844
\\
 & GPFM & 0.444(-0.013)& 0.433(+0.002)&0.844(+0.000)\\
  \multirow{2}{*}{CRC-MSI}& DINOv2 
& 0.679& 0.655& 0.777
\\
 & GPFM & 0.733(+0.054)& 0.672(+0.023)&0.812(+0.035)\\
  \multirow{2}{*}{ESCA}& DINOv2 
& 0.705& 0.705& 0.900
\\
 & GPFM & 0.732(+0.027)& 0.734(+0.029)&0.902(+0.002)\\
  \multirow{2}{*}{\pmb{Average}}& DINOv2 
& 0.835& 0.830& 0.949
\\
 & GPFM & 0.853(+0.018)& 0.848(+0.018)&0.955(+0.006)\\
 \hline
\end{tabular}
\end{table*}
\begin{table*}
    \centering
    \caption{\textbf{The configuration of different foundation models used for comparison.} The details of the datasets used in GPFM are shown in \textbf{Extended Data} Table \ref{tab:dataset_distributation}. UDK represents Unified Knowledge Distillation}
    \resizebox{\linewidth}{!}{\begin{tabular}{lllllll}
        \hline
         Model&  Data Source& WSIs & Patches & Model arch.  & Model size & Pretraining\\
         \hline
         ResNet50 \cite{resnet_paper}& ImageNet&  NA&  NA&  ResNet50&  25M&  Supervised\\
         Ctranspath \cite{ctranspath_paper}& TCGA+PAIP &32K& 4.2M& SwinTrans.&28M&MoCoV3 \cite{mocov3} \\
         Phikon \cite{filiot2023scaling}&TCGA&6K& 43M& ViT-B&86M& iBOT \cite{iBOT} \\
         UNI \cite{uni_paper}& Private+GTEx&100K& 100M& ViT-L & 307M&DINOv2 \cite{oquab2023dinov2} \\
         PLIP \cite{huang2023visual}& OpenPath & NA & 200K & ViT-B& 86M & CLIP\cite{clip_paper} \\
         CONCH \cite{lu2024visual}& PMC-Path +EDU & NA & 1.2M & ViT-B & 86M & CoCa \cite{coca} \\
 CHIEF \cite{wang2024pathology}& Public+Private& 60K& 15M& SwinTrans.& 28M&MoCoV3+CLIP\\
 Prov-Gigapath \cite{xu2024whole}& Private& 171K& 1.3B& Vit-g& 1.1B&DINOv2+MAE\\
         GPFM (our)& 33 Public datasets& 72K & 190M & ViT-L& 307M & UDK \\
         \hline
    \end{tabular}}
    \label{tab:sota_config}
\end{table*}
\begin{table*}
    \centering
    \caption{\textbf{The hyper parameters for pretraining the proposed foundation model.} The pretraining is conducted on 2 DGX nodes with 16$\times$80GB H800 GPUs.}
    %

    \label{tab:hyperparameter}
\end{table*}
\begin{table*}
\centering
\caption{\textbf{The architecture of ABMIL model and training details for WSI classification and survival analysis.}}
\label{tab:abmil_config}
\begin{tabular}{ll}
\hline
Architecture & Two-layer ABMIL \\
Embedding Dimension & 512 \\
Hidden Dimensions & 128\\
Dropout Rates & 0.25 \\
\hline
Optimizer & AdamW \\
Learning Rate & 2e-4 \\
WSI Classification Loss& Cross-entropy \\
 Survival Analysis Loss& NLL loss\\
Maximum Epochs & 100 \\
Early Stopping &  Yes\\
\hline
\end{tabular}
\end{table*}
\begin{table*}
    \centering
        \caption{The datasets used for survival analysis.}
    \begin{tabular}{ccc}
    \hline
         Dataset&  Cases&  WSIs\\
         \hline
         TCGA-BRCA&  1,023&  1,089\\
         TCGA-BLCA&  376&  446\\
 TCGA-KIRC& 498& 504\\
 TCGA-KIRP& 261& 285\\
 TCGA-STAD& 363& 389\\
 TCGA-CESC& 250& 260\\
 TCGA-LUAD& 455& 518\\
 TCGA-LUSC& 452& 484\\
 TCGA-COADREAD& 579& 588\\
 TCGA-GBM& 372 & 856 \\
 TCGA-LGG& 462 & 843 \\
 TCGA-SKCM& 415& 456\\
  TCGA-HNSC& 443 & 472 \\
         HANCOCK& 749 & 1078 \\
         \hline
    \end{tabular}

    \label{tab:survival_data_info}
\end{table*}
\begin{table*}
    \centering
\caption{\textbf{The number of slides and processed patches of 33 datasets used for pretraining foundation models.} "-" represents the dataset only providing ROIs.}
    %

    \label{tab:dataset_distributation}
\end{table*}
\begin{table*}
    \centering
    \caption{\textbf{The primary site of tissues used for pretraining foundation models and downstream tasks evaluation.}}
    %
    \label{tab:primariy_site}
\end{table*}

\begin{table*}
    \centering
\caption{\textbf{The public codes used in this study.} Please note that the pretrained weights of UNI and CONCH need to be permitted before downloading.
}
\label{tab:code_source}
    %

\end{table*}
\clearpage
\begin{table}
\caption{\textbf{The public datasets used in this study.} Please note that some datasets may need permission before downloading.}
    %
    \label{tab:data_links}
\end{table}

\end{appendices}


\end{document}